\documentclass[fleqn,usenatbib]{mnras}

\usepackage{newtxtext,newtxmath}

\usepackage[T1]{fontenc}
\usepackage{ae,aecompl}
\usepackage{xcolor}

\usepackage{graphicx}	
\usepackage{amsmath}	
\usepackage{amssymb}	

\newcommand{\garciamunoz}{{Garc\'ia Mu\~noz}}

   \title{Exoplanet phase curves at large phase angles. 
   Diagnostics for extended hazy atmospheres.}

\author[A. {\garciamunoz} \& J. Cabrera]{
A. {\garciamunoz}$^{1}$\thanks{E-mail: garciamunoz@astro.physik.tu-berlin.de; tonhingm@gmail.com}
and J. Cabrera$^{2}$
\\
$^{1}$Zentrum f\"ur Astronomie und Astrophysik, Technische Universit\"at Berlin, D-10623 Berlin, Germany\\   
$^{2}$Institut f\"ur Planetenforschung, Deutsches Zentrum f\"ur Luft- und Raumfahrt, D-12489, Berlin, Germany
}

\date{September 2017. Accepted for publication in MNRAS. 
}

\pubyear{2017}

\begin{document}
\label{firstpage}
\pagerange{\pageref{firstpage}--\pageref{lastpage}}
\maketitle

\begin{abstract}
At optical wavelengths,
Titan's brightness for large Sun-Titan-observer phase angles significantly exceeds its 
  dayside brightness.
  The brightening that occurs near back-illumination  
  is due to moderately large haze particles in the moon's extended atmosphere 
  that forward-scatter the incident sunlight.
  Motivated by this phenomenon, here we investigate the forward scattering 
  from currently known exoplanets, {its diagnostics possibilities,} 
  the observational requirements to resolve it, and potential implications.
  An analytical expression is derived for the amount of starlight
  forward-scattered by an exponential atmosphere {that 
  takes into account the finite angular size of the star.
  We use this expression to tentatively estimate how prevalent 
  this phenomenon may be.
  }
  {Based on numerical calculations that consider exoplanet visibility,}
  we identify numerous planets with predicted out-of-transit forward scattering signals of up to tens of parts-per-million 
  provided that aerosols of $\gtrsim$1 $\mu$m size form over an 
  extended vertical region near the optical radius level.
  We propose that the interpretation of available optical
  phase curves should be revised to constrain the strength of this phenomenon that might
  provide insight into aerosol scale heights and particle sizes. 
  {
  For the relatively general atmospheres considered here, forward scattering reduces the 
  transmission-only transit depth by typically less than the equivalent to a scale height. 
  For short-period exoplanets 
  the finite angular size of the star severely affects the amount of radiation 
  scattered towards the observer at mid-transit.
  } 
\end{abstract}




\section{Introduction}

\begin{figure}
   \centering
   \includegraphics[width=9.cm]{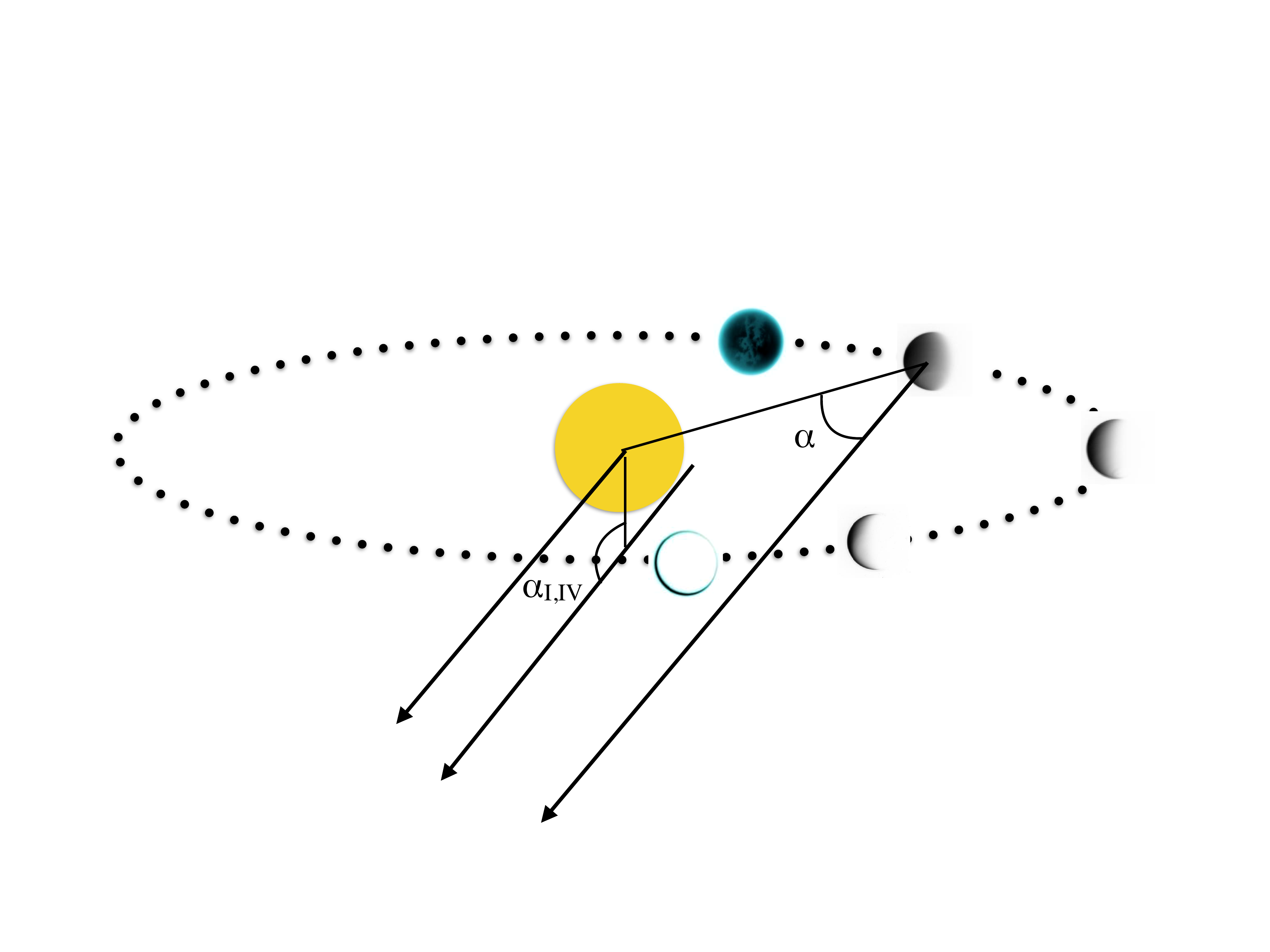}
   \includegraphics[width=9.cm]{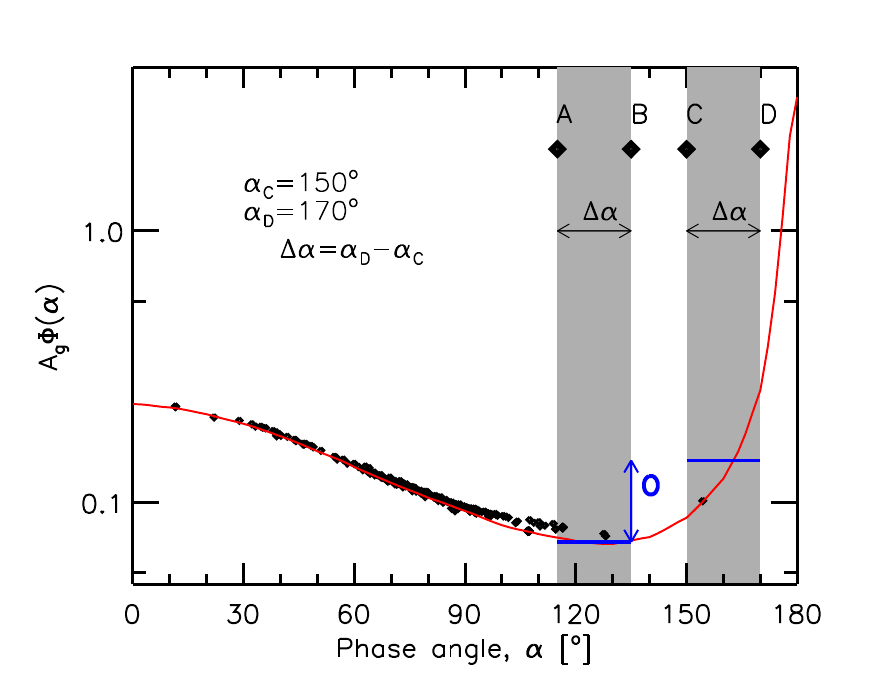}
      \caption{\label{cartoon_fig} \textbf{Top.} Orbit sketch.  
      $\alpha$ is the star-planet-observer phase angle 
      and $\alpha_{\rm{I,IV}}$ the angle at first or fourth contacts, i.e.
      immediately before or after transit. 
      The arrows indicate the direction towards the observer. 
      \textbf{Bottom.} Titan's phase curve from Cassini/ISS 
       white-light photometry (effective wavelength of 650 nm) (black symbols) and best-fit model
       based on solving the radiative transfer problem with a realistic description of the
       atmosphere 
      (red curve) \citep{garciamunozetal2017}. 
      The two grey intervals span an identical range $\Delta\alpha$ of phase angles, and 
      indicate the forward scattering and control intervals (see text). The observable \textbf{O} is
      defined as the difference in the mean values of the planet brightness over the two
      intervals. In the example, $\alpha_{\rm{D}}$=170$^{\circ}$ and $\Delta\alpha$=20$^{\circ}$.
      {
      In the examples with real planets, we always assume $\alpha_{\rm{D}}$=$\alpha_{\rm{I,IV}}$.}
     }
   \end{figure}

Given the limited possibilities that exist for the remote sensing of exoplanet atmospheres, 
it is crucial to understand the information contained in each observing technique
and the synergies between them.  
In that setting, 
this work aims to show that brightness measurements at 
large star-planet-observer phase angles potentially inform on
atmospheric properties such as the scale height and scattering properties of aerosols in 
the uppermost atmospheric layers. 
Our investigation is motivated by recent work on 
Saturn's moon Titan \citep{garciamunozetal2017} 
showing that Titan brightens up at phase angles $\alpha$$>$150$^{\circ}$
and that, when back-illuminated, it becomes brighter than in full illumination
by a {wavelength-dependent} factor of 10--200. 
The presence of forward-scattering haze in Titan's 
 extended atmosphere is key to the occurrence of this optical phenomenon.
Its prospective detection at an exoplanet will allow us to infer 
the occurrence of haze and, more importantly, will provide insight into its vertical distribution and 
particle size near the optical radius level.
Figure (\ref{cartoon_fig}) sketches the phenomenon.


The effect of forward scattering on the measured radius of transiting planets
has been considered before 
\citep{brown2001,hubbardetal2001,garciamunozetal2012,dekokstam2012,robinson2017}.
{
In particular, 
\citet{dekokstam2012} note that it may bias the 
transit radius by up to a few scale heights in specific cases, and
\citet{robinson2017} observes that the bias can be of hundreds of parts-per-million (ppm) 
for hot Jupiters. None of these works provide an easy way to quantitatively estimate the effect, or
its connection with the stratification and size of the dominating atmospheric particles.}
{As shown later, the finite angular size of the star as viewed from the planet
limits the amount of starlight forward-scattered towards the observer during the transit, and  
forward scattering will affect the measured transit radius by less than a scale height
in typical configurations with scattering particles of up to a few $\mu$m in size.}
Our treatment here differs from the above works in that we focus preferentially on 
orbital phases immediately before or after transit. 
This configuration is better suited to identify the forward scattering contribution.

To date, the best evidence for forward scattering from exoplanets
{comes from 
ultra-short period planets on orbits of less than one day.} 
In a few such systems, the  shape of the pre-/post-transit brightness curve 
is attributed to starlight scattered from dust clouds surrounding the planets 
\citep[e.g.][]{budaj2013,devoreetal2016}.
Since the dust is plausibly of planetary origin, these planets are thought to be
disintegrating.
Refraction may also produce shoulders in the pre-/post-transit brightness curve
\citep{huiseager2002,sidissari2010,garciamunozmills2012,garciamunozetal2012,misrameadows2014}.
Refraction lensing of starlight by the planet atmosphere
competes with extinction within the atmosphere. 
As a result, a brightness surge due to refraction 
will be prominent only on planets with clear, aerosol-free envelopes
\citep{garciamunozmills2012,misrameadows2014}.
{Contrary to forward scattering, 
refraction lensing becomes significant for planets on relatively long-period orbits
\citep{sidissari2010, misrameadows2014}. This distinction should make it posible to identify whether
the brightness surge is due to refraction or to forward scattering.}

The paper is structured as follows. 
In \S\ref{scatflash_sec} we summarize the findings on Titan 
that motivate this study and generalize them for application to exponential atmospheres. 
In \S\ref{sec:hazy_sec} we describe, 
through combined analytical and numerical work, 
the planet properties more favourable for forward scattering.  
Based on a zeroth-order characterization, we attempt to classify 
the known exoplanets according to their potential for forward scattering. 
In \S\ref{sec:detectability_sec}  
we elaborate further on the detectability of this phenomenon out of transit.
In \S\ref{sec:otherterms_sec} we comment on the blending 
with the brightness modulation due to stellar tides, and on the impact 
upon the measured transit radius.
Finally, \S\ref{sec:summary_sec} summarizes the main conclusions and presents avenues for
follow-up studies. 
{In the appendices, we derive an analytical expression for the amount of
starlight forward scattered by a planet with an exponential atmosphere, 
comment on the accuracy of the single scattering approximation, and describe the modifications
to our numerical radiative transfer model to take into account the finite angular size 
of the star.}
 
\section{Forward scattering}
\label{scatflash_sec}

\subsection{Titan}

The brightness phase curve of Titan is quite unique. 
Titan dims as it passes from  phase angles $\alpha$=0 to $\alpha$$\sim$120$^{\circ}$ 
due to the decreasing area of the dayside visible to the observer
\citep{tomaskosmith1982,westetal1983}.
For larger phase angles, however, the diminishing size of the visible dayside 
is compensated by forward scattering from the abundant upper-atmosphere haze 
and the whole-disk brightness increases again. 
Observations made with the Cassini Imaging Science Subsystem 
have revealed that at $\alpha$$\sim$165$^{\circ}$ Titan becomes as 
bright as in full illumination \citep{garciamunozetal2017}. 
An empirically-constrained prediction of that study is that 
for $\alpha$$\rightarrow$180$^{\circ}$, 
Titan's twilight appears brighter than its dayside 
by a factor of $\sim$10 at wavelengths of
$\sim$1 $\mu$m and by factors of up to 200 at wavelengths of $\sim$300 nm.

This behaviour is due to the facts that 
Titan has an atmosphere that is both extended and hazy, 
and that the haze particles are moderately large
(equal-projected-area radii $\sim$2--3 $\mu$m) and thus efficient at forward scattering
\citep{ragesetal1983,westsmith1991}. 
The haze is produced photochemically through reactions initiated in the upper
atmosphere \citep{lavvasetal2010}.
Forward scattering from Titan originates within a
few scale heights from the level at which the atmosphere is optically thick when viewed
through the limb. 
This is similar to the optical radius level 
probed during a hypothetical transit of Titan across the
solar disk \citep{karkoschkalorenz1997, lecavelierdesetangsetal2008}. 
Near that level, 
the number densities of the gas and haze drop in altitude with
comparable scale heights, $H$$\sim$$H_{\rm{a}}$$\sim$45 km.
If $R_{\rm{T}}$ stands for Titan's optical radius ($\sim$3,000 km, dependent on wavelength), 
$H_{\rm{a}}$/$R_{\rm{T}}$$\sim$1.5$\times$10$^{-2}$.

\subsection{Exponential atmospheres}

Next, we identify the key planet properties that result in strong forward scattering. 
Appendix \ref{sec:appendixa} elaborates on exponential atmospheres described in
terms of an average scattering particle and a single scale height. 
We will refer to the average scattering particles as aerosols, 
although they may actually represent a mix of gases and condensates in the atmosphere. 
For an exponential atmosphere, the aerosol number density decays as 
$n_{\rm{a}}$($r$)=$n_{\rm{a}}$($R_{\rm{0}}$)$\exp{(-(r-R_{\rm{0}})/H_{\rm{a}})}$,
where $r$ is the radial distance to the planet centre and 
$R_{\rm{0}}$ is a reference level. 
$H_{\rm{a}}$ is the aerosol scale height. 

In this idealized scenario and $\alpha$$\rightarrow$180$^{\circ}$, 
single scattering dominates {(Appendix \ref{sec:appendixb})} 
and the planet-to-star contrast is approximately {(Appendix \ref{sec:appendixa})}:
\begin{equation}
\frac{F_{\rm{p}}}{F_{\star}} (\alpha=180^{\circ}) \approx 
\underbrace{2\pi p_{\rm{a}}(\theta=0) \varpi_{0,\rm{a}} \frac{H_{\rm{a}}}{R_{\rm{p}}}} \left(\frac{R_{\rm{p}}}{a} \right)^2.
\label{FpFstar_eq}
\end{equation}
Here, $p_{\rm{a}}$$(\theta$=0) and $\varpi_{0,\rm{a}}$ refer to the aerosol 
scattering phase function in the forward direction (scattering angle $\theta$=0) and the aerosol 
single scattering albedo, respectively. 
{($p_{\rm{a}}$$(\theta$=0) is the relevant phase function when
the angular size of the star as viewed from the planet is small, i.e. in the
point-like star limit; otherwise a
generalized form $\mbox{<}p_{\rm{a}}\mbox{>}$$(\Theta$=0) should be used, Appendix
\ref{sec:appendixa}.)
}
$H_{\rm{a}}$/$R_{\rm{p}}$ is the ratio of the aerosol scale height to the
planet radius, and $R_{\rm{p}}$/$a$ the ratio of the planet radius to the orbital distance. 
The geometrical terms in Eq. (\ref{FpFstar_eq}) can be re-arranged into
2$\pi$$R_{\rm{p}}$$H_{\rm{a}}$/$a^2$, the numerator of which is
the projected area of a ring of radius $R_{\rm{p}}$ and width $H_{\rm{a}}$. 
{This ring, which concentrates most of the forward-scattered starlight,
is seen in large phase angle images of Titan \citep{garciamunozetal2017}.}
Equation (\ref{FpFstar_eq}) is analogous to the usual representation of the 
planet-to-star brightness contrast in full illumination ($\alpha$=0)
 if the underlined terms are replaced by the geometric albedo, $A_{\rm{g}}$ 
 (see Eq. \ref{FpFstar2_eq} below). 
Equation (\ref{FpFstar_eq}) enables the direct comparison of the
brightness of a planet when it is fully illuminated ($\alpha$=0) and back-illuminated 
($\alpha$=180$^{\circ}$). 
 
The single scattering albedo $\varpi_{0,\rm{a}}$ is of order one for many
plausible condensates in exoplanet atmospheres \citep{budajetal2015, wakefordsing2015}. 
However, both $p_{\rm{a}}$$(\theta$=0) and $H_{\rm{a}}$/$R_{\rm{p}}$ are likely to 
differ by orders of magnitude amongst different planets depending on the specifics of their 
atmospheres. {(However, for short-period planets 
the finite angular size of the star will limit the
effective scattering phase function to $\mbox{<}p_{\rm{a}}\mbox{>}$$(\Theta$=0), which may
be much smaller than $p_{\rm{a}}$$(\theta$=0); Appendix \ref{sec:appendixa}.)}
$H_{\rm{a}}$/$R_{\rm{p}}$ is a measure of how puffy the atmosphere is,
meaning that large ($\sim$10$^{-2}$) values are associated with extended envelopes. 
It is seen from Eq. (\ref{FpFstar_eq}) that 
for a given $R_{\rm{p}}$/$a$ (measurable for transiting systems), 
the strength of forward scattering depends on $p_{\rm{a}}$$(\theta$=0)$H_{\rm{a}}$/$R_{\rm{p}}$.
Also according to Eq. (\ref{FpFstar_eq}), 
the strength of this effect depends on orbital distance as $a^{-2}$. 
This interpretation is however likely oversimplistic 
as the orbital distance will foreseeably affect the planet temperature
and therefore $p_{\rm{a}}$($\theta$) through
the microphysics that enables aerosol formation. 
{Also, for small orbital distances, the finite size of the star 
tends to reduce the relevant $\mbox{<}p_{\rm{a}}\mbox{>}$$(\Theta$=0)
with respect to $p_{\rm{a}}$$(\theta$=0).}

\begin{figure}
   \centering
   \includegraphics[width=9.cm]{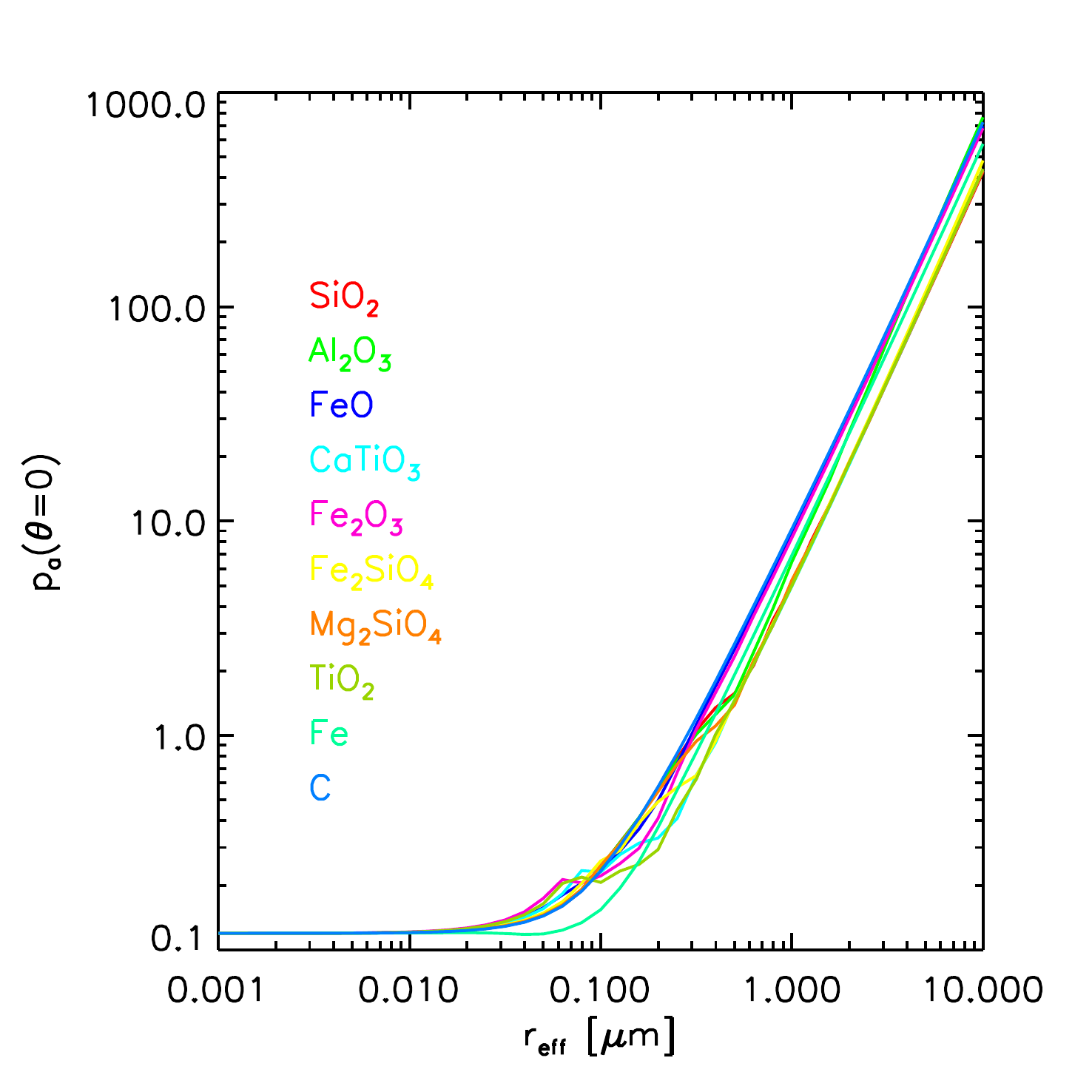}
   \includegraphics[width=9.cm]{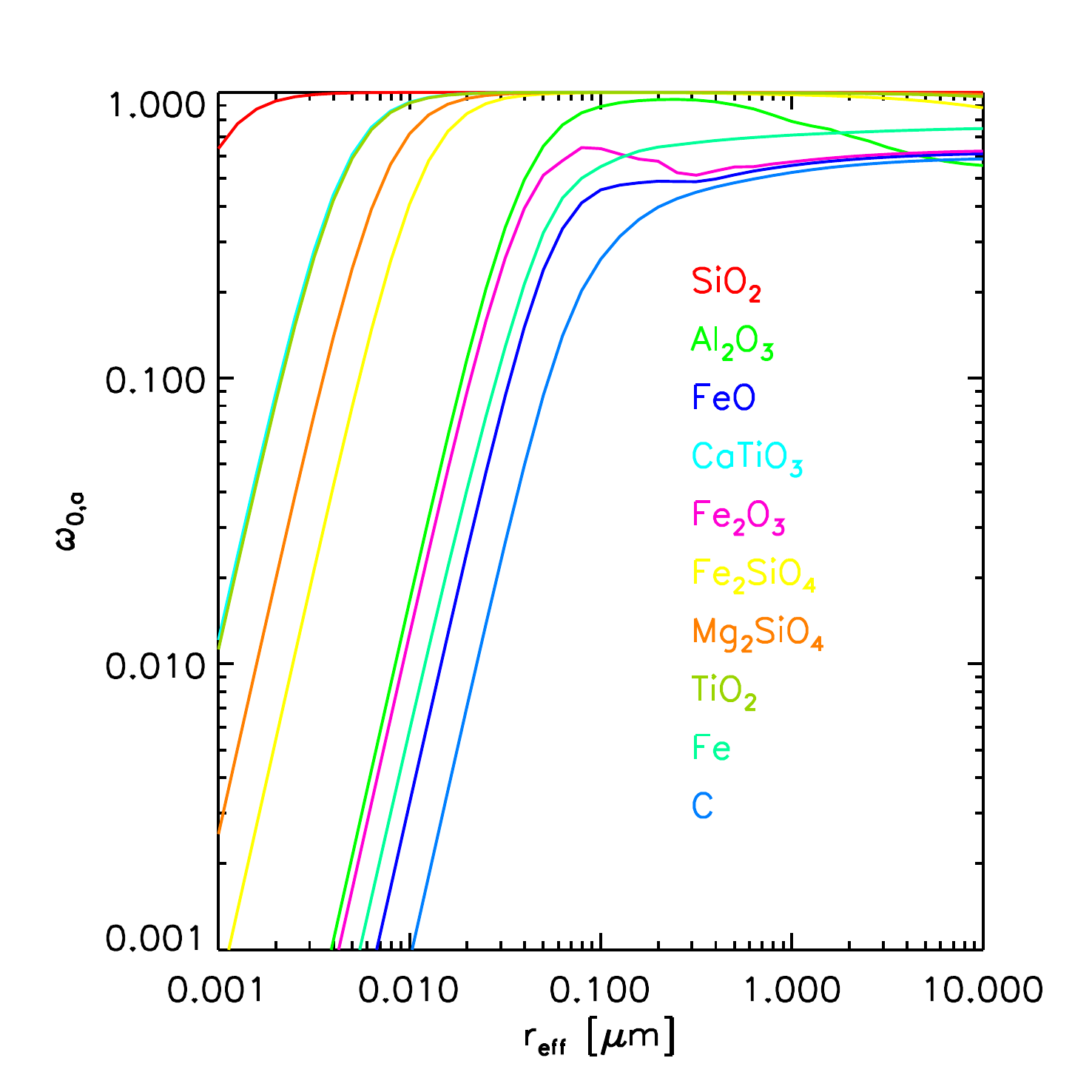}
      \caption{\label{passa_fig} \textbf{Top.} Scattering phase function in forward
      scattering $p_{\rm{a}}$($\theta$=0) for the condensates of Table (\ref{composition_table}). 
      {(The finite angular size of the star reduces the effective
       scattering phase function $<$$p_{\rm{a}}$($\Theta$=0)$>$ with respect to 
      $p_{\rm{a}}$($\theta$=0); see Fig. (\ref{paconv_fig}).)}
      {We normalize $p_{\rm{a}}$($\theta$) so that its integral over 4$\pi$ is
      equal to one (Appendix \ref{sec:appendixa}).}
      \textbf{Bottom.} For the same condensates, their corresponding single scattering 
      albedo, $\varpi_{\rm{0,a}}$. 
     }
   \end{figure}

To illustrate how forward scattering affects the planet brightness 
at configurations other than $\alpha$=180$^{\circ}$, 
we produced numerical solutions to the problem of multiple
scattering in spherical,  exponential atmospheres. 
We generally assumed that the aerosols scatter following Mie theory, 
and that the photon wavelength is $\lambda_{\rm{eff}}$=0.65 $\mu$m. 
For the aerosol particles we assumed a power law size distribution with
effective radii $r_{\rm{eff}}$ ranging from 0.01 to 10 $\mu$m  
and a fixed effective variance $v_{\rm{eff}}$=0.1 \citep{hansentravis1974}. 
We adopted refractive indices ($n$=$n_{\rm{r}}$+i$n_{\rm{i}}$) specific to a few plausible 
condensates listed in Table (\ref{composition_table}).
The selection of condensates does not rank them by relevance in the context of
exoplanet atmospheres. Rather, it simply tries to include a variety 
of refractive indices.
In Mie theory, $p_{\rm{a}}$($\theta$) depends on all three properties: 
$x_{\rm{eff}}$=2$\pi$$r_{\rm{eff}}$/$\lambda_{\rm{eff}}$, $v_{\rm{eff}}$ and $n$. 
More specifically, $p_{\rm{a}}$($\theta$=0) 
depends strongly on $x_{\rm{eff}}$ but weakly on $v_{\rm{eff}}$ and $n$, 
and typically increases as $x_{\rm{eff}}$ increases, 
which establishes a diagnostic connection between the particle size 
and the strength of forward scattering. 
The implemented single scattering albedos $\varpi_{0,\rm{a}}$ were also calculated from Mie theory. 
Figure (\ref{passa_fig}) shows the calculated $p_{\rm{a}}$($\theta$=0) and $\varpi_{0,\rm{a}}$. 
For comparison, we also produced phase curves based on atmospheres with Titan-like haze 
at an effective wavelength $\lambda_{\rm{eff}}$=600 nm.
In these cases, we adopted $p_{\rm{a}}$($\theta$) as reported in Table 1 of 
\citet{tomaskoetal2008}, and for $\varpi_{0,\rm{a}}$ we simply experimented 
with values between 0.2 and 1.

\begin{table}
\caption{
Refractive indices of condensates 
at $\lambda_{\rm{eff}}$=0.65 $\mu$m,  based on 
\citet{budajetal2015} and \citet{wakefordsing2015}. 
{(The latter quote condensation temperatures at 1 mbar pressure.)} 
Note: 
DOCDD: http://www.astro.uni-jena.de/Laboratory/OCDB/. 
For the calculation of optical properties, 
we used Mie theory (www.giss.nasa.gov/staff/mmishchenko/t$\_$matrix.htmlS).}             
\label{composition_table}
\centering                          
\begin{tabular}{c c c c}        
\hline                        
Composition & $n_{\rm{r}}$ & $n_{\rm{i}}$ & Ref. \\
\hline                        
SiO$_2$ & 1.5 & 10$^{-7}$ & see \citet{garciamunozisaak2015} \\
Al$_2$O$_3$ & 1.56 & 1.3$\times$10$^{-2}$ & \citet{koikeetal1995} \\
FeO & 2.42 & 0.60 & DOCDD \\
CaTiO$_3$ & 2.25 & 10$^{-4}$ & see \citet{garciamunozisaak2015} \\
Fe$_2$O$_3$ & 2.84 & 0.23 & DOCDD \\
Fe$_2$SiO$_4$ & 1.85 & 7.7$\times$10$^{-4}$ & DOCDD \\
Mg$_2$SiO$_4$//MgSiO$_3$ & 1.6 & 10$^{-4}$ & see \citet{garciamunozisaak2015} \\  
TiO$_2$ & 2.57 & 1.8$\times$10$^{-4}$ & DOCDD \\
Fe & 2.92 & 3.10 & \citet{johnsonchristy1974}\\
C & 1.59 & 0.73 & DOCDD \\
\hline                        
\end{tabular}
\end{table}

For the smaller orbital distances, the star appears as an extended object as viewed from 
the planet, a fact that must be considered in the implementation of the aerosols 
scattering phase function going into the multiple scattering calculations. 
The way to deal with this is to 
convolve $p_{\rm{a}}$($\theta$) with the star disk brightness
\citep{budajetal2015,devoreetal2016} in the evaluation of the starlight entering the atmosphere 
{(\S\ref{lammer_sec}, and Appendices \ref{sec:appendixa} and \ref{sec:appendixc})}. 
The {resulting} effective scattering phase function $\mbox{<}p_{\rm{a}}\mbox{>}$($\Theta$)
 depends on the limb-darkening law and the angular size of the star. 
For $p_{\rm{a}}$($\theta$) functions associated with strong forward scattering,
the effective scattering phase function for deflections larger than the angular radius of the
star, i.e. $\Theta$$>$$\theta_{\star}$$=$$\arcsin{R_{\star}/a}$, is usually 
larger than the non-convolved $p_{\rm{a}}$($\theta$) (Fig. 3, \citet{budajetal2015};
Fig. \ref{paconv_fig} in Appendix \ref{sec:appendixa}). 
{
The opposite is generally true for $\Theta$$<$$\theta_{\star}$.} 
In contrast, 
for $p_{\rm{a}}$($\theta$) functions associated with mild forward scattering,
the convolution process has little impact on the effective scattering phase function
$\mbox{<}p_{\rm{a}}\mbox{>}$($\Theta$) 
(Fig. 3, \citet{budajetal2015}; Figs. 3 and A2, \citet{devoreetal2016}).
{
For simplicity, as the convolution process is specific to each planet-star system,
we have omitted this effect from most of the multiple scattering calculations
done here. 
Its omission will tend to increase the forward scattering signal towards the observer
at out-of-transit orbital phases.} 
{Therefore, the calculations presented here in the point-like star limit
for pre-/post-transit configurations generally underestimate the actual 
forward scattering signal received by the observer. In \S\ref{lammer_sec}, we provide examples of how the finite 
angular size of the star will impact the brightness phase curve in the specific case of the
exoplanet CoRoT-24b.
}

It is convenient to present the planet-to-star contrast in a manner that
separates the various geometric and non-geometric factors:
\begin{equation}
\frac{F_{\rm{p}}}{F_{\star}}(\alpha) = A_{\rm{g}} \Phi(\alpha) \left(\frac{R_{\rm{p}}}{a} \right)^2.
\label{FpFstar2_eq}
\end{equation}
Here, $A_{\rm{g}}$ is the geometric albedo and $\Phi$($\alpha$) ($\Phi$($\alpha$$\equiv$0)=1)
the planet phase law.
The definition of $R_{\rm{p}}$ is somewhat arbitrary for planets with extended atmospheres. 
Because we are mainly interested in gas planets {with large scale heights}, 
we will use for $R_{\rm{p}}$ the optical radius,
which is based on the limb-viewing optical thickness of the atmosphere, $\tau_{\rm{limb}}$.
The optical radius at an effective wavelength $\lambda_{\rm{eff}}$
is calculated from the condition
\citep{karkoschkalorenz1997,lecavelierdesetangsetal2008}:
\begin{equation}
\tau_{\rm{limb}}(R_{\rm{p}})=\tau_{\rm{nadir}}(R_{\rm{p}}\rightarrow {\rm{TOA}}) 
(2\pi R_{\rm{p}}/H_{\rm{a}})^{1/2}=0.56, 
\label{optrad1_eq}
\end{equation}
and the optical thickness from the optical radius level $R_{\rm{p}}$ to the top of the
atmosphere (TOA):
\begin{equation}
\tau_{\rm{nadir}}(R_{\rm{p}}\rightarrow {\rm{TOA}})=\tau_{\rm{nadir}, 0} 
\exp{(-(R_{\rm{p}}-R_{\rm{0}})/H_{\rm{a}})}.
\label{optrad2_eq}
\end{equation}
The square root term in Eq. (\ref{optrad1_eq})
is the approximate conversion factor
between limb- and nadir-integrated columns in exponential atmospheres. 
$\tau_{\rm{nadir}, 0}$ is the nadir optical thickness upwards of the $R_{\rm{0}}$ 
reference level. 
We could take $R_{\rm{0}}$ deep enough into the planet and 
$\tau_{\rm{nadir}, 0}$ large enough so that the exponential description of the atmosphere
effectively reaches to all depths. 
Instead, and to alleviate the computational
cost of the calculations, we implemented finite values for 
$R_{\rm{0}}$ (equal to a Jupiter radius, $R_{\rm{J}}$) and $\tau_{\rm{nadir}, 0}$ (=10).
Also, the atmosphere below the $R_{\rm{0}}$ level was replaced by a black surface.
The truncation of the atmosphere at $R_{\rm{0}}$ will affect the planet's overall 
reflectance at the
smaller phase angles, but not at large phase angles because in the latter
viewing configuration the stellar photons will not penetrate to such depths.
$R_{\rm{p}}$ {($>$$R_{\rm{0}}$)} can be solved numerically from Eqs. (\ref{optrad1_eq})--(\ref{optrad2_eq})
for a given scale height $H_{\rm{a}}$. 

\begin{figure*}
   \centering
   \includegraphics[width=4.9cm]{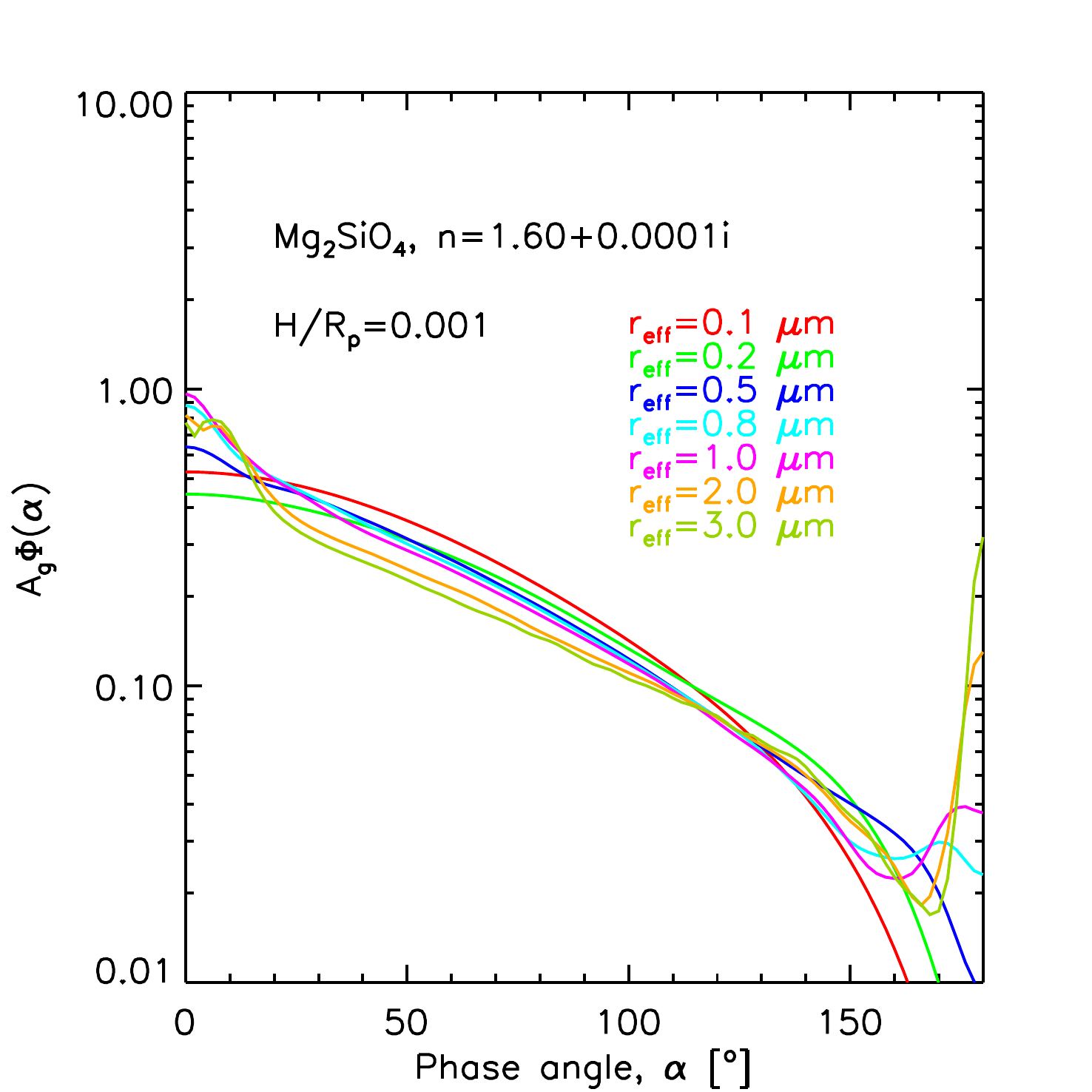} \hspace{-0.6cm} \includegraphics[width=4.9cm]{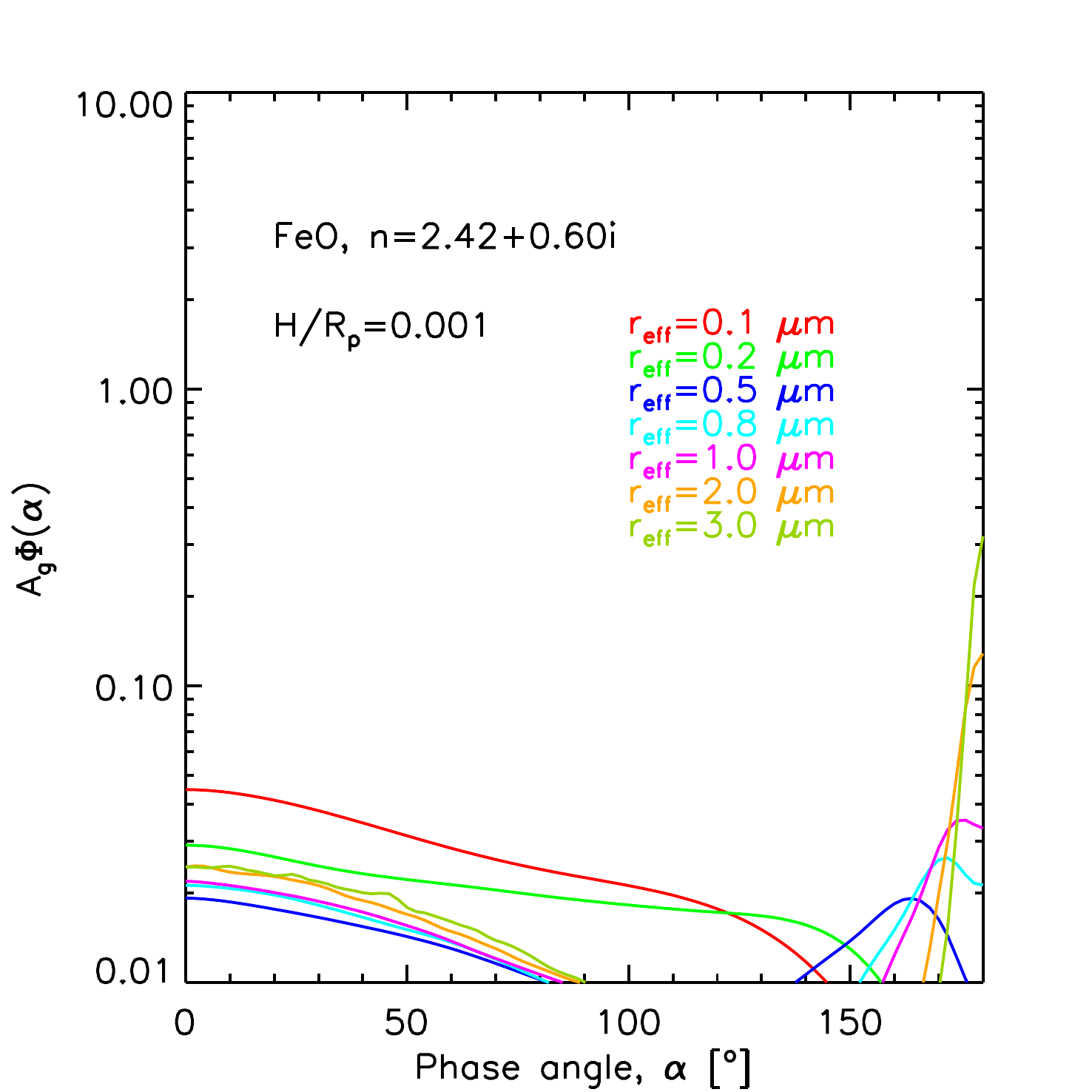} \hspace{-0.6cm} \includegraphics[width=4.9cm]{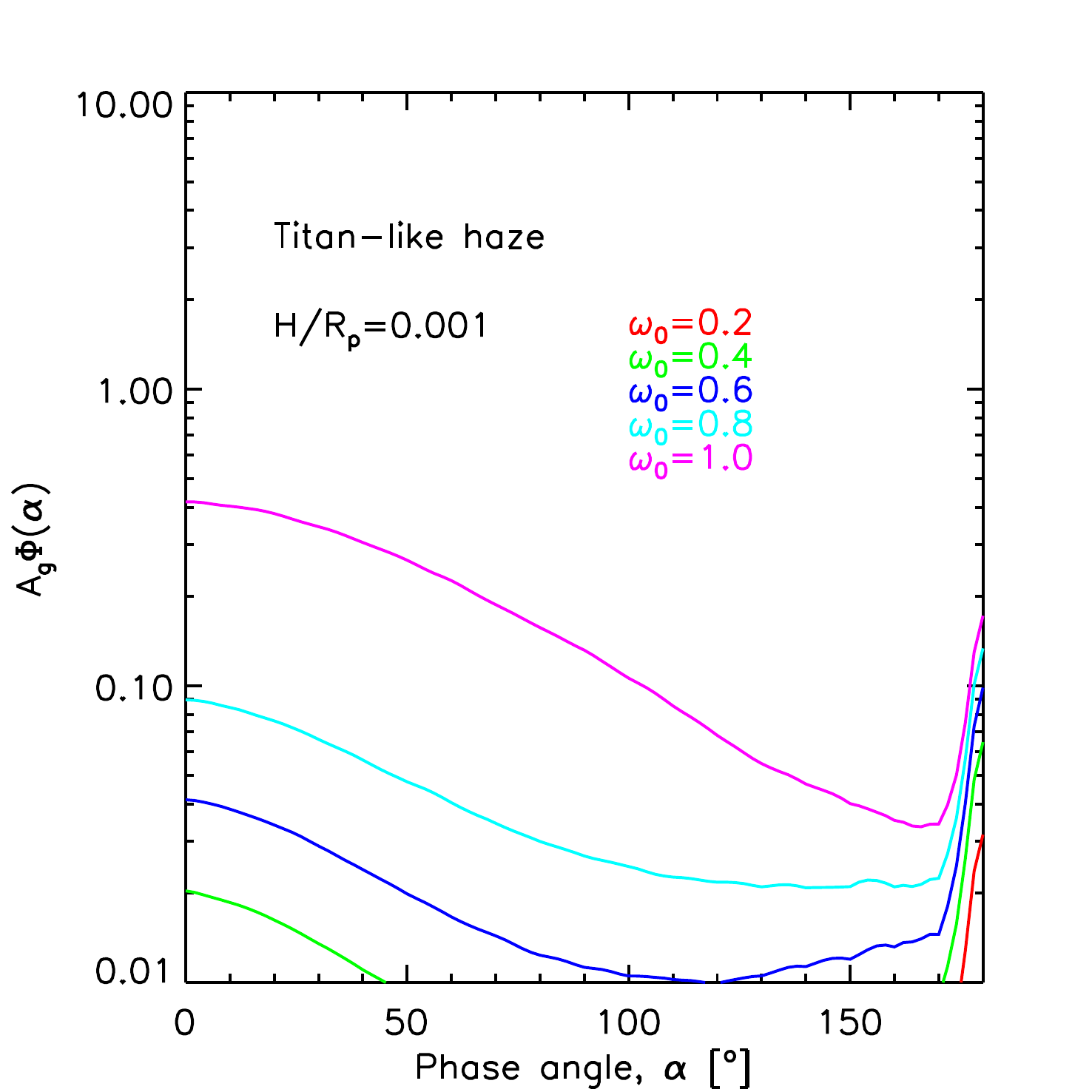} \\
   \includegraphics[width=4.9cm]{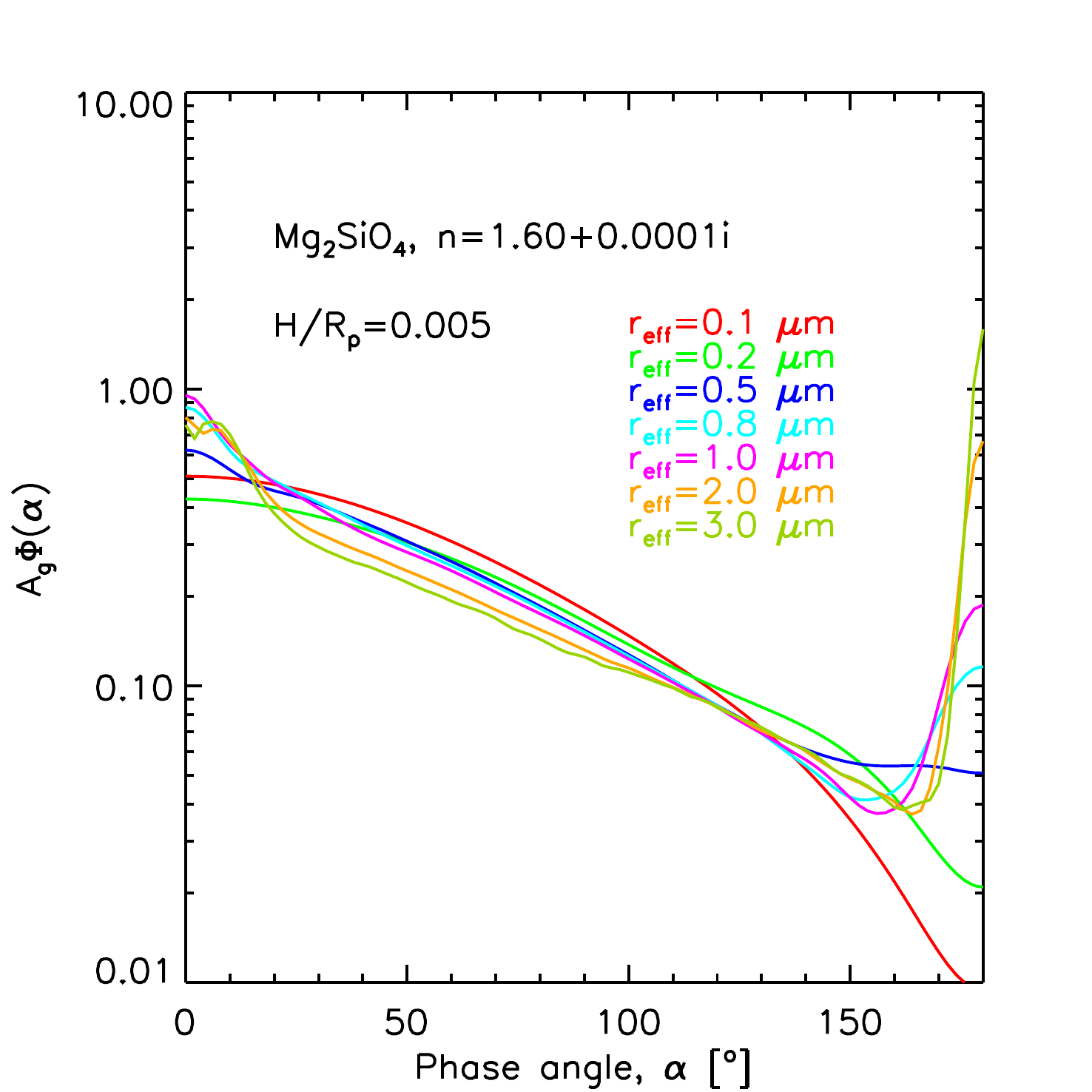} \hspace{-0.6cm} \includegraphics[width=4.9cm]{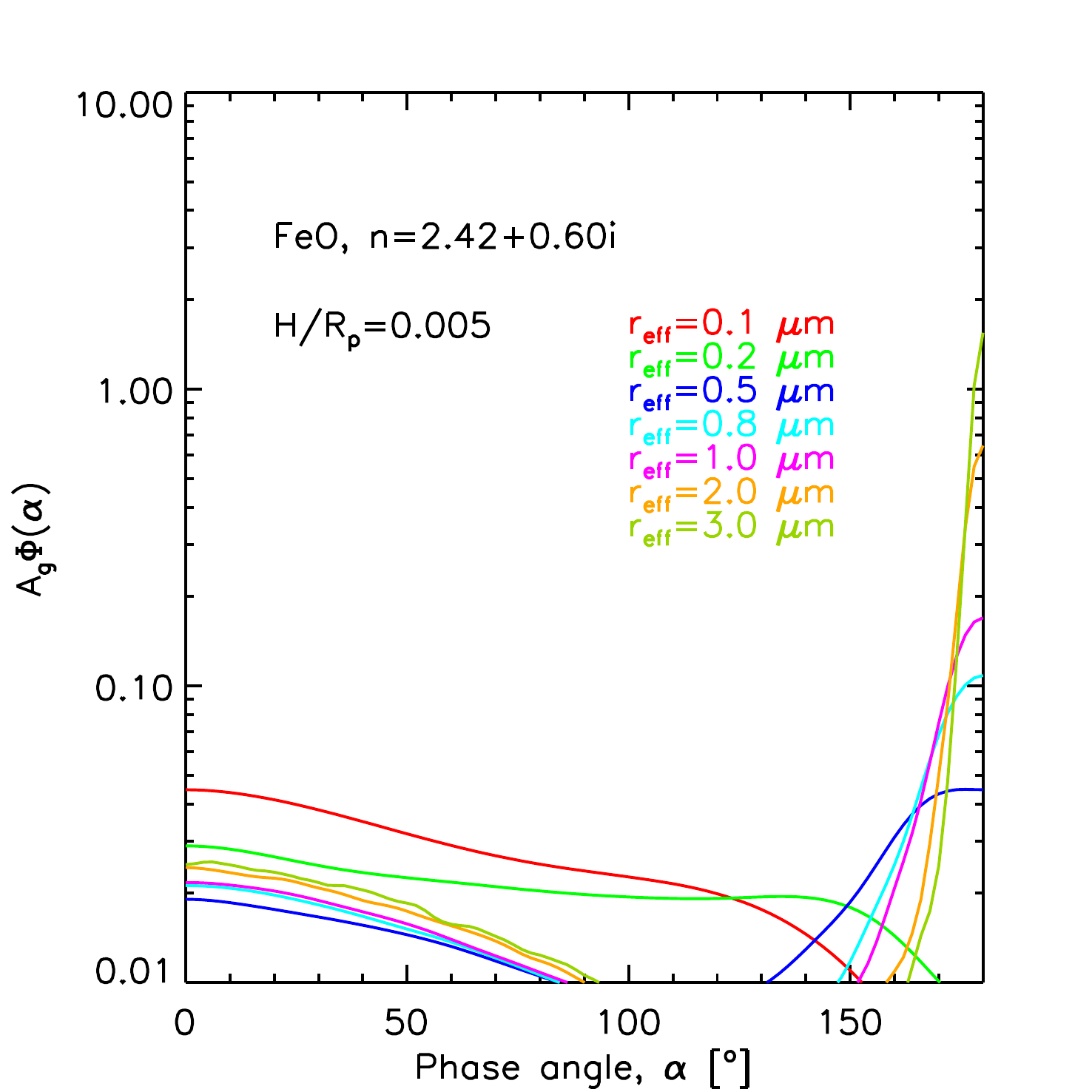} \hspace{-0.6cm} \includegraphics[width=4.9cm]{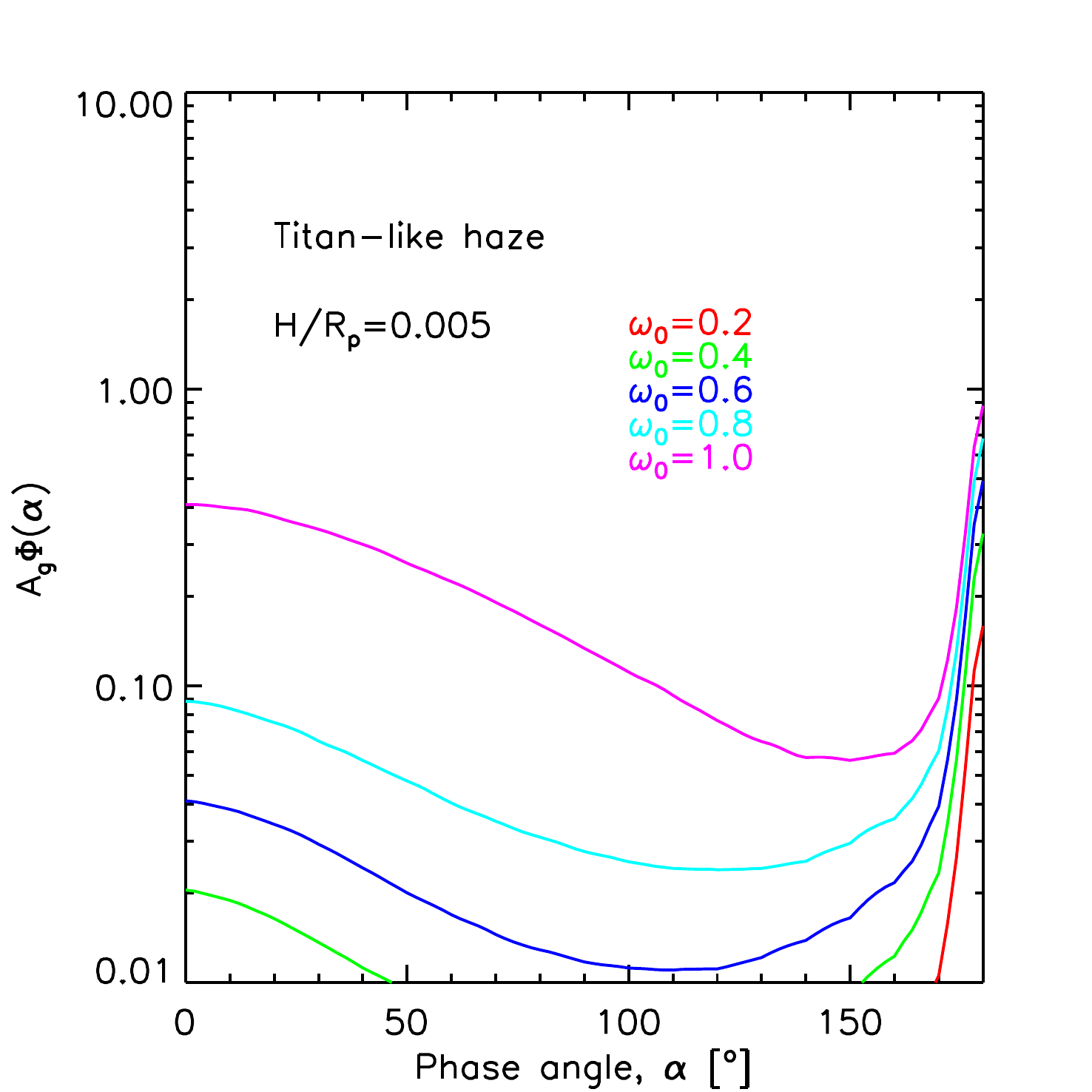} \\
   \includegraphics[width=4.9cm]{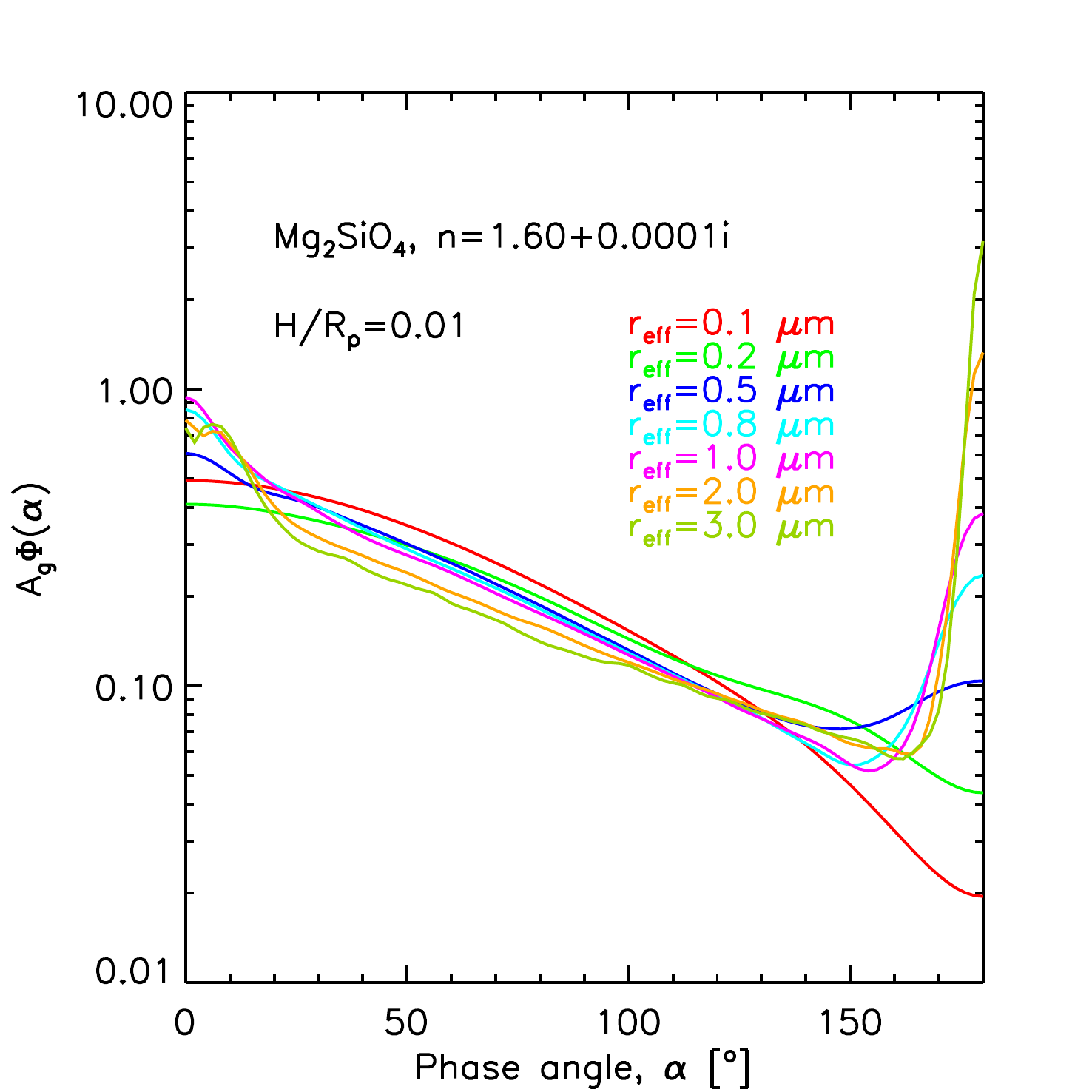} \hspace{-0.6cm} \includegraphics[width=4.9cm]{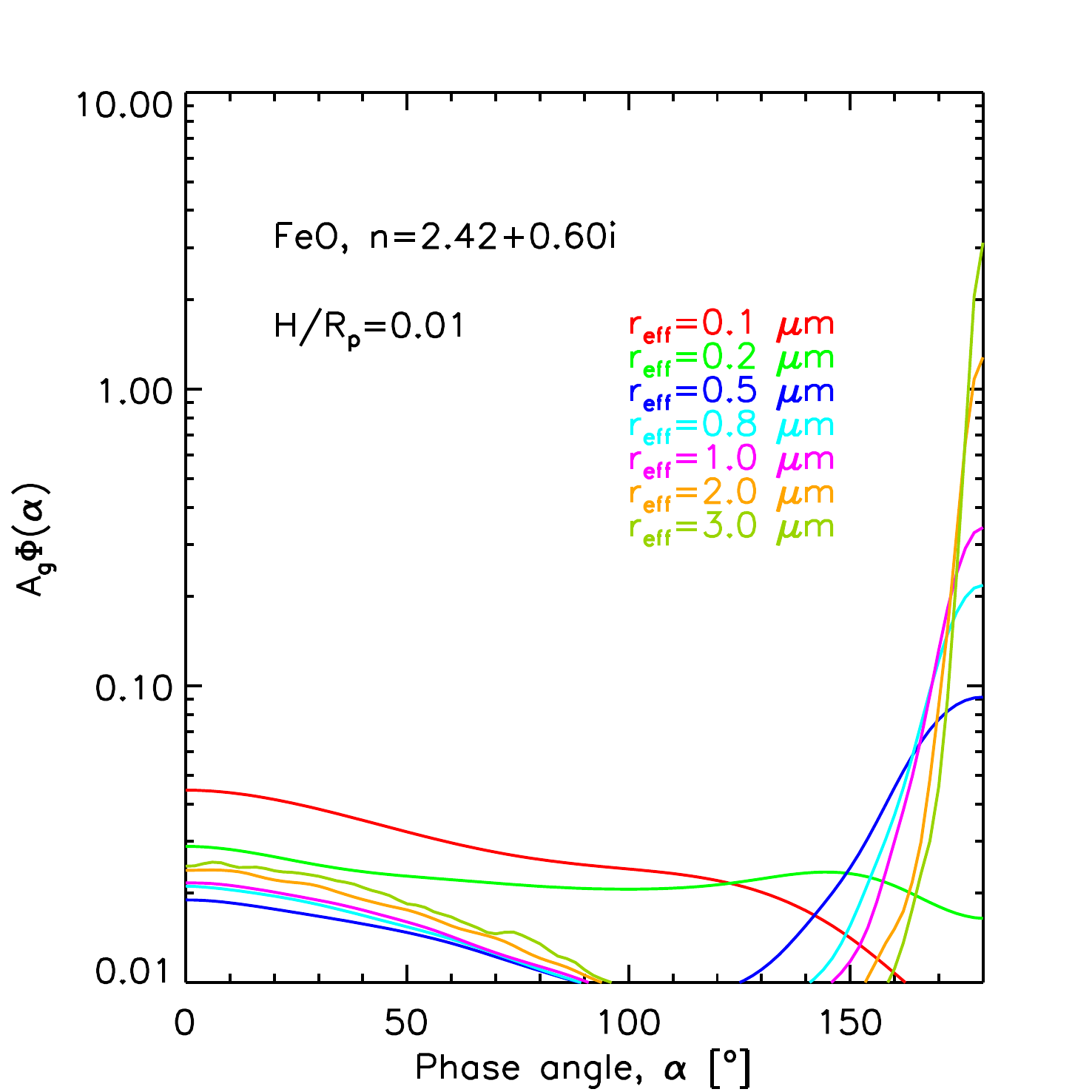} \hspace{-0.6cm} \includegraphics[width=4.9cm]{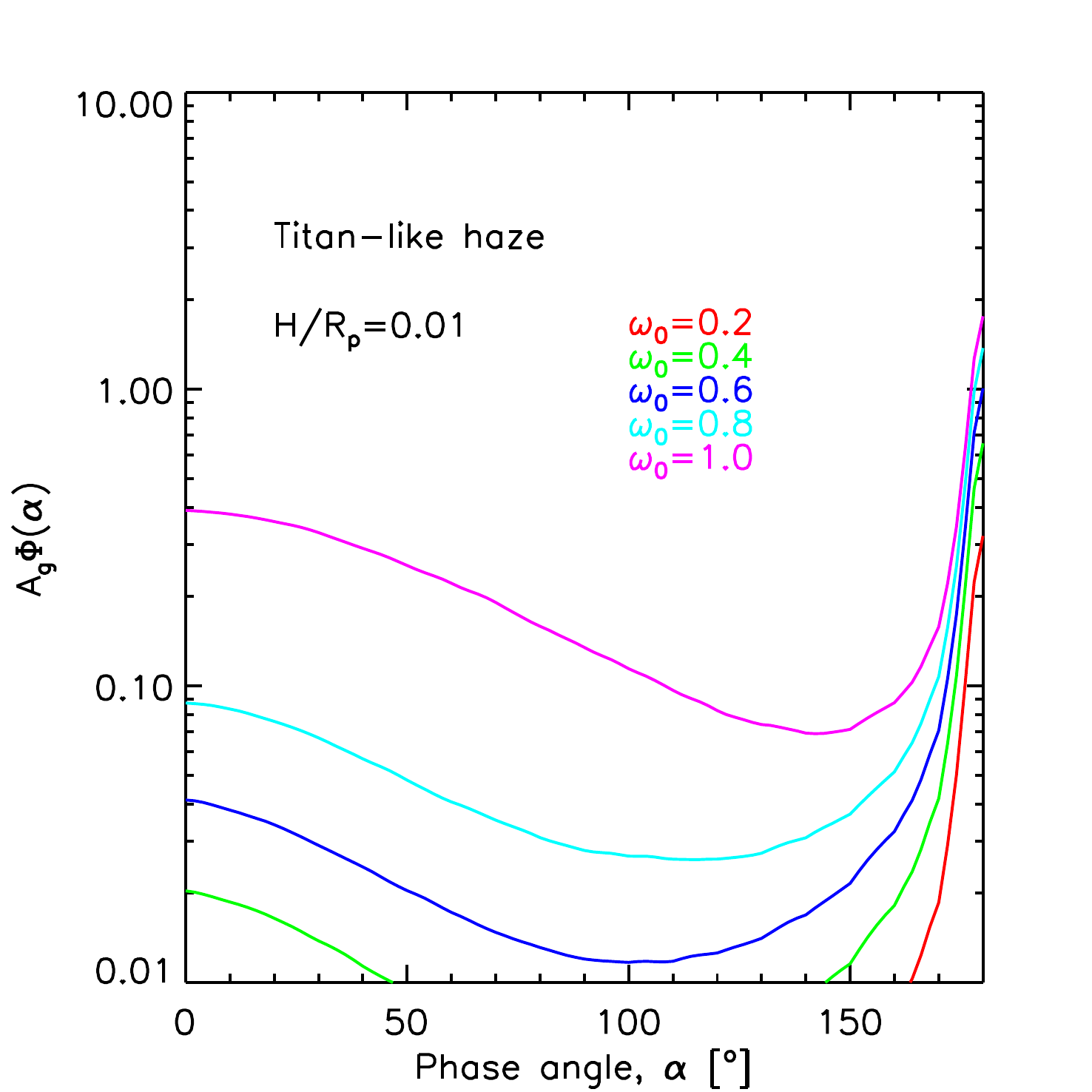} \\
   \includegraphics[width=4.9cm]{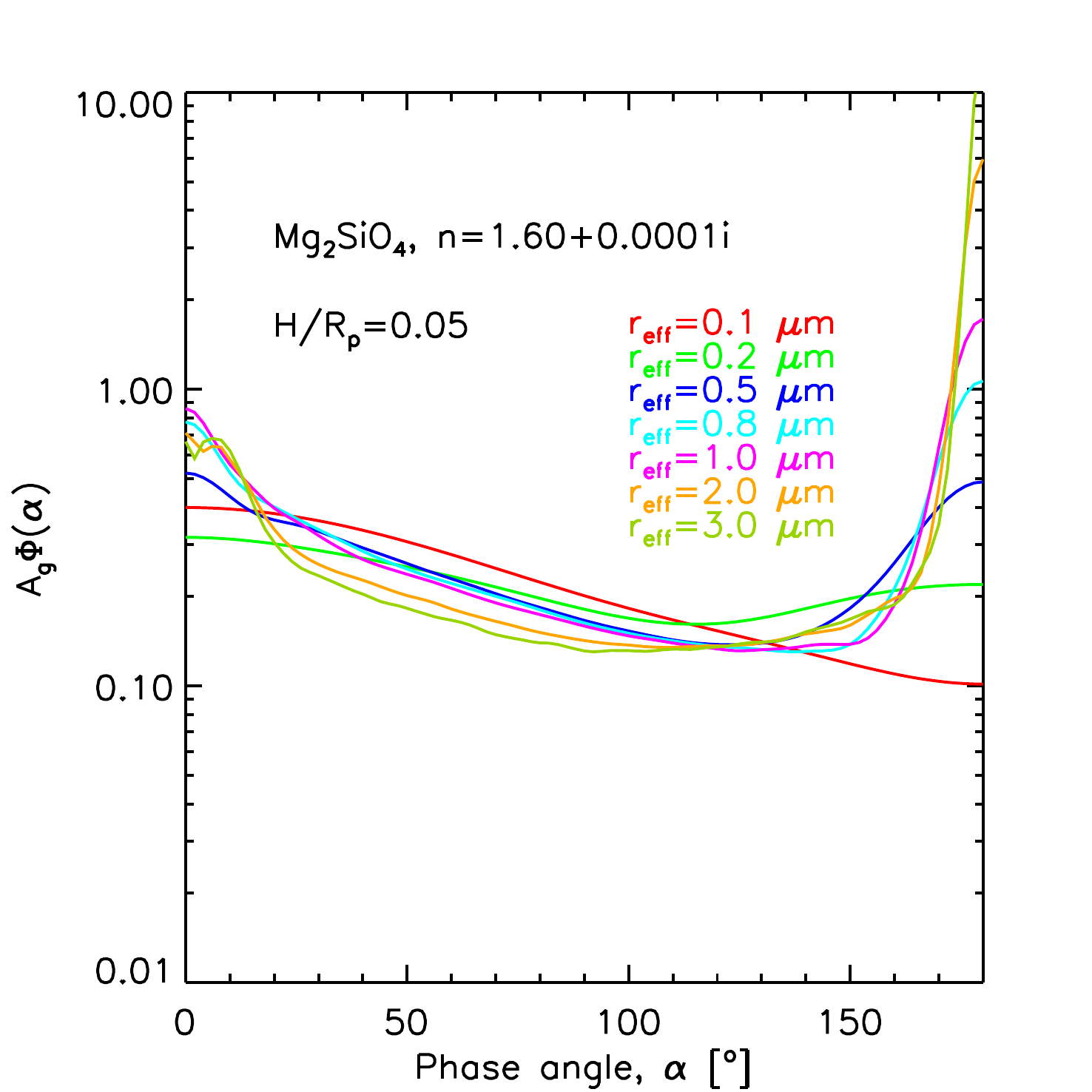} \hspace{-0.6cm} \includegraphics[width=4.9cm]{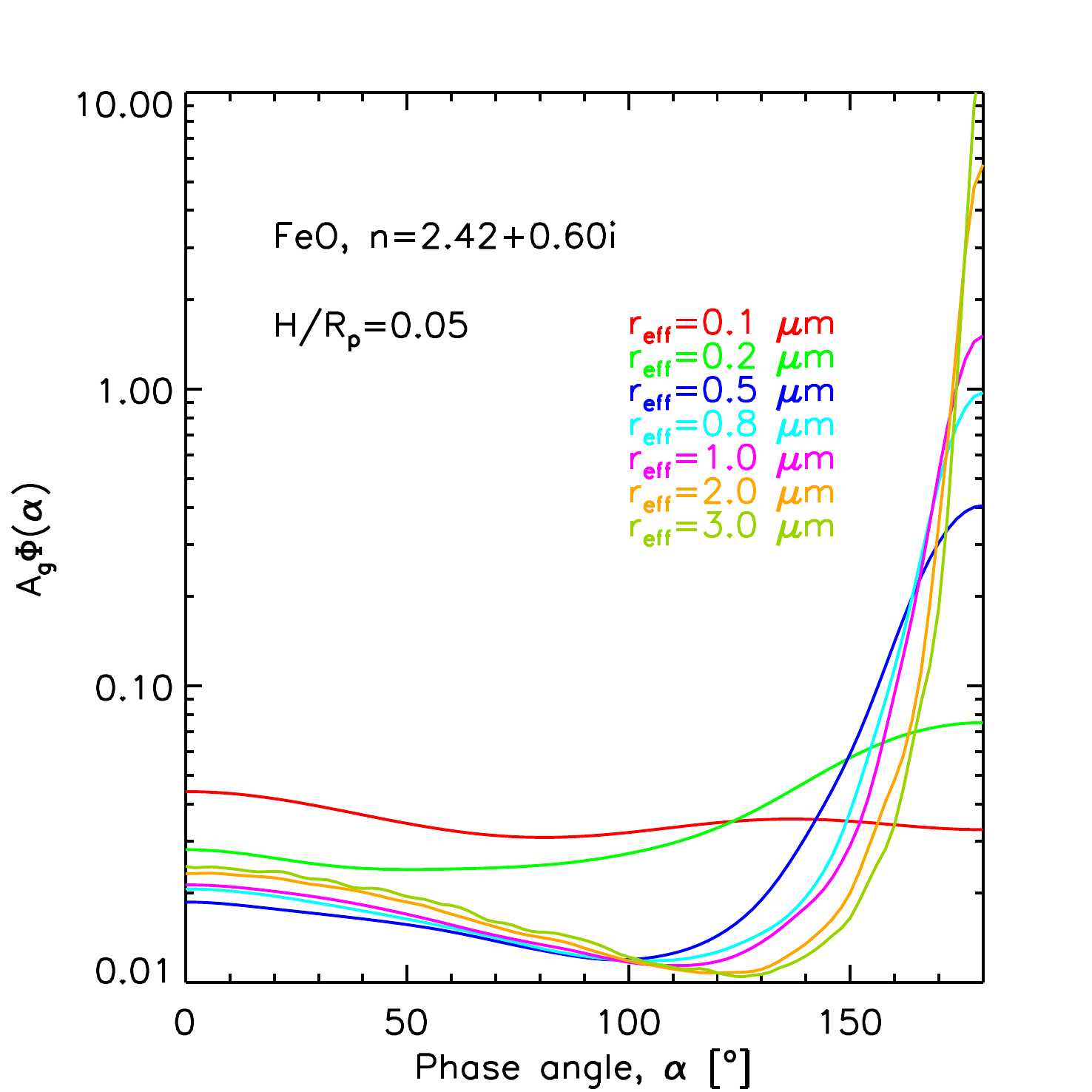} \hspace{-0.6cm} \includegraphics[width=4.9cm]{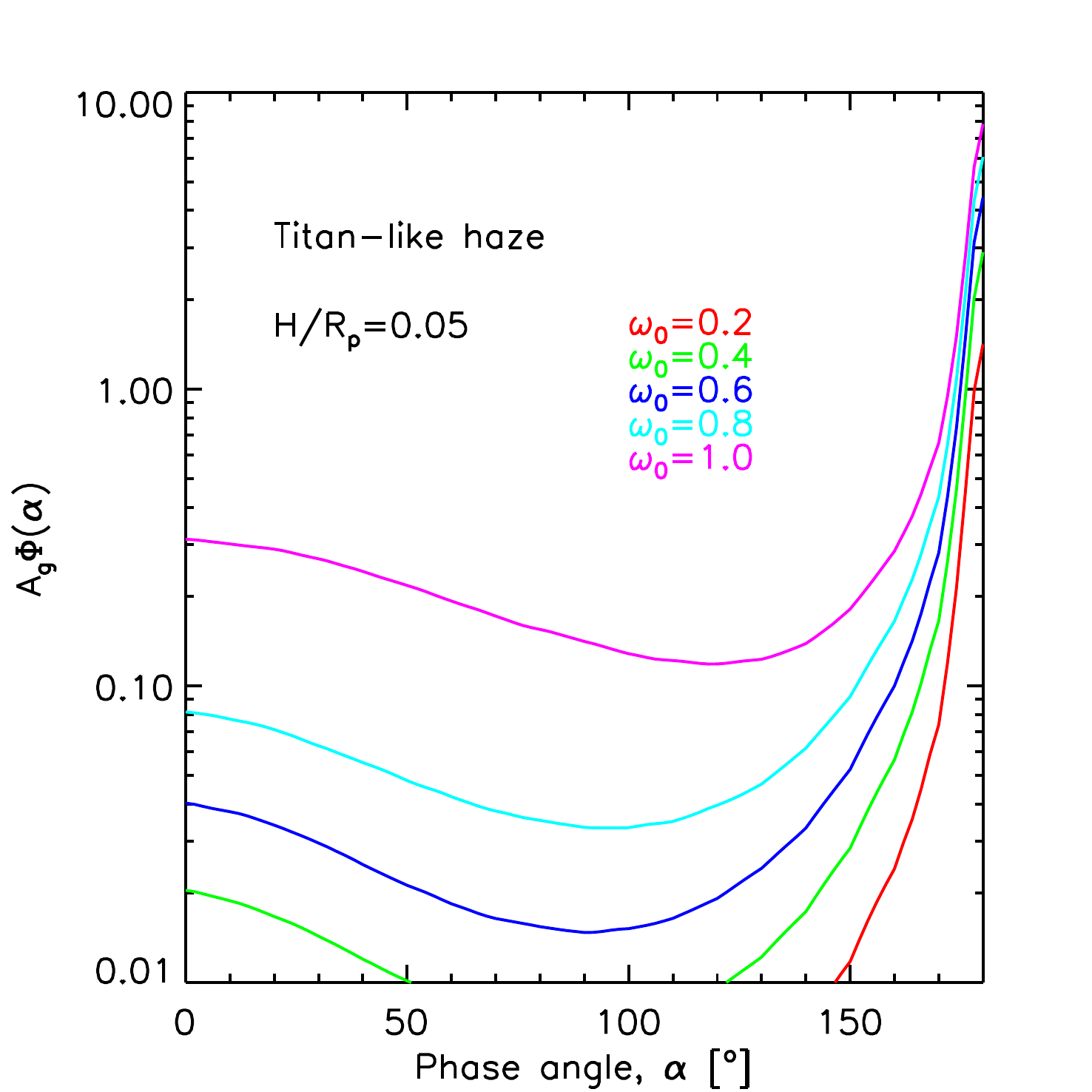}
   \caption{\label{phasecurvediversity_fig}
   Reflected starlight phase curves for spherical, exponential atmospheres. Three types of scattering 
   aerosols are considered: Mg$_2$SiO$_4$ (left), FeO (middle) and Titan-like haze (right). 
   From top to bottom, the graphs explore increasing values of $H_{\rm{a}}$/$R_{\rm{p}}$.
   The phase curves are normalized as in Eq. (\ref{FpFstar2_eq}), 
    with $R_{\rm{p}}$ being the optical radius. 
    {The simulations are based on the point-like star limit, thereby assuming
    that the angular size of the star is small. A non-negligible angular size will
    modify the phase curves by leaking some of the forward-scattered radiation to smaller
    phase angles (see \S\ref{lammer_sec}).
    }
}
   \end{figure*}

In total, we produced about 800 phase curves for different combinations of 
aerosol composition, particle radius and 
ratio of the aerosol scale height to the planet optical radius. 
Figure (\ref{phasecurvediversity_fig}) shows a subset of them in the dimensionless 
form $A_g$$\Phi(\alpha)$.  
From top to bottom, the graphs are arranged
by increasing $H_{\rm{a}}$/$R_{\rm{p}}$. 
It is apparent that puffy
planets with large $H_{\rm{a}}$/$R_{\rm{p}}$ ratios exhibit stronger forward scattering.
{$A_g$$\Phi(\alpha)$$>$1 
is possible at large phase angles, especially for puffy atmospheres rich in large aerosol particles. 
This physically consistent result confirms that the overall planet brightness
mimics to some extent the behaviour of the aerosols scattering phase function when 
the planet is back-illuminated}.

The graphs in the left and central columns show the impact of 
particle size for two compositions 
that result in more reflective (Mg$_2$SiO$_4$) or absorbing (FeO) aerosols.
The effective radius of the particles
$r_{\rm{eff}}$ 
affects both $p_{\rm{a}}(\theta)$ and $\varpi_{0,\rm{a}}$.
Larger $r_{\rm{eff}}$ values typically lead to 
$p_{\rm{a}}$($\theta$) functions with a stronger diffraction peak focused on a narrower range of 
scattering angles. This behaviour is mimicked by the planet phase curve, which tends to
exhibit a brightness surge closer to $\alpha$=180$^{\circ}$. 
{The finite angular size of the star, an effect 
omitted in the calculations of Fig. (\ref{phasecurvediversity_fig}), 
will smear the forward scattering peak and leak it into smaller angles (see \S\ref{lammer_sec}).}

At small phase angles, the planet brightness is strongly dependent on $\varpi_{0,\rm{a}}$.
The simulations show that atmospheres with Mg$_2$SiO$_4$ aerosols 
($\varpi_{0,\rm{a}}$$\sim$0.997 for $r_{\rm{eff}}$=1 $\mu$m)
result in brighter planets than if they are rich in FeO aerosols
($\varpi_{0,\rm{a}}$$\sim$0.555 for $r_{\rm{eff}}$=1 $\mu$m).
At large phase angles however the dependence of the planet brightness with the 
assumed $\varpi_{0,\rm{a}}$ is almost linear (Eq. \ref{FpFstar_eq}) and the 
difference between the Mg$_2$SiO$_4$- and FeO-aerosol atmospheres is reduced.

The graphs in the right column of Fig. (\ref{phasecurvediversity_fig}) 
show phase curves calculated with Titan-like 
haze scattering phase functions $p_{\rm{a}}$($\theta$) 
at $\lambda_{\rm{eff}}$=600 nm \citep{tomaskoetal2008}. 
To explore the impact of $\varpi_{0,\rm{a}}$, we ran this set of simulations with 
$\varpi_{0,\rm{a}}$ values between 0.2 and 1, as indicated in the graphs. 
This battery of simulations confirms that 
the planet brightness is very sensitive to $\varpi_{0,\rm{a}}$ at the small phase angles,
but much less so at large phase angles.

As a corollary, the analytical expression of Eq. (\ref{FpFstar_eq}) 
{(and its generalization to finite angular size stars) together with}
the phase curves of Fig. (\ref{phasecurvediversity_fig}) indicate that:
extended hazy atmospheres result in significant forward scattering {at large phase angles}; 
the aerosol size partly dictates the strength of the phenomenon and 
whether it occurs on either a narrow or broad range of phase angles 
{(at least in the point-like star limit)}; 
the aerosol composition is not critical at large phase angles. 
The arguments presented above suggest that 
the detection of such an optical phenomenon at an exoplanet will lead to a joint constraint on 
its aerosol scale height and particle size. 

\section{ Extended hazy atmospheres}
\label{sec:hazy_sec}

Diverse theoretical approaches to the formation of condensates in exoplanet atmospheres 
of different complexity and scope have been presented {in the literature}
\citep[e.g.][]{marleyetal1999,sudarskyetal2000,ackermanmarley2001,morleyetal2012,
hellingfomins2013,parmentieretal2013,leeetal2016,lavvaskoskinen2017}. 
Their predictive capacity however remains uncertain.
In an attempt to develop a few guidelines, 
\citet{sudarskyetal2000} proposed five broadly-defined 
regimes in the formation of condensates on substellar gas objects 
depending on the objects temperature. 
This classification is not comprehensive, but is useful because reveals part of the complexity of
the problem.
Specifically, 
the authors indicate that the low-gravity ($g$$<$10 m s$^{-2}$),
very hot ($>$1,500 K) 
planets of their Class V are likely to have silicate condensates lofted high in their (extended) 
atmospheres and therefore appear as highly reflective during occultations and 
hazy during transits. If the prediction is correct, 
these planets would be potential candidates for {strong
forward scattering provided that the condensate particles are of the appropriate size}.

The increasing number of exoplanet 
data has also led to phenomenological approaches that seek to
correlate the empirical evidence for condensates with properties such as planet 
gravity or temperature \citep{stevenson2016}, water absorption \citep{singetal2016}, 
or the muting of alkali features in the visible and near infrared \citep{heng2016}. 
In particular, \citet{stevenson2016} suggests on the basis of 
near-infrared observations for 14 exoplanets that 
condensates form preferentially in 
low gravity ($g$$<$16 ms$^{-2}$), low temperature ($T_{\rm{eq}}$$<$750 K) environments. 
On the other hand, \citet{barstowetal2017}
note that from their sample of 10 hot Jupiters the
planets with $T_{\rm{eq}}$$<$1,300 or $>$1,700 K seem to exhibit 
Rayleigh extinction at short wavelengths attributable to small condensates.
In contrast, the planets of their sample with temperatures in the 1,300--1,700 K range seem to exhibit 
wavelength-independent extinction suggestive of larger condensates.
The predictive capacity of phenomenological
approaches remains to be confirmed with additional targets and tested with 
robust interpretation tools \citep{stevensonetal2016}.

The uncertainties in the occurrence of condensates in exoplanet atmospheres translate
into uncertainties in their vertical distribution, composition and particle size. 
\citet{singetal2016} note that if the pressure-temperature 
profile of an atmosphere runs (nearly) parallel to the condensation 
curve of a potential condensate, disturbances in the temperature-pressure profile
may cause that the planet atmosphere shows itself as either hazy or essentially clear. 
The condensate composition will depend on the material available for condensation 
and on the local chemistry if the haze is formed photochemically. 
The particle size will depend on these effects, but also on competing 
microphysical processes that may either favour or disrupt the growth of small 
aerosols onto larger ones. Atmospheric dynamics, and its capacity to keep the 
particles suspended against gravitational settling, will also play a role. 

A number of exoplanets reveal continuum extinction that increases towards ultraviolet 
wavelengths. 
The usual interpretation of this behaviour is that the wavelength-dependent 
extinction cross sections of small, weakly-absorbing particles ($\sigma$$\propto$$\lambda^{-\alpha}$ 
with $\alpha$$\sim$4) cause the so-called Rayleigh slopes
in the planet transmission spectra \citep{lecavelierdesetangsetal2008}. 
Evidence for small condensates has also been found in the interpretation of 
reflected starlight. 
Planets with large-particle clouds will likely exhibit more structure in their 
brightness variation with orbital phase than if the particles are small
\citep{seageretal2000,jenkinsdoyle2003}. 
This idea lies at the core of a recent analysis of Kepler-7b's optical phase curve that shows that
the measurements are consistent with morning-side clouds made of poorly absorbing, 
submicron-size particles \citep{garciamunozisaak2015}. 

Other planets show no detectable colour dependence in their transit depths.
Two well-known cases of this grey behaviour are the sub-Neptunes
GJ1214b \citep{kreidbergetal2014} and GJ436b \citep{knutsonetal2014}. 
In both cases, the bulk atmospheric composition is {possibly} dominated by hydrogen and 
thus they may have non-negligible scale heights.
If so, grey transits are suggestive of moderately large particles lofted to 
mbar--$\mu$bar pressures. 
In spite of multiple degeneracies in the interpretation of grey transits, it is possible 
to constrain the location of the effective cloud level, defined as  
an artifical cutoff between two distinct altitude ranges:
 one opaque and one aerosol free. 
It has not been possible though to
gain insight into the vertical profile of the condensates.
Achieving this calls for more elaborate treatments of the aerosol vertical 
distribution that may not be justifiable given the multiple degeneracies 
already identified in the interpretation of transmission spectra.
Regardless of the various uncertainties that exist in the nature of condensates
and their distribution, grey transits suggest
the possibility of atmospheres containing moderately large particles. If the condensates
are distributed over a sufficiently broad range of altitudes, 
such planets might exhibit forward scattering to some extent.

The fact that some exoplanets have anomalously large radii for their age is well
documented. Such inflated, low-density planets occur amongst the population of hot Jupiters
\citep{demoryseager2011} and sub-Neptunes \citep{lammeretal2016,cubillosetal2017}. 
The inflation mechanisms that 
sustain their interior structure have not been fully elucidated
\citep{spiegeletal2014}, but it is possible that there are multiple
at play \citep{tremblinetal2017}. 
Low-density exoplanets may represent good candidates for showing forward scattering
provided that their extended atmospheres are accompanied by extended aerosol layers. 
In what follows, we derive expressions that allow us to guess when an exoplanet 
has suitable conditions for forward scattering.
These expressions incorporate 
a few necessary simplifying assumptions on the envelopes.

For that purpose, we first 
write $H$/$R_{\rm{p}}$ in terms of measurable quantities. 
$H$=$k T$/$\mu g$ is the gas pressure scale height, where $k$ is the Boltzmann constant, 
$T$ stands for temperature, $\mu$ is the 
atmospheric molecular mass, and $g$ the gravitational acceleration. 
The relevant 
$H$ must be estimated near the optical radius level, which is also the level probed during transit.
Since $g$=$GM_p$/$R_p^2$, where $G$ is the gravitational constant, and
$M_{\rm{p}}$ is the planet mass, we obtain:
\begin{equation}
\frac{H}{R_{\rm{p}}}= \frac{k T/\mu}{G M_{\rm{p}}/R_{\rm{p}}}.
\label{hrp1_eq}
\end{equation}

In our treatment 
we assume that the atmospheric pressure scale height and the aerosol scale height
are equal, i.e. $H$=$H_{\rm{a}}$. This is very approximately the case for Titan
\citep{tomaskoetal2008}. 
For other solar system planets, 
$H_{\rm{a}}$ is a fraction of $H$ \citep{sanchezlavegaetal2004,perezhoyosetal2016},
with the exact $H_{\rm{a}}$/$H$ ratio depending on the range of altitudes being considered and on whether 
the aerosols include the high-altitude haze that occur in most atmospheres. 
If the conditions in the atmosphere are such that $H_{\rm{a}}$$\ll$$H$, forward scattering
will be negligible {and therefore undetectable}.

Measurements of temperature at the optical radius level are not available. 
Instead, we will use for our estimates the planet equilibrium temperature 
$T_{\rm{eq}}$=$T_{\rm{eff}}$($R_{\star}$/2$a$)$^{1/2}$ 
that assumes that the incident stellar flux
(effective temperature $T_{\rm{eff}}$) is balanced by thermal radiation
from a rapidly-rotating dark planet. 
$T_{\rm{eq}}$ does not pertain to a specific altitude  
and thus it is possible that the temperature at the optical radius level 
will differ from it. 

Re-arranging Eq. (\ref{hrp1_eq}) with the expression for $T_{\rm{eq}}$:
$$
\frac{H}{R_{\rm{p}}}= 
4.69\times10^{-6} \frac{T_{\rm{eq}} [{\rm{K}}] R_{\rm{p}}/R_{\rm{J}}}
{{\mu [\rm{a.m.u.}}]M_{\rm{p}}/M_{\rm{J}}}=
$$
\begin{equation}
=2.26\times10^{-7} 
T_{\rm{eff}} [{\rm{K}}] \left(\frac{R_{\star}/R_{\sun}}{a [\rm{AU}]} \right)^{1/2}
\frac{  R_{\rm{p}}/R_{\rm{J}}}
{{\mu [\rm{a.m.u.}}]M_{\rm{p}}/M_{\rm{J}}}.
\label{Eq_HRp_final_eq}
\end{equation}
Finally, if Eqs. (\ref{FpFstar_eq}) and (\ref{Eq_HRp_final_eq}) are combined:
\begin{equation}
\frac{F_{\rm{p}}}{F_{\star}} = 3.23\times10^{-13} 
T_{\rm{eff}} (R_{\star}/R_{\sun})^{1/2}
\frac{p_{\rm{a}}(\theta=0)\varpi_{0,\rm{a}}} {\mu [\rm{a.m.u.}] \rho_{\rm{p}}/\rho_{\rm{J}}
{(a [\rm{AU}])^{5/2}}}, 
\label{FpFstar_eq2}
\end{equation}
where $\rho_{\rm{p}}$/$\rho_{\rm{J}}$ is the planet density { relative to Jupiter's
and $R_{\star}/R_{\sun}$ is the stellar radius relative to the Sun's.} 
The appeal of Eq. (\ref{FpFstar_eq2}) is that the information needed for its evaluation 
is available for many systems. 
It suggests that 
low-density planets at small orbital distances are good candidates for the occurrence
of forward scattering. 
{The reality may be more complex than that, 
because it is unclear how these and other parameters will
affect the occurrence of aerosols and their optical properties.}

In a {zeroth-order} approximation,  
the amount of forward scattering from a planet can 
be ranked on the basis of the planet-to-star contrast at $\alpha$=180$^{\circ}$.
{This simplified treatment avoids elaborate calculations such as those 
presented in Fig. (\ref{phasecurvediversity_fig}).}
According to Eq. (\ref{FpFstar_eq}), the amount of forward-scattered starlight depends on the product of 
2$\pi H_{\rm{a}}/R_{\rm{p}}(R_{\rm{p}}/a)^2$, which is essentially a geometric factor, 
and $p_{\rm{a}}(\theta=0)\varpi_{0,\rm{a}}$, which depends on the aerosol optical properties
{(and possibly, the star angular size)}. 

We have searched the exoplanets.org \citep{hanetal2014} and 
exoplanetarchive.ipac.caltech.edu archives 
and collected the information needed in Eqs. (\ref{FpFstar_eq}), 
(\ref{Eq_HRp_final_eq})--(\ref{FpFstar_eq2}). 
As of the time of writing (April 2017), 
this information is available for a total of 462 exoplanets.
Then, we calculated 
${H}/{R_{\rm{p}}}$, $(R_{\rm{p}}/a)^2$, $T_{\rm{eq}}$, $\rho_{\rm{p}}$/$\rho_{\rm{J}}$,
$g$/$g_{\rm{J}}$ and $2\pi (H/R_{\rm{p}}) (R_{\rm{p}}/a)^2$. 
For simplicity, $a$ is taken to be the semi-major axis, also for planets on eccentric orbits. 
Since most planets of interest have densities consistent with hydrogen-helium envelopes, 
we adopted $\mu$=2.3 a.m.u. 
Table (\ref{derivedparameters_table}) shows a selection of the planets investigated arranged by
decreasing $2\pi (H/R_{\rm{p}}) (R_{\rm{p}}/a)^2$. 
{
Table (\ref{derivedparameters2_table}) shows the same information specific to the Kepler 
planets discussed in \citet{angerhausenetal2015} and \citet{estevesetal2015}. 
}

\begin{figure}
\centering
\includegraphics[width=9.cm,clip]{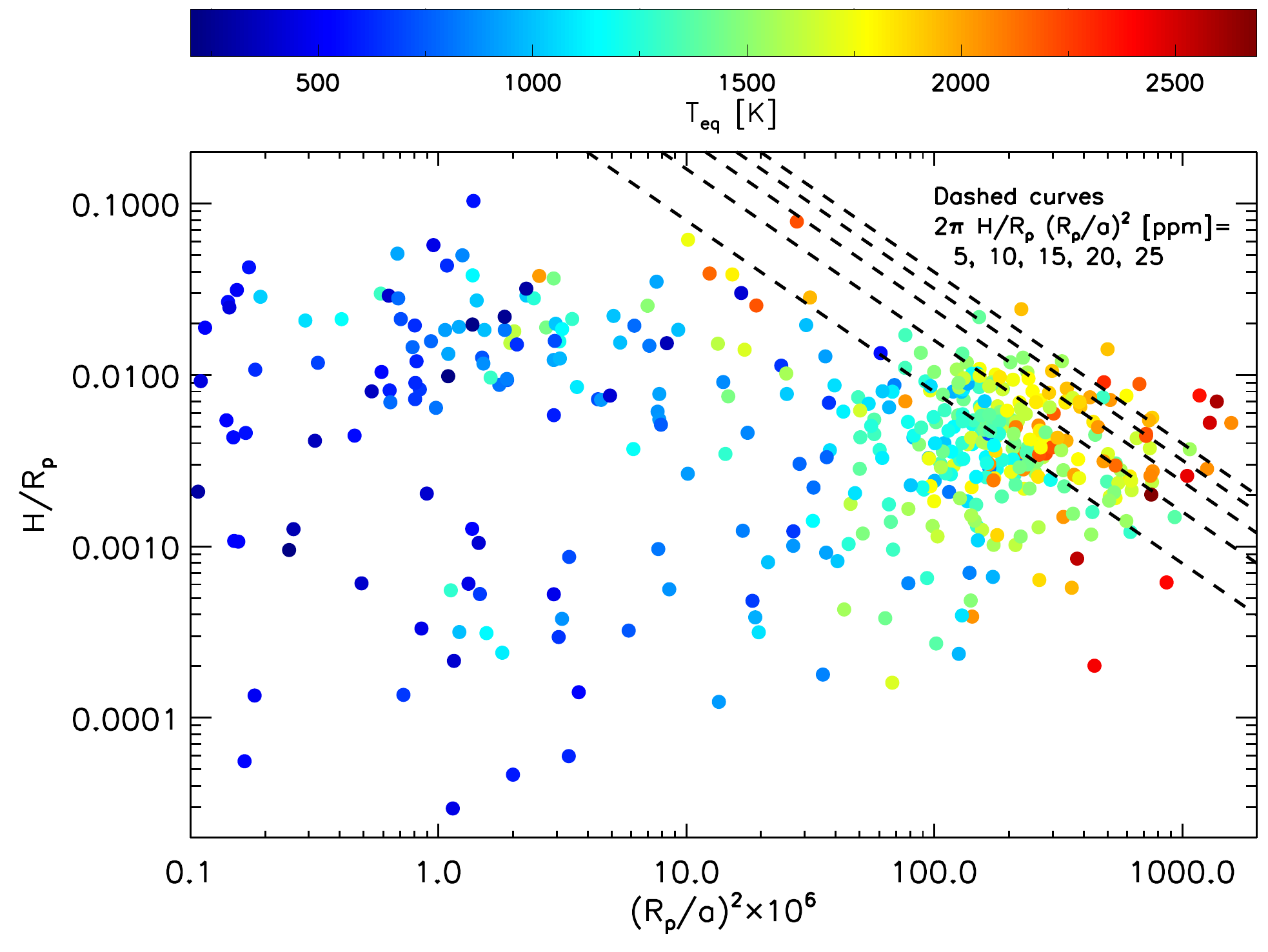}
\vspace{0.1cm}\\
\includegraphics[width=9.cm,clip]{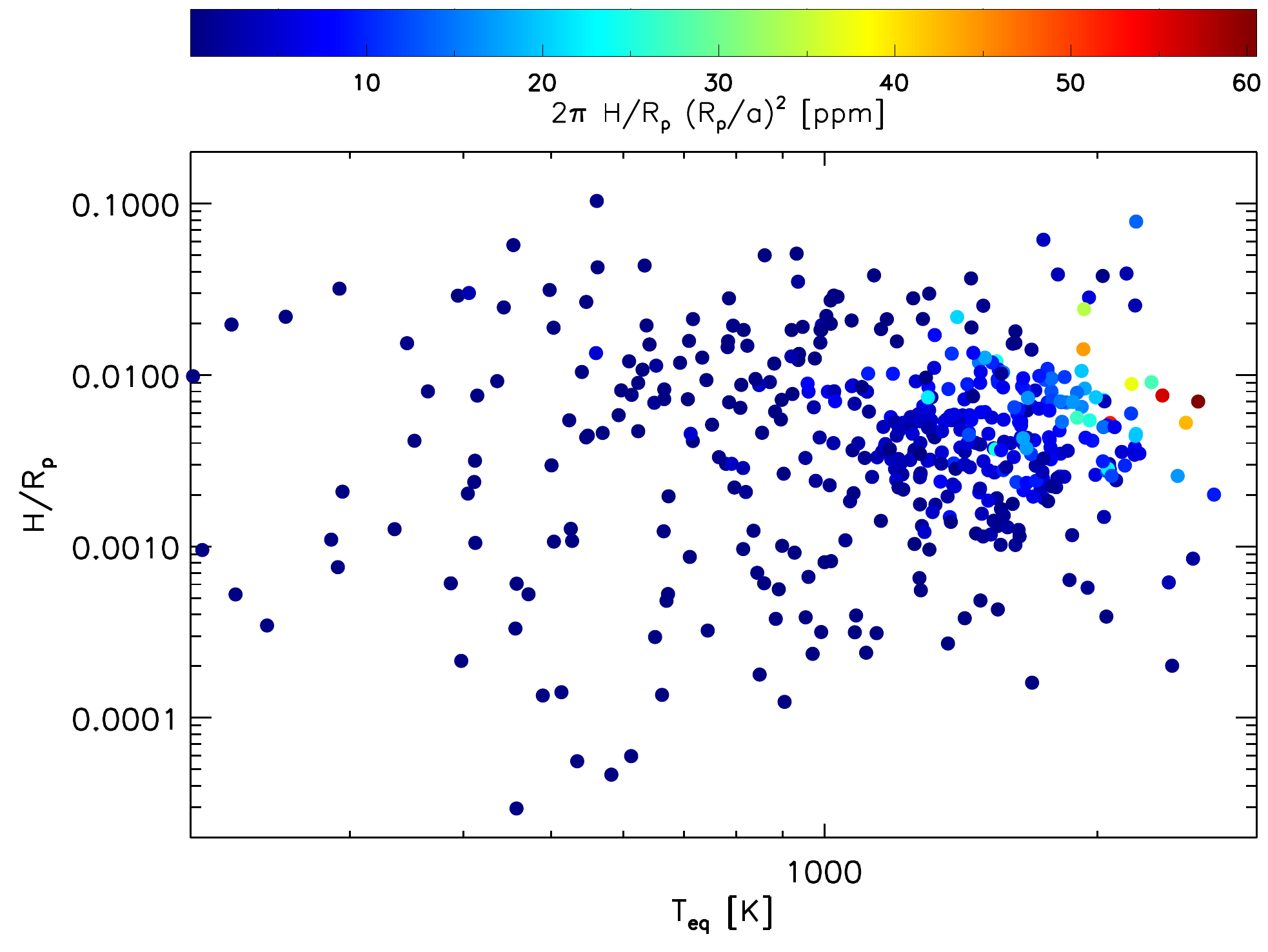}   
\vspace{0.1cm}\\
\includegraphics[width=9.cm,clip]{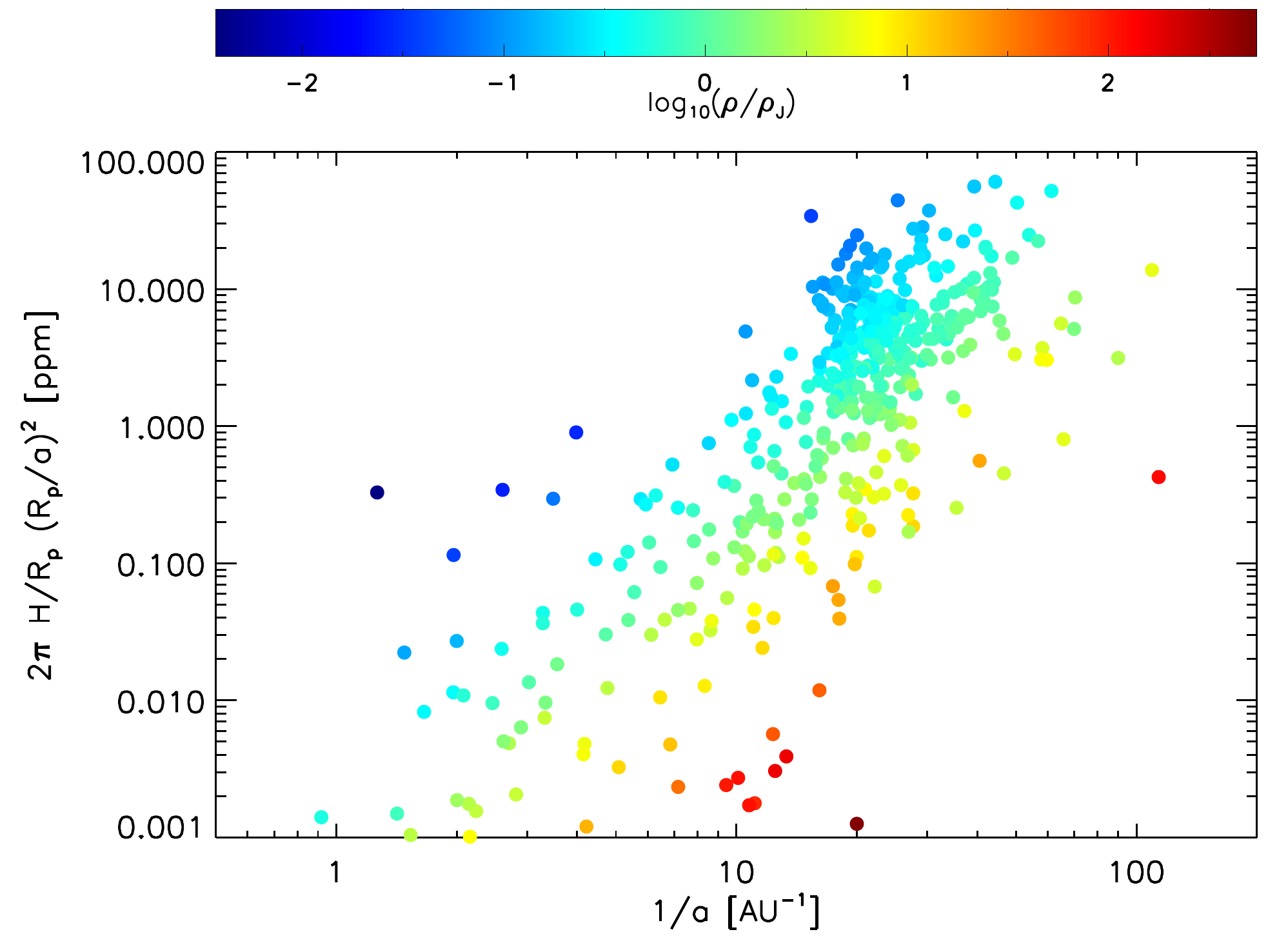}   

\caption{\label{planetparam_fig}
Estimated parameters for the sample of 462
exoplanets considered in the study. In the top graph, the dashed lines divide the 
parameter space with 2$\pi$$H$/$R_{\rm{p}}$$(R_{\rm{p}}/a)^2$ $>$ or $<$ the
quoted ppms. From left to right, the dashed lines represent 5, 10, 15, 20 and 25
parts per million. The planet and stellar parameters were extracted from
exoplanets.org and complemented from exoplanetarchive.ipac.caltech.edu.
}
   \end{figure}

The estimated $H$/$R_{\rm{p}}$, $(R_{\rm{p}}/a)^2$, $T_{\rm{eq}}$ and $2\pi (H/R_{\rm{p}}) (R_{\rm{p}}/a)^2$
are displayed in Fig. (\ref{planetparam_fig}).
The dashed lines in Fig. (\ref{planetparam_fig}, Top) divide the region of the parameter
space with $2\pi (H/R_{\rm{p}}) (R_{\rm{p}}/a)^2$ $>$ or $<$ 5, 10, 15, 20 and 25 
parts per million (ppm). 
According to Eq. (\ref{FpFstar_eq}), 
on top of each dashed line, aerosols with $p_{\rm{a}}$($\theta$=0)$\varpi_{0,\rm{a}}$=1 
(or more generally $\mbox{<}p_{\rm{a}}\mbox{>}$($\Theta$=0)$\varpi_{0,\rm{a}}$=1) will
produce the quoted planet-to-star contrast at $\alpha$=180$^{\circ}$. 
{This is not an unrealistic situation, as the aerosol optical properties graphed
in Fig. (\ref{passa_fig}) and Fig. (\ref{paconv_fig}) suggest.}

\begin{table*}
\caption{
Partial list of 
discovered exoplanets ordered by decreasing 2$\pi$$H_{\rm{a}}$$R_{\rm{p}}$/$a^2$. 
The full list is available through the journal website. 
The list excludes KOI-55 b and KOI-55 c, both with $T_{\rm{eq}}$$\sim$7000 K,
but densities $\rho_{\rm{p}}$/$\rho_{\rm{J}}$$\sim$4.5 incompatible with hydrogen-helium envelopes.
\textcolor{black}{For simplicity, we assumed that the atmospheres of all planets are dominated
by hydrogen-helium, although this is not necessarily the case for many of them, including
the well-studied GJ1214b. 
\textcolor{black}{The observable \textbf{O} has been estimated from phase curves calculated in
the point-like star limit. Therefore, the quoted \textbf{O}s likely underestimate the
starlight forward-scattered and reaching the observer.
Missing fields are due to the absence of the stellar magnitude from the consulted catalogues,
or because $\alpha_{\rm{D}}$$<$160$^{\circ}$. 
\textcolor{black}{The ratio (2$\pi$$H_{\rm{a}}$$R_{\rm{p}}$/$a^2$)/PN is an indicator of 
the potential forward scattering strength vs. photon noise.}
}} 
}             
\label{derivedparameters_table}      
\begin{small}
\centering          
\begin{tabular}{c | c c c c c c | c c c c c c c c c}     
\hline\hline       
             Planet  &    $H_{\rm{a}}$/$R_{\rm{p}}$   &  ($R_{\rm{p}}$/$a$)$^2$   &    $T_{\rm{eq}}$  & 2$\pi$$H_{\rm{a}}$$R_{\rm{p}}$/$a^2$ & $\rho_{\rm{p}}$/$\rho_{\rm{J}}$ & $g$/$g_{\rm{J}}$ &
             $\alpha_{\rm{D}}$ & $t_{\Delta\alpha}$ &  $m_V$ & PN &
             2$\pi$$H_{\rm{a}}$$R_{\rm{p}}$/$a^2$ &
             \textbf{O}$^{0.5\mu\rm{m}}$ & \textbf{O}$^{1\mu\rm{m}}$ & \textbf{O}$^{2\mu\rm{m}}$ \\
                     &           &  [ppm]        &    [K]  &    [ppm]  & & & [$^{\circ}$] & [s] &   &  [ppm]     &   /PN    & [ppm] & [ppm] & [ppm] \\

\hline\hline       
           WASP-12 b  &  0.0070  &   1378.  &   2584.  &             60.6  &     0.24  &   0.42  &        160.7  &          188.  &        11.7  &     96.2  &                0.63  &            37.3  &            24.9  &             9.6 \\
          WASP-121 b  &  0.0076  &   1173.  &   2360.  &             55.9  &     0.18  &   0.34  &        164.7  &         1439.  &        10.4  &     19.6  &                2.85  &            38.5  &            31.2  &            11.6 \\
           WASP-19 b  &  0.0052  &   1576.  &   2066.  &             51.9  &     0.42  &   0.58  &        163.8  &          724.  &        12.6  &     71.8  &                0.72  &            38.4  &            30.1  &            11.1 \\
          HAT-P-65 b  &  0.0141  &    499.  &   1931.  &             44.3  &     0.08  &   0.15  &        167.5  &         4659.  &        13.1  &     37.1  &                1.19  &            29.0  &            27.6  &            11.4 \\
          WASP-103 b  &  0.0053  &   1291.  &   2505.  &             42.7  &     0.42  &   0.64  &        160.7  &          163.  &        12.1  &    123.9  &                0.34  &            28.8  &            19.4  &             7.4 \\
           WASP-76 b  &  0.0089  &    671.  &   2183.  &             37.4  &     0.15  &   0.27  &        166.0  &         2617.  &         9.5  &      9.4  &                3.98  &            26.0  &            22.6  &             8.6 \\
          HAT-P-67 b  &  0.0242  &    224.  &   1934.  &             34.1  &     0.04  &   0.08  &        169.6  &        11025.  &        10.1  &      6.0  &                5.71  &            22.0  &            24.7  &            11.9 \\
          HAT-P-32 b  &  0.0076  &    595.  &   1786.  &             28.3  &     0.15  &   0.27  &        170.5  &         5426.  &        11.4  &     15.9  &                1.78  &            22.7  &            28.7  &            14.2 \\
           WASP-78 b  &  0.0091  &    483.  &   2295.  &             27.5  &     0.18  &   0.30  &        163.8  &         1970.  &        12.0  &     33.4  &                0.82  &            17.6  &            13.4  &             5.0 \\
           CoRoT-1 b  &  0.0056  &    756.  &   1900.  &             26.7  &     0.31  &   0.46  &        168.3  &         3011.  &        13.6  &     57.2  &                0.47  &            21.6  &            22.9  &             9.5 \\
           KELT-14 b  &  0.0054  &    734.  &   1962.  &             25.1  &     0.24  &   0.42  &        166.5  &         2675.  &        11.0  &     18.2  &                1.38  &            19.6  &            18.2  &             6.9 \\
             WTS-2 b  &  0.0037  &   1072.  &   1544.  &             24.8  &     0.51  &   0.66  &        169.1  &         2217.  &        15.9  &    182.8  &                0.14  &            22.2  &            25.7  &            10.9 \\
           WASP-17 b  &  0.0120  &    326.  &   1549.  &             24.7  &     0.07  &   0.14  &        173.6  &        12187.  &        11.6  &     11.5  &                2.14  &            19.9  &            32.5  &            23.3 \\
           WASP-48 b  &  0.0071  &    512.  &   2034.  &             23.0  &     0.21  &   0.35  &        166.5  &         3331.  &        11.7  &     22.2  &                1.03  &            16.9  &            15.2  &             5.8 \\
           HATS-18 b  &  0.0028  &   1258.  &   2056.  &             22.4  &     0.83  &   1.11  &        164.6  &          917.  &        14.1  &    126.6  &                0.18  &            20.7  &            17.2  &             5.6 \\
           WASP-52 b  &  0.0074  &    478.  &   1301.  &             22.2  &     0.22  &   0.28  &        172.2  &         5140.  &        12.0  &     19.9  &                1.12  &            18.5  &            27.4  &            16.2 \\
          WASP-127 b  &  0.0217  &    151.  &   1401.  &             20.7  &     0.07  &   0.10  &        172.9  &        12910.  &        10.1  &      5.6  &                3.72  &            15.1  &            22.9  &            15.1 \\
        OGLE-TR-56 b  &  0.0045  &    714.  &   2207.  &             20.3  &     0.53  &   0.73  &        164.8  &         1383.  &        16.6  &    331.1  &                0.06  &            16.1  &            13.3  &             4.8 \\
       OGLE-TR-056 b  &  0.0044  &    714.  &   2204.  &             19.8  &     0.55  &   0.75  &        164.8  &         1392.  &        15.3  &    184.7  &                0.11  &            15.8  &            13.1  &             4.7 \\
           HATS-26 b  &  0.0106  &    298.  &   1922.  &             19.8  &     0.12  &   0.21  &        168.5  &         6749.  &        13.0  &     28.5  &                0.69  &            14.4  &            15.0  &             6.4 \\
          WASP-142 b  &  0.0074  &    424.  &   1993.  &             19.7  &     0.23  &   0.36  &        167.4  &         3647.  &        12.3  &     28.6  &                0.69  &            14.8  &            14.3  &             5.7 \\
         WASP-94 A b  &  0.0126  &    227.  &   1504.  &             18.0  &     0.08  &   0.14  &        173.2  &        12506.  &        10.1  &      5.6  &                3.20  &            14.3  &            22.5  &            15.2 \\
          HAT-P-41 b  &  0.0083  &    342.  &   1938.  &             17.9  &     0.17  &   0.28  &        169.5  &         6119.  &        11.4  &     14.5  &                1.23  &            13.8  &            15.7  &             7.1 \\
           HATS-23 b  &  0.0043  &    654.  &   1657.  &             17.6  &     0.23  &   0.42  &        170.6  &         5488.  &        13.9  &     48.3  &                0.36  &            15.4  &            20.3  &            10.0 \\
            WASP-4 b  &  0.0037  &    739.  &   1671.  &             17.4  &     0.51  &   0.68  &        169.4  &         3027.  &        12.5  &     33.7  &                0.52  &            15.5  &            18.6  &             8.2 \\
           KELT-16 b  &  0.0026  &   1046.  &   2453.  &             16.9  &     0.97  &   1.37  &        162.3  &          526.  &        11.9  &     63.1  &                0.27  &            15.7  &            11.4  &             3.6 \\
           WASP-92 b  &  0.0070  &    384.  &   1880.  &             16.8  &     0.26  &   0.38  &        169.7  &         5081.  &        13.2  &     36.7  &                0.46  &            13.4  &            15.9  &             7.4 \\
            KELT-8 b  &  0.0073  &    361.  &   1677.  &             16.7  &     0.13  &   0.25  &        170.3  &         7989.  &        10.8  &      9.7  &                1.71  &            13.4  &            16.4  &             7.9 \\
           WASP-74 b  &  0.0065  &    387.  &   1923.  &             15.9  &     0.25  &   0.39  &        168.2  &         4202.  &         9.7  &      8.0  &                1.97  &            12.4  &            12.8  &             5.3 \\
           WASP-31 b  &  0.0103  &    237.  &   1575.  &             15.4  &     0.13  &   0.20  &        172.8  &        10497.  &        11.7  &     12.9  &                1.19  &            12.6  &            19.4  &            12.4 \\
         Kepler-12 b  &  0.0119  &    202.  &   1481.  &             15.1  &     0.09  &   0.15  &        172.9  &        13734.  &        13.8  &     29.4  &                0.51  &            12.0  &            18.4  &            11.9 \\
          KELT-4 A b  &  0.0070  &    338.  &   1822.  &             14.9  &     0.18  &   0.31  &        170.1  &         7261.  &        10.0  &      7.0  &                2.12  &            12.0  &            14.6  &             7.0 \\
            WASP-1 b  &  0.0069  &    338.  &   1849.  &             14.7  &     0.24  &   0.36  &        169.5  &         5747.  &        11.8  &     18.1  &                0.81  &            11.7  &            13.6  &             6.2 \\
        Kepler-412 b  &  0.0053  &    437.  &   1829.  &             14.6  &     0.40  &   0.53  &        168.4  &         3473.  &        13.7  &     56.1  &                0.26  &            12.0  &            12.8  &             5.4 \\
           WASP-54 b  &  0.0095  &    239.  &   1781.  &             14.4  &     0.14  &   0.23  &        170.2  &         9070.  &        10.4  &      7.7  &                1.87  &            11.1  &            13.5  &             6.5 \\
          HAT-P-66 b  &  0.0079  &    290.  &   1900.  &             14.3  &     0.19  &   0.31  &        168.5  &         6073.  &        13.0  &     30.5  &                0.47  &            10.9  &            11.5  &             4.9 \\
           HATS-35 b  &  0.0050  &    457.  &   2032.  &             14.3  &     0.39  &   0.57  &        168.1  &         3533.  &        12.6  &     33.1  &                0.43  &            11.8  &            12.3  &             5.0 \\
         Kepler-78 b  &  0.0784  &     27.  &   2208.  &             13.8  &     5.37  &   0.56  &        158.4  &           ---  &        11.7  &      ---  &                 ---  &             0.0  &             0.0  &             0.0 \\
          HAT-P-23 b  &  0.0027  &    759.  &   2051.  &             13.1  &     0.82  &   1.12  &        166.2  &         1798.  &        11.9  &     34.4  &                0.38  &            12.5  &            11.6  &             3.9 \\
          HAT-P-33 b  &  0.0080  &    249.  &   1780.  &             12.6  &     0.16  &   0.27  &        171.3  &         9390.  &        11.0  &     10.1  &                1.25  &            10.2  &            13.6  &             7.3 \\
           HATS-34 b  &  0.0045  &    445.  &   1444.  &             12.5  &     0.32  &   0.46  &        171.7  &         5939.  &        13.8  &     44.5  &                0.28  &            11.0  &            16.2  &             9.0 \\
          HAT-P-39 b  &  0.0094  &    207.  &   1751.  &             12.2  &     0.15  &   0.24  &        171.5  &         9779.  &        12.4  &     18.7  &                0.65  &             9.8  &            13.3  &             7.2 \\
           HATS-24 b  &  0.0026  &    744.  &   2074.  &             12.1  &     0.74  &   1.10  &        167.8  &         2510.  &        12.8  &     44.5  &                0.27  &            11.9  &            12.6  &             4.6 \\
            TrES-4 b  &  0.0071  &    267.  &   1786.  &             12.0  &     0.16  &   0.29  &        170.5  &         8951.  &        11.6  &     13.3  &                0.90  &             9.7  &            12.1  &             6.0 \\
           WASP-81 b  &  0.0065  &    292.  &   1620.  &             11.9  &     0.25  &   0.36  &        171.3  &         7347.  &        12.3  &     20.1  &                0.59  &             9.9  &            13.4  &             7.1 \\
           WASP-82 b  &  0.0060  &    303.  &   2180.  &             11.3  &     0.27  &   0.45  &        167.0  &         4569.  &        10.1  &      9.4  &                1.20  &             8.8  &             8.4  &             3.3 \\
            TrES-3 b  &  0.0024  &    755.  &   1629.  &             11.2  &     0.79  &   1.05  &        170.5  &         3283.  &        12.4  &     31.2  &                0.36  &            11.3  &            15.5  &             7.2 \\
           WASP-90 b  &  0.0097  &    183.  &   1841.  &             11.2  &     0.15  &   0.24  &        170.6  &         9971.  &        11.7  &     13.3  &                0.84  &             8.8  &            11.0  &             5.5 \\

\hline                        
\hline                        
\end{tabular}
\end{small}
\end{table*}

\begin{table*}
\caption{
Same as Table (\ref{derivedparameters_table}) for the planets discussed in
\citet{angerhausenetal2015} and \citet{estevesetal2015}.
}             
\label{derivedparameters2_table}      
\begin{small}
\centering          
\begin{tabular}{c | c c c c c c | c c c c c c c c c}     
\hline\hline       
             Planet  &    $H_{\rm{a}}$/$R_{\rm{p}}$   &  ($R_{\rm{p}}$/$a$)$^2$   &    $T_{\rm{eq}}$  & 2$\pi$$H_{\rm{a}}$$R_{\rm{p}}$/$a^2$ & $\rho_{\rm{p}}$/$\rho_{\rm{J}}$ & $g$/$g_{\rm{J}}$ &
             $\alpha_{\rm{D}}$ & $t_{\Delta\alpha}$ &  $m_V$ & PN &
             2$\pi$$H_{\rm{a}}$$R_{\rm{p}}$/$a^2$ &
             \textbf{O}$^{0.5\mu\rm{m}}$ & \textbf{O}$^{1\mu\rm{m}}$ & \textbf{O}$^{2\mu\rm{m}}$ \\
                     &           &  [ppm]        &    [K]  &    [ppm]  & & & [$^{\circ}$] & [s] &   &  [ppm]     &   /PN    & [ppm] & [ppm] & [ppm] \\

\hline\hline       
         Kepler-12 b  &  0.0119  &    202.7  &   1481.  &             15.1  &     0.09  &   0.15  &        172.9  &        13734.  &       13.8  &     29.4  &                0.51  &            12.0  &            18.4  &            11.9 \\
        Kepler-412 b  &  0.0053  &    437.7  &   1829.  &             14.6  &     0.40  &   0.53  &        168.4  &         3473.  &       13.7  &     56.1  &                0.26  &            12.0  &            12.8  &             5.4 \\
         Kepler-76 b  &  0.0030  &    538.0  &   2145.  &             10.0  &     0.80  &   1.09  &        167.2  &         2654.  &       13.3  &     54.1  &                0.19  &             9.4  &             9.4  &             3.4 \\
          Kepler-8 b  &  0.0083  &    188.3  &   1662.  &              9.8  &     0.20  &   0.29  &        171.8  &         9979.  &       13.6  &     31.7  &                0.31  &             8.0  &            11.3  &             6.4 \\
          Kepler-7 b  &  0.0107  &    123.2  &   1557.  &              8.3  &     0.14  &   0.20  &        172.1  &        14195.  &       12.9  &     19.1  &                0.44  &             6.6  &             9.5  &             5.6 \\
          Kepler-6 b  &  0.0061  &    183.8  &   1504.  &              7.0  &     0.29  &   0.38  &        171.9  &         9215.  &       13.3  &     28.1  &                0.25  &             5.9  &             8.6  &             4.9 \\
         Kepler-17 b  &  0.0019  &    538.2  &   1745.  &              6.5  &     1.05  &   1.40  &        169.6  &         3406.  &       13.8  &     58.0  &                0.11  &             7.1  &             9.0  &             3.7 \\
         Kepler-41 b  &  0.0055  &    184.3  &   1577.  &              6.3  &     0.83  &   0.70  &        171.1  &         4946.  &       14.5  &     66.7  &                0.10  &             5.4  &             7.4  &             3.9 \\
           HAT-P-7 b  &  0.0035  &    281.5  &   2226.  &              6.2  &     0.70  &   0.96  &        166.1  &         3223.  &       10.5  &     13.2  &                0.47  &             5.4  &             4.9  &             1.7 \\
            TrES-2 b  &  0.0031  &    258.5  &   1498.  &              5.1  &     0.65  &   0.80  &        172.5  &         7403.  &       11.4  &     13.3  &                0.38  &             4.7  &             7.7  &             4.6 \\
         Kepler-44 b  &  0.0040  &    162.3  &   1605.  &              4.1  &     0.53  &   0.66  &        171.1  &         8643.  &       14.7  &     55.6  &                0.07  &             3.6  &             5.0  &             2.6 \\
         Kepler-77 b  &  0.0057  &     99.3  &   1248.  &              3.5  &     0.49  &   0.47  &        174.1  &        12143.  &       14.1  &     35.2  &                0.10  &             3.1  &             5.6  &             4.3 \\
         Kepler-91 b  &  0.0070  &     76.4  &   2037.  &              3.4  &     0.32  &   0.43  &        157.0  &           ---  &       12.5  &      ---  &                 ---  &             0.0  &             0.0  &             0.0 \\
         Kepler-10 b  &  0.0390  &     12.4  &   2154.  &              3.0  &     7.05  &   0.89  &        163.2  &          644.  &       11.0  &     36.9  &                0.08  &             1.4  &             1.1  &             0.4 \\
          Kepler-5 b  &  0.0025  &    174.4  &   1807.  &              2.7  &     0.72  &   1.03  &        170.6  &         8995.  &       13.5  &     32.0  &                0.09  &             2.7  &             3.7  &             1.8 \\
            KOI-13 b  &  0.0008  &    376.6  &   2551.  &              2.0  &     2.68  &   4.06  &        167.3  &         3073.  &       10.0  &     11.0  &                0.18  &             3.5  &             3.7  &             1.1 \\
          Kepler-4 b  &  0.0152  &     13.4  &   1614.  &              1.3  &     1.70  &   0.61  &        171.3  &         8717.  &       12.2  &     17.6  &                0.07  &             0.9  &             1.2  &             0.7 \\
         Kepler-15 b  &  0.0033  &     61.6  &   1108.  &              1.3  &     0.75  &   0.72  &        175.4  &        18235.  &       13.8  &     25.0  &                0.05  &             1.2  &             2.4  &             2.3 \\
         Kepler-43 b  &  0.0012  &    155.8  &   1638.  &              1.2  &     1.86  &   2.23  &        171.6  &         8401.  &       14.0  &     41.3  &                0.03  &             1.5  &             2.5  &             1.2 \\
          HAT-P-11 b  &  0.0091  &     14.1  &    871.  &              0.8  &     1.10  &   0.46  &        176.2  &        18998.  &        9.6  &      3.3  &                0.24  &             0.7  &             1.4  &             1.6 \\
         Kepler-40 b  &  0.0018  &     45.9  &   1613.  &              0.5  &     1.36  &   1.59  &        173.0  &        21381.  &       14.8  &     37.3  &                0.01  &             0.5  &             1.0  &             0.6 \\
         Kepler-14 b  &  0.0004  &     43.3  &   1554.  &              0.1  &     5.73  &   6.51  &        173.2  &        21567.  &       12.0  &     10.4  &                0.01  &             0.2  &             0.5  &             0.3 \\
\hline                        
\hline                        
\end{tabular}
\end{small}
\end{table*}

Atmospheric temperature will surely play a key role in the occurrence and optical 
properties of aerosols. It is thus interesting that the equilibrium temperature of planets with 
$2\pi (H/R_{\rm{p}}) (R_{\rm{p}}/a)^2$$>$1 ppm covers a broad range from 
410 to 2600 K. 
The main conclusion of Table (\ref{derivedparameters_table}) and Fig. (\ref{planetparam_fig})
is that there is a significant number of 
exoplanets with sufficiently extended atmospheres to potentially exhibit 
\textcolor{black}{strong} forward scattering. 
Many of these planets have been targets of transit observations.

\section{Detectability of forward scattering}
\label{sec:detectability_sec}

To assess how feasible it is the detection of forward scattering, 
one must also consider the favourable range of phase angles from the observer's vantage point, 
in addition to the detailed shape of the planet phase curve and the stellar brightness.
To simplify this task, 
we will assume that: \textit{1}) 
All orbits are circular of radius equal to the orbital semi-major axis;
\textit{2}) The planet phase curve is binned over two equal-size intervals:
[$\alpha_{\rm{A}}$, $\alpha_{\rm{B}}$] and [$\alpha_{\rm{C}}$, $\alpha_{\rm{D}}$], with
$\Delta\alpha=\alpha_{\rm{D}}-\alpha_{\rm{C}}=\alpha_{\rm{B}}-\alpha_{\rm{A}}$
(Fig. \ref{cartoon_fig}).
Forward scattering is strongest over the [$\alpha_{\rm{C}}$, $\alpha_{\rm{D}}$] bin, 
and we take
$\alpha_{\rm{D}}=\alpha_{\rm{I,IV}}\approx\pi-(\rm{R}_{\rm{p}}+R_{\rm{\star}})/a\approx\pi-R_{\rm{\star}}/a$, 
and $\alpha_{\rm{C}}$=160$^{\circ}$.  
The scattering signal is weak over the properly selected
[$\alpha_{\rm{A}}$, $\alpha_{\rm{B}}$] control bin, which sets a valid comparison baseline. 
With these simplifications, we calculated the time elapsed over each
interval, $t_{\Delta\alpha} = P \Delta \alpha / 2\pi$, where $P$ is
the orbital period; 
\textit{3}) The stellar radiation is approximated by a black body at
the star effective temperature $T_{\rm{eff}}$ between two wavelengths 
[$\lambda_{\rm{1}}$, $\lambda_{\rm{2}}$]. 

Note that the planet does not need to transit in order to produce forward scattering.
However, only planets that reach closer to the star than $\alpha$$\sim$160$^{\circ}$ 
will produce a measurable effect (see Figs. \ref{cartoon_fig} and \ref{phasecurvediversity_fig}). 
Highly inclined orbits will have maximum phase
angles $\alpha$$<$$160^{\circ}$ resulting in negligible forward scattering \textcolor{black}{towards the observer}.
For each planet (and specific $H_{\rm{a}}$, $R_{\rm{p}}$ and $a$;  
Table \ref{derivedparameters_table}), we
calculated the planet-to-star contrast $F_{\rm{p}}/F_{\star}$ over all phases 
by interpolating in $H_{\rm{a}}/R_{\rm{p}}$ 
from our battery of synthetic phase curves. 
Since the aerosol size is a key parameter \textcolor{black}{that we prescribe but} do not predict,
we explored the sizes $r_{\rm{eff}}$=0.5, 1 and 2 $\mu$m. 
We define the observable \textbf{O} as the difference 
in the average planet-to-star contrast over the forward scattering and control bins:  
\textbf{O}=
$<$$F_{\rm{p}}/F_{\star}$$>$$_{\rm{C\rightarrow D}}$$-$$<$$F_{\rm{p}}/F_{\star}$$>$$_{\rm{A\rightarrow B}}$
(see Fig. \ref{cartoon_fig}). 

Photon noise (PN) sets a floor for the detection of the forward scattering signal, 
and it will be the only term considered in \textcolor{black}{our} noise budget.
We estimate PN=1/$\sqrt{N_{\Delta\alpha}}$, where $N_{\Delta\alpha}$
is the number of photons collected at the telescope over
$t_{\Delta\alpha}$. 
The expression for PN takes into account two canceling $\sqrt{2}$ and
1/$\sqrt{2}$ factors, one arising from the differential definition of
\textbf{O} and one arising from the assumption that the pre-
 and post-transit observations can be phase-folded to improve PN.
  We base our calculation of the number of photons at the telescope
   from an $m_V$-magnitude star on the stellar radiated power per unit of area and time. 
(For completeness, we looked up the visible magnitudes of a few host stars
(OGLE-TR-211, WASP-53, -81, -104, WTS-2)
on TEPCat (http://www.astro.keele.ac.uk/jkt/tepcat/), 
and (LUPUS-TR-3) on the Extrasolar Planets Encyclopaedia (exoplanet.eu). 
For CoRoT-24, we estimated $m_{\rm{V}}$=16 based on a magnitude in R-band of
 $m_{\rm{R}}$=15.6. 
For a few Kepler targets (4--8, 10, 15, 43--44, 98, 447), 
we used their quoted Kepler magnitudes (http://archive.stsci.edu/kepler/kepler$\_$fov/search.php).
This ensured that PN could be estimated for all planets with a priori better conditions
for forward scattering.)
  We integrate the Planck function over the specified spectral 
  interval [$\lambda_1, \lambda_2$], and divide by the energy at the given wavelength. 
  Then, using as reference the measured flux of an $m_V$=0 star,
  we estimate the rate:
  $$
  \dot{N}( m_V) = 10^{-0.4 m_V} f( m_V=0; \lambda=550 \mbox{ nm}) \times
  $$
  $$  
  \int_{\lambda_1}^{\lambda_2}
    \frac{\exp{ \frac{ hc}{(550 \mbox{ nm})kT_{\rm{eff}}}} - 1}{\exp{ \frac{hc}{\lambda kT_{\rm{eff}}}} - 1}
    \left(\frac{ 550 \mbox{ nm}}{\lambda} \right)^{5} \frac{ \lambda}{hc}
    d\lambda
$$
Here, $h$ is the Planck constant and $c$ the speed of light. 
The flux calibration factor 
  $f$($m_V$=0; $\lambda$=550 nm)=3.6182$\times$10$^{-12}$ W cm$^{-2}$ $\mu$m$^{-1}$
 is taken from \citet{casagrandevandenberg2014}. As a check, we confirmed the validity
 of our estimated photon rates by comparing them to those calculated from star distances
 and temperatures.

 Finally, the number of photons received at the telescope over a time $t_{\Delta\alpha}$
between $\lambda_1$=400 nm and $\lambda_2$=900 nm is calculated as:
 $$N_{\Delta\alpha}=\dot{N} t_{\Delta\alpha} \eta_{\rm{tel}} A_{\rm{tel}}.$$ 
 For the telescope collecting area, we adopt $A_{\rm{tel}}$=$\pi/4$ m$^2$, and 
 include an overall instrument efficiency $\eta_{\rm{tel}}$=0.75 that accounts 
 for the CCD quantum efficiency and possible tranmission/reflection losses. 
  This instrument configuration is loosely related to the Kepler
  telescope \citep{boruckietal2003}.

In our feasibility analysis, we also assumed that the composition of the prevailing aerosols is FeO.
This condensate absorbs significantly at wavelengths between 0.5 and 2 $\mu$m 
(Fig. \ref{passa_fig}). 
The motivation for this choice is to show that planets that appear 
dark at small phase angles  
may result in strong forward scattering. For this optical phenomenon to occur, particle size 
is more critical than the single scattering albedo. 
Thus, FeO may be regarded as a proxy for the effect of dark haze particles 
that might exist in the upper atmosphere of some exoplanets.

Setting $\alpha_{\rm{D}}$$=$$\alpha_{\rm{I,IV}}$, 
the optimal $\Delta$$\alpha$ depends on the trade-off between the brightness curve shape 
and associated PN. 
{In the point-like star limit considered in the preparation of Table (\ref{derivedparameters_table})}, 
large aerosols tend to focus most of the scattered starlight on a narrow range
of phase angles near $\alpha$=180$^{\circ}$. 
In turn,
small particle sizes generally result in forward scattering 
that is less pronounced but spreads over a broader range of phase angles. 
Implementing a small $\Delta$$\alpha$ 
enhances the planet-to-star contrast over the forward scattering bin 
at the cost of reducing the integration time and therefore worsening PN. 
For simplicity, we adopted 
$\Delta$$\alpha$=$\alpha_{\rm{I,IV}}$$-$160$^{\circ}$ (i.e. $\alpha_{\rm{C}}$=160$^{\circ}$)
in all cases, but note that this
choice may be sub-optimal and therefore leaves room for improvement of the \textbf{O}/PN ratio.
The choice of $\alpha_{\rm{A}}$ and $\alpha_{\rm{B}}$ is such that
$<$$F_{\rm{p}}/F_{\star}$$>$$_{\rm{A\rightarrow B}}$$\approx$0. 

Table (\ref{derivedparameters_table}) summarizes the estimated \textbf{O}$^{0.5\mu\rm{m}}$,
\textbf{O}$^{1\mu\rm{m}}$ and \textbf{O}$^{2\mu\rm{m}}$ 
{(each exploring the quoted aerosol particle radius)} and PN.
A few comments are due. {Obviously,} the process of averaging over $\Delta$$\alpha$ and having 
$\alpha_{\rm{D}}$$<$180$^{\circ}$ dilutes the observable \textbf{O} below the
predicted forward scattering {peak} at $\alpha$=180$^{\circ}$.
Particles that are small result in little forward scattering. 
Particles that are large result in significant forward scattering, but 
most of the scattered starlight focuses on phase angles that are unobservable 
(at least in the point-like star limit)
and therefore do not contribute towards \textbf{O}.  
As a result, the highest \textbf{O}s  often occur for the intermediate $r_{\rm{eff}}$$\sim$1 $\mu$m.
In a few cases, \textbf{O}$\sim$20--30 ppm values are predicted. 
{We emphasize that, as shown below for the specific case of 
CoRo-T-24b, considering the finite angular size of the star
will tend to increase the predicted \textbf{O}s by factors of up to a few
from the values quoted in Tables (\ref{derivedparameters_table})--(\ref{derivedparameters2_table}).}
The comparison of \textbf{O} and PN shows that photon noise should not be critical 
for a number of planets provided that multiple orbits can be stacked to improve the
\textbf{O}/PN ratio. 
Some of the planets listed in Table (\ref{derivedparameters_table}) have been observed
at out-of-transit phases 
with precisions comparable to the quoted \textbf{O}s, in particular the Kepler planets
(Table \ref{derivedparameters2_table}).
It is left for future work the re-analysis of their phase curves 
in the specific search for forward scattering.

\subsection{\label{lammer_sec} Low-mass, low-density planets}
We next turn our attention to the low-mass, low-density, 
sub-Neptune CoRoT-24b ($T_{\rm{eq}}$=935 K, 
$M_{\rm{p}}$/$M_{\rm{J}}$$<$0.018, $\rho_{\rm{p}}$/$\rho_{\rm{J}}$$<$0.5). 
Recent work \citep{lammeretal2016} has proposed that the measured transit 
radius probes a low-pressure region high in the atmosphere,
 and that the opacity is due to an undetermined condensate capable of continuum extinction. 
The hypothesis of high-altitude aerosols is in line with some of the interpretations for the
transmission spectra of e.g. GJ1214b and GJ436b. 
What makes CoRoT-24b stand out with respect to better characterized sub-Neptunes is its 
large $H$/$R_{\rm{p}}$$\sim$0.035. 
CoRoT-24b may be one of a population of planets in similar conditions
\citep{cubillosetal2017,fossatietal2017}.
 
The ratio $H$/$R_{\rm{p}}$ (as given by Eq. \ref{hrp1_eq}) is the inverse of the
parameter $X$ that appears in thermal evaporation theory and represents 
the squared ratio of the escape velocity and the most probable velocity of the gas Maxwellian 
distribution \citep{chamberlainhunten1987}. Small $X$ values represent
favourable conditions for escape.
In thermal evaporation theory, though,
$X$ is evaluated at the exobase and thus well above the optical radius level
at which $H$/$R_{\rm{p}}$ is evaluated. 
The coincidental structure of $H$/$R_{\rm{p}}$ and $X^{-1}$ suggests
that puffy planets also offer good conditions for thermal escape. 
  
We have explored the possibilities offered by CoRoT-24b's 
large $H$/$R_{\rm{p}}$ for its characterization at large phase angles. 
We estimate $(R_{\rm{p}}/a)^2$=7.6 ppm, 
2$\pi$$R_{\rm{p}}$$H_{\rm{a}}$/$a^2$=1.7 ppm, and $\alpha_{\rm{I,IV}}$=175.9$^{\circ}$. 
At small phase angles, assuming a geometric albedo $A_{\rm{g}}$$\sim$0.3 
\citep{demory2014} {(which would preclude an envelope dominated by a
dark condensate)}, 
the planet-to-star contrast is $\sim$2.3 ppm. 
Correspondingly, {at $\alpha$=180$^{\circ}$ the contrast can be as high as 
1.7$\times$$\mbox{<}p_{\rm{a}}\mbox{>}(\Theta=0) \varpi_{0,\rm{a}}$ ppm,} thereby exceeding
the contrast at small phase angles if micron-size or larger aerosols prevail at the 
optical radius level (Fig. \ref{paconv_fig}).
 
We produced synthetic phase curves for CoRoT-24b with $H$/$R_{\rm{p}}$=0.035.
To emphasize the possibilities of large versus small phase angles,
we adopted a dark condensate (FeO), and tested $r_{\rm{eff}}$ values 
of 0.5, 1, 2, 3, 5 and 10 $\mu$m. 
As usual, we adopted $\alpha_{\rm{D}}$$=$$\alpha_{\rm{I,IV}}$ but
unlike in the preparation of Table (\ref{derivedparameters_table}) 
we explored various bin sizes $\Delta$$\alpha$. 
Again, the control bin was defined 
at phases for which the planet appears dark and 
$<$$F_{\rm{p}}/F_{\star}$$>$$_{\rm{A\rightarrow B}}$$\approx$0. 
{A set of phase curves was calculated in the point-like star limit. 
We produced another set of phase curves that consider the finite angular size of the host star. 
The needed 
modifications to the original radiative transfer algorithm \citep{garciamunozmills2015}
are described in Appendix \ref{sec:appendixc}.}

{Figure (\ref{corot24_thetas_fig}) shows the two sets of curves with
emphasis on the large phase angles. At the top, we show simulations for the various
particle sizes in both the
point-like star limit (solid) and in the finite angular size star approach (dashed) 
for the stellar angular radius specific to CoRoT-24b, $\theta_{\star}$$\approx$4.1$^{\circ}$. 
To convert from $A_{\rm{g}}$$\Phi(\alpha)$ to planet-to-star contrasts, the  
scaling factor is 7.6 ppm.
For illustration purposes, the Middle and Bottom plots
show additional calculations for $\theta_{\star}$$\approx$10 and 20$^{\circ}$, 
and $H_{\rm{a}}$/$R_{\rm{p}}$=0.035 as in the initial configuration.}

{The most obvious effect of considering the finite angular size of the 
star is that the out-of-transit brightness of the planet ($\alpha$$\lesssim$180$^{\circ}$$-$$\theta_{\star}$)
 tends to become larger than 
in the point-like star limit. This a direct consequence of the convolution of 
$p_{\rm{a}}$$(\theta$) over the stellar disk brightness to produce $\mbox{<}p_{\rm{a}}\mbox{>}$$(\Theta$). 
Thus, near transit the planet terminator forward-scatters photons 
with deflection angles within a $\pm$$\theta_{\star}$ range  as 
for photons coming from the stellar centre.
At mid-transit ($\alpha$$\equiv$180$^{\circ}$) the amount of starlight 
forward scattered by the atmosphere is lower in the finite angular size limit because
the planet sees stellar photons arriving from a range of directions, some of them not
overlapping with the peak of the $p_{\rm{a}}$$(\theta$) function (Fig. \ref{paconv_fig}).
The attenuating effect of the finite angular size of the star for radiation scattered
at $\alpha$$\equiv$180$^{\circ}$ 
is more pronounced for the large scattering particles associated with 
a narrow forward scattering peak $p_{\rm{a}}$$(\theta$=0). 
}

Table (\ref{corot24_table}) summarizes our estimates for the observable \textbf{O}, 
{in both the point-like (regular typeface) and 
finite angular size (bold typeface) treatments of the star. 
Their inter-comparison indicates that the point-like star treatment
can underestimate the observable by factors of up to a few 
depending on the combination of $\alpha_{\rm{C}}$, $\theta_{\star}$ and $r_{\rm{eff}}$.
}
{For CoRoT-24b, \textbf{O} reaches
up to $\sim$10--20 ppm when $\Delta$$\alpha$ is small enough that the steepest part
of the brightness surge is resolved.} 
The table also shows the photon noise PN per orbit, which 
goes from $\sim$64 ppm for $\alpha_{\rm{C}}$=160$^{\circ}$ to $\sim$186 ppm for 
$\alpha_{\rm{C}}$=174$^{\circ}$. Thus, \textbf{O}/PN$<<$1
over the $\Delta$$\alpha$ bin sizes explored.
Although signals weaker than $\sim$10 ppm have been detected with Kepler, 
improving the \textbf{O}/PN ratio to detectable levels
calls for one or more of the following strategies: 
accumulating data from numerous orbits; focusing on planets around bright stars; 
stacking observations from multiple planets with similar characteristics 
\citep[e.g.][]{sheetsdeming2014}. 
Ultimately, Table (\ref{corot24_table}) 
suggests that there is a chance for low-density
exoplanets that are too small or dark for detection 
in occultation to be detected through forward scattering.

   \begin{figure}
   \centering
   \includegraphics[width=7.cm]{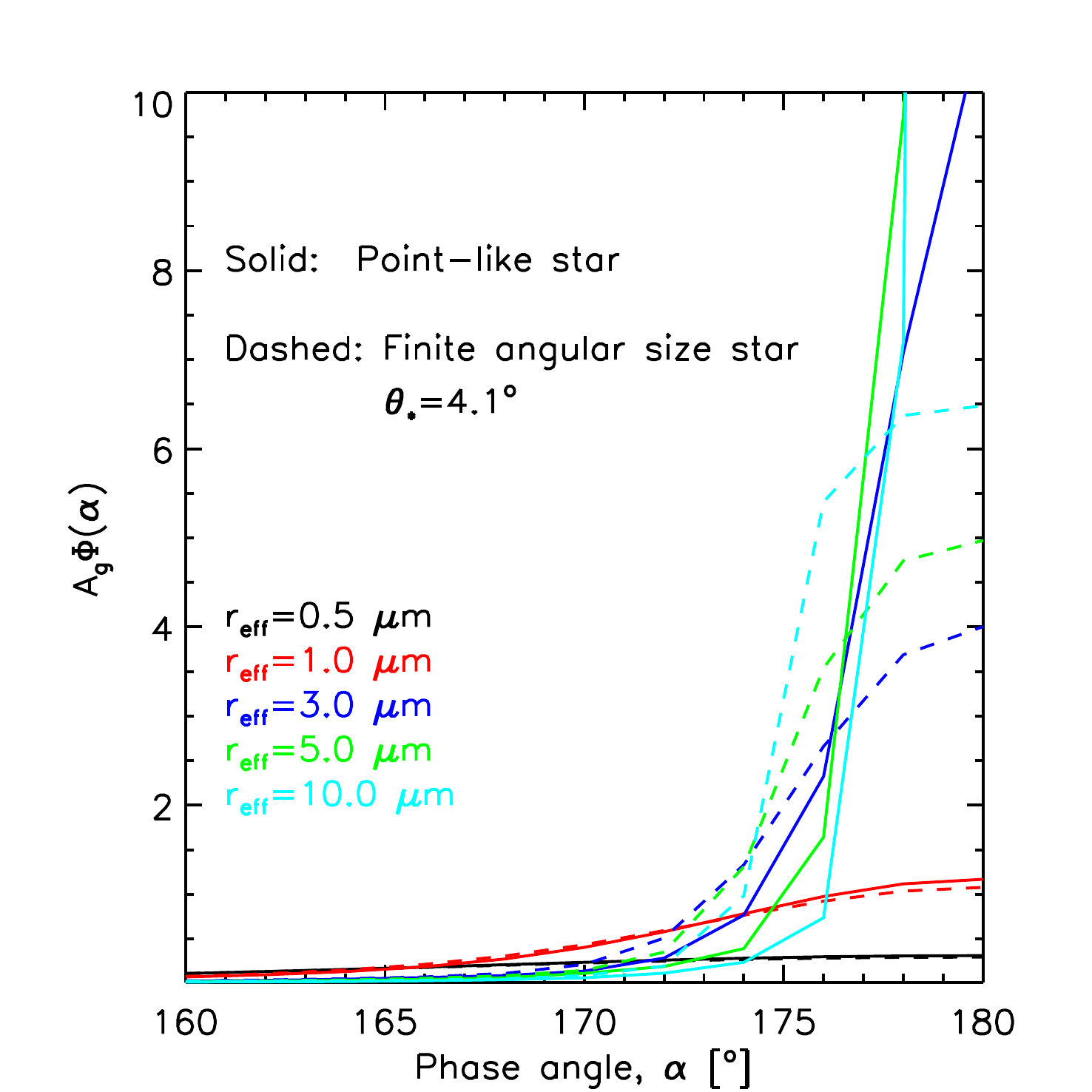}\\
   \includegraphics[width=7.cm]{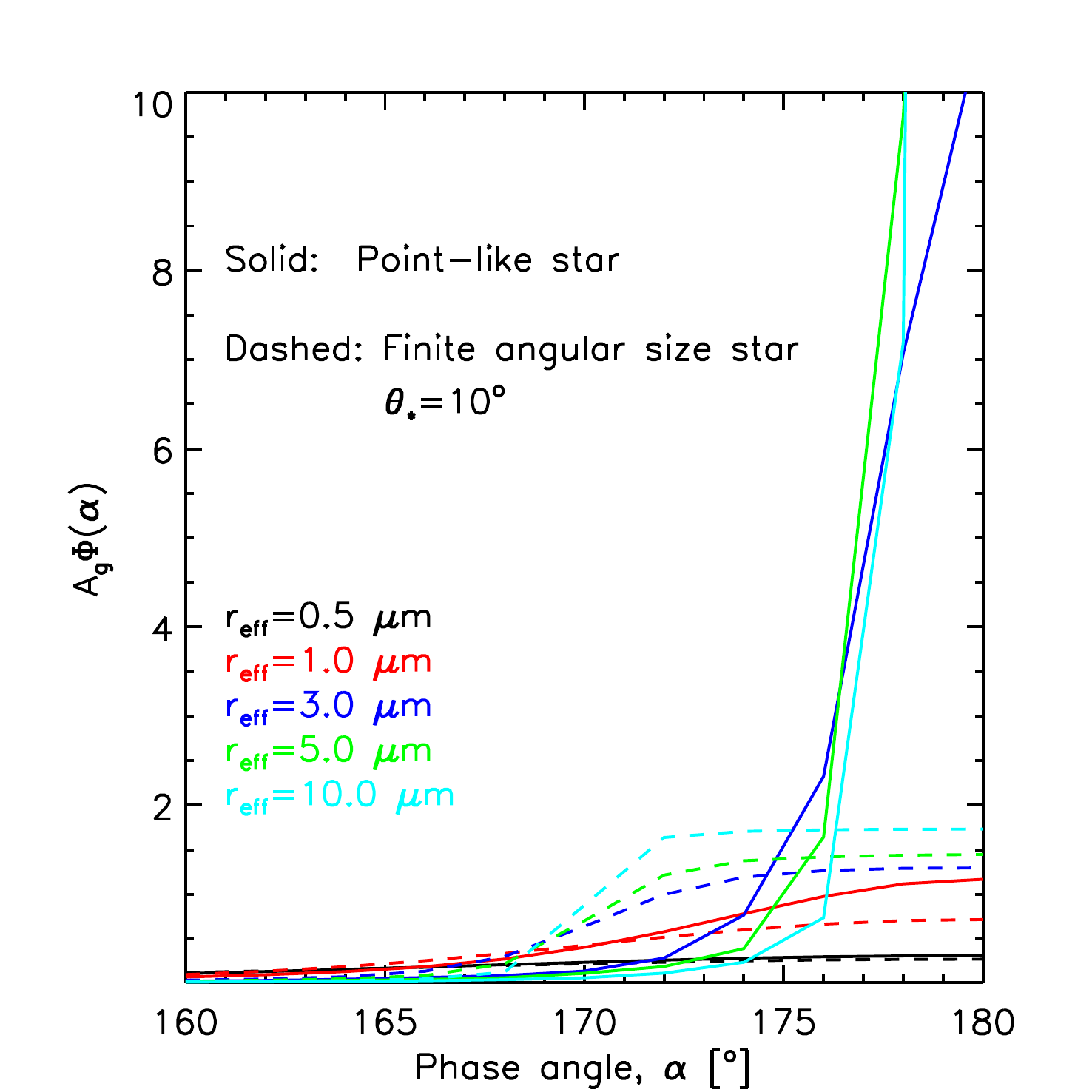}\\
   \includegraphics[width=7.cm]{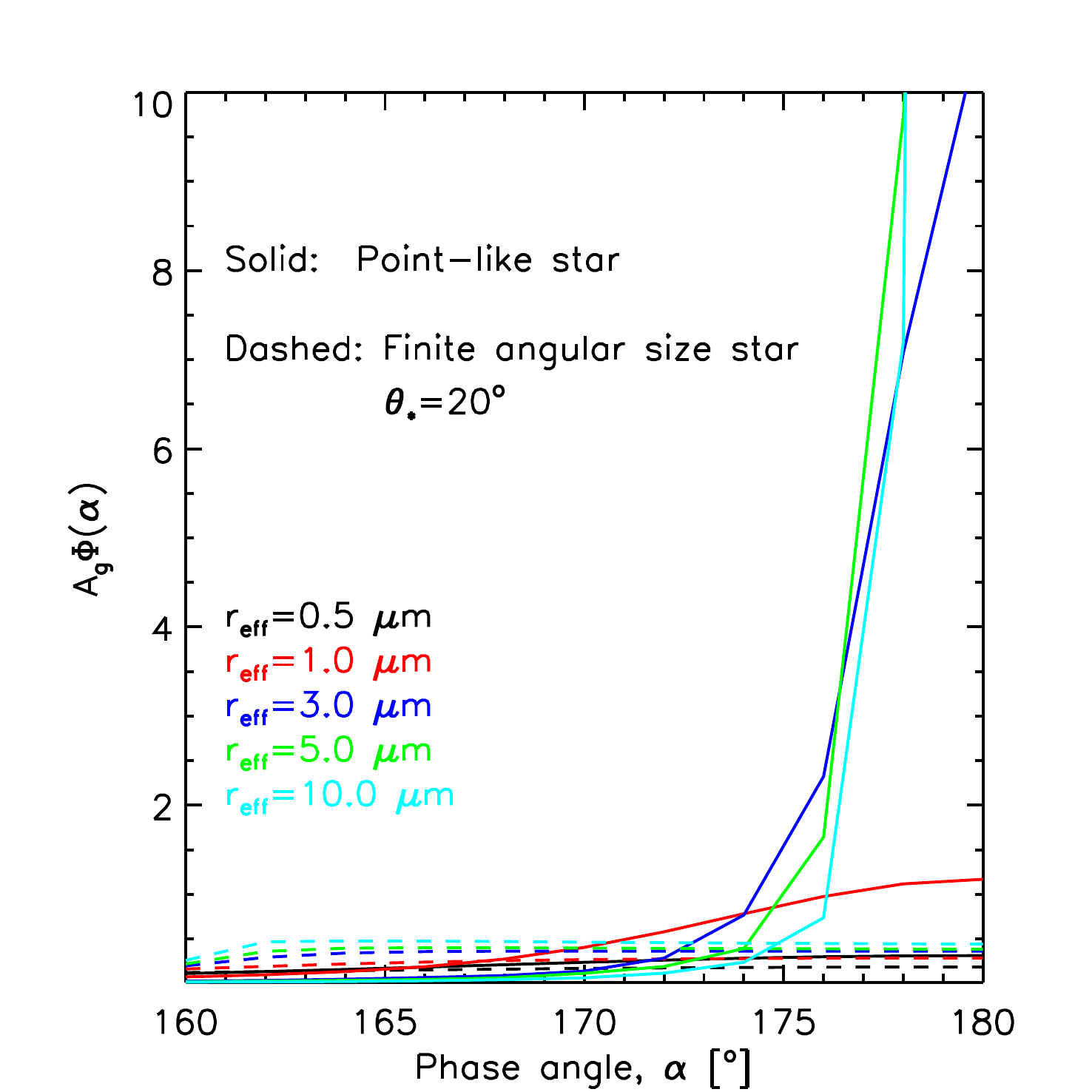}
      \caption{\label{corot24_thetas_fig} 
      \textbf{Top.} Phase curve simulations of CoRoT-24b
     ($H_{\rm{a}}$/$R_{\rm{p}}$$\sim$0.035, see text). Taking into account the finite angular
     size of the star in $<$$p_{\rm{a}}$($\Theta$)$>$ (Fig. (\ref{paconv_fig}))
     results in the leakage of brightness from the larger to the smaller phase angles.
     To convert to planet-to-star contrasts, the quoted
     $A_{\rm{g}}$$\Phi$($\alpha$) should be multiplied by $(R_{\rm{p}}/a)^2$=7.6 ppm.
     \textbf{Middle and Bottom.} Phase curves for $H_{\rm{a}}$/$R_{\rm{p}}$$\sim$0.035, but with 
     the star spanning larger fractions of the sky as viewed from the planet.      
     }
   \end{figure}

\begin{table}
\caption{\label{corot24_table}
Forward scattering from CoRoT-24b. We have varied both $\Delta$$\alpha$ 
(by varying $\alpha_{\rm{C}}$) and $r_{\rm{eff}}$. The tabulated values provide the
estimated photon noise PN and the observable \textbf{O} for each aerosol size. {
The quoted \textbf{O}s in regular typeface refer to calculations in the point-like star
limit. Bold typeface refers to calculations taking into account the finite angular size 
of the star.}
}             
\begin{tabular}{c c c c c c c c}   
\hline                       
$\alpha_{\rm{C}}$  & PN  & \textbf{O}$^{0.5\mu\rm{m}}$  &  \textbf{O}$^{1\mu\rm{m}}$  &  \textbf{O}$^{2\mu\rm{m}}$  & \textbf{O}$^{3\mu\rm{m}}$   & \textbf{O}$^{5\mu\rm{m}}$   &  \textbf{O}$^{10\mu\rm{m}}$ \\
{[$^{\circ}$]} & [ppm] & [ppm] & [ppm] & [ppm] & [ppm] & [ppm] & [ppm] \\
\hline                        
160. &  64.3  &  1.4  &    2.7 &   3.0 &    2.3 &    1.5 &    0.8 \\
 &    &  \textbf{1.4} &   \textbf{2.6} &    \textbf{3.3} &    \textbf{3.5} & \textbf{3.4} & \textbf{3.7} \\

162. &  68.8  &  1.5  &  3.0 &    3.4 &    2.6 &    1.7 &    0.9 \\
 &   &  \textbf{1.4} &   \textbf{2.9} &    \textbf{3.7} &    \textbf{3.9} & \textbf{3.9} & \textbf{4.2} \\

164. &  74.2  &  1.6 &    3.4 &    3.9 &    3.0 &    1.9 &    1.0 \\
 &    &  \textbf{1.5} &   \textbf{3.3} &    \textbf{4.3} &    \textbf{4.6} & \textbf{4.5} & \textbf{4.9} \\

166. &  81.4  &  1.6 &    3.8 &    4.6 &    3.6 &    2.3 &    1.2 \\
 &    &  \textbf{1.6} &   \textbf{3.7} &    \textbf{5.1} &    \textbf{5.4} & \textbf{5.3} & \textbf{5.8} \\
 
168. &  91.1  &  1.7  &   4.4  &  5.5 &    4.4 &    2.8 &    1.4 \\
 &   &  \textbf{1.7} &   \textbf{4.2} &    \textbf{6.1} &    \textbf{6.6} & \textbf{6.5} & \textbf{7.2} \\

170. &  105.4  &  1.8  &    5.0 &    6.9 &    5.6 &    3.5 &    1.8 \\
 &    &  \textbf{1.7} &  \textbf{4.7} &    \textbf{7.5} &    \textbf{8.4} & \textbf{8.5} & \textbf{9.5} \\

172. &  129.6  &  1.9   & 5.7  &    8.9 &    7.6 &    4.7 &    2.4 \\
 &    &  \textbf{1.8} &   \textbf{5.3} &    \textbf{9.3} &   \textbf{11.2} & \textbf{11.8} & \textbf{13.8} \\

174. & 185.6  &  1.9   &    6.5 &   12.2 &   11.3 &    7.4 &    3.5 \\
 &   &  \textbf{1.9} &   \textbf{5.9}  & \textbf{11.7}  & \textbf{15.4} &  \textbf{17.7} &   \textbf{23.3} \\
\hline                        
\end{tabular}
\end{table}

\subsection{Pre-ingress and post-ingress forward scattering}

{A number of Kepler planets exhibit brightness peaks that occur at 
phases somewhat displaced from full illumination
 \citep[e.g.][]{demoryetal2013,angerhausenetal2015,estevesetal2015}. 
For the less strongly irradiated planets, this finding is often explained as caused by clouds 
forming on the nightside that move onto the dayside and then evaporate, thereby
causing an asymmetry in the phase curve \citep{garciamunozisaak2015,shporerhu2015}. 
If the non-uniform cloud reaches to the altitudes probed by forward scattering, it is 
conceivable that the pre-ingress and post-ingress brightness curves will 
have dissimilar slopes.
}

\begin{figure}
   \centering
   \includegraphics[width=9.cm]{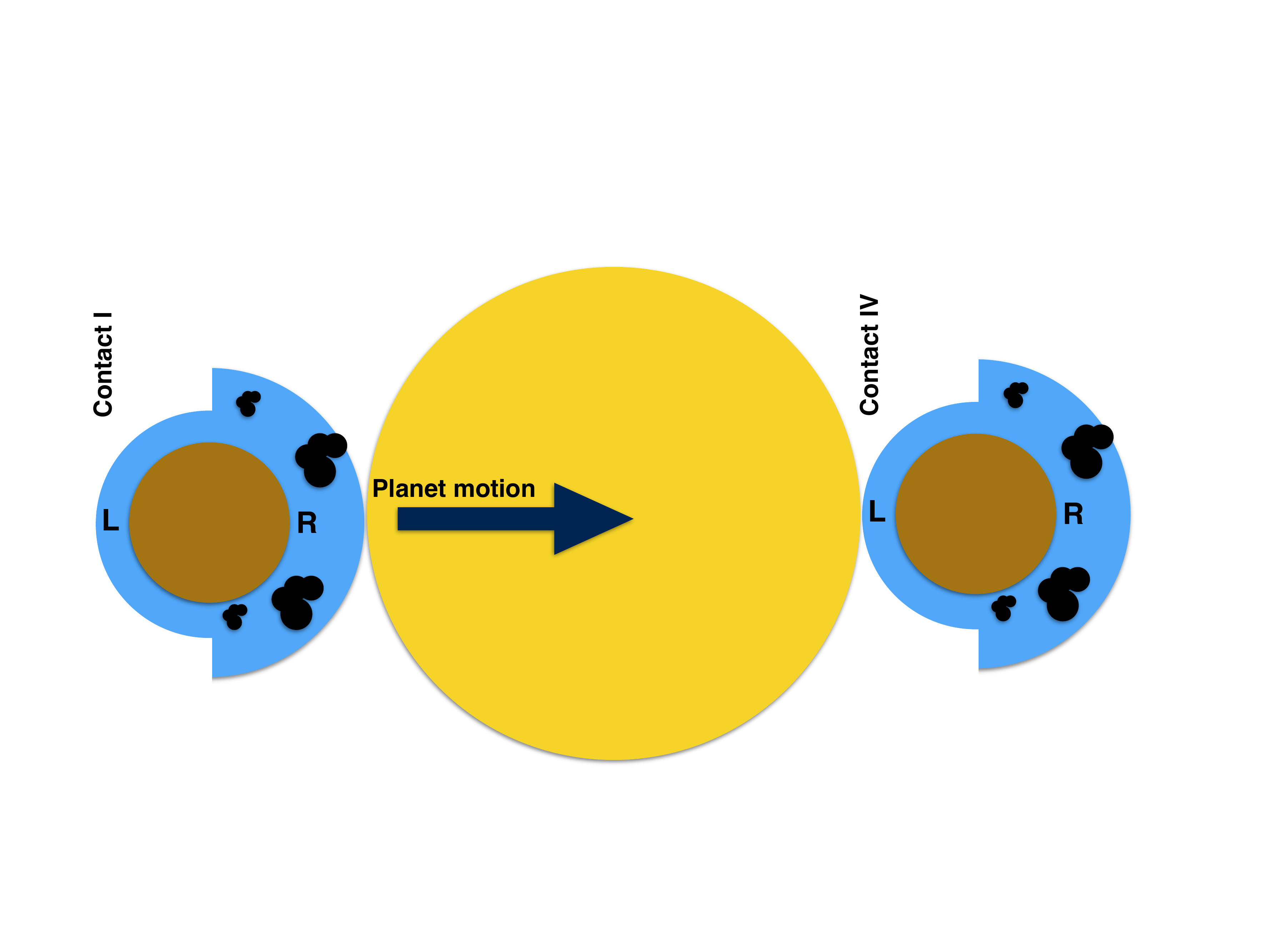}
   \includegraphics[width=9.cm]{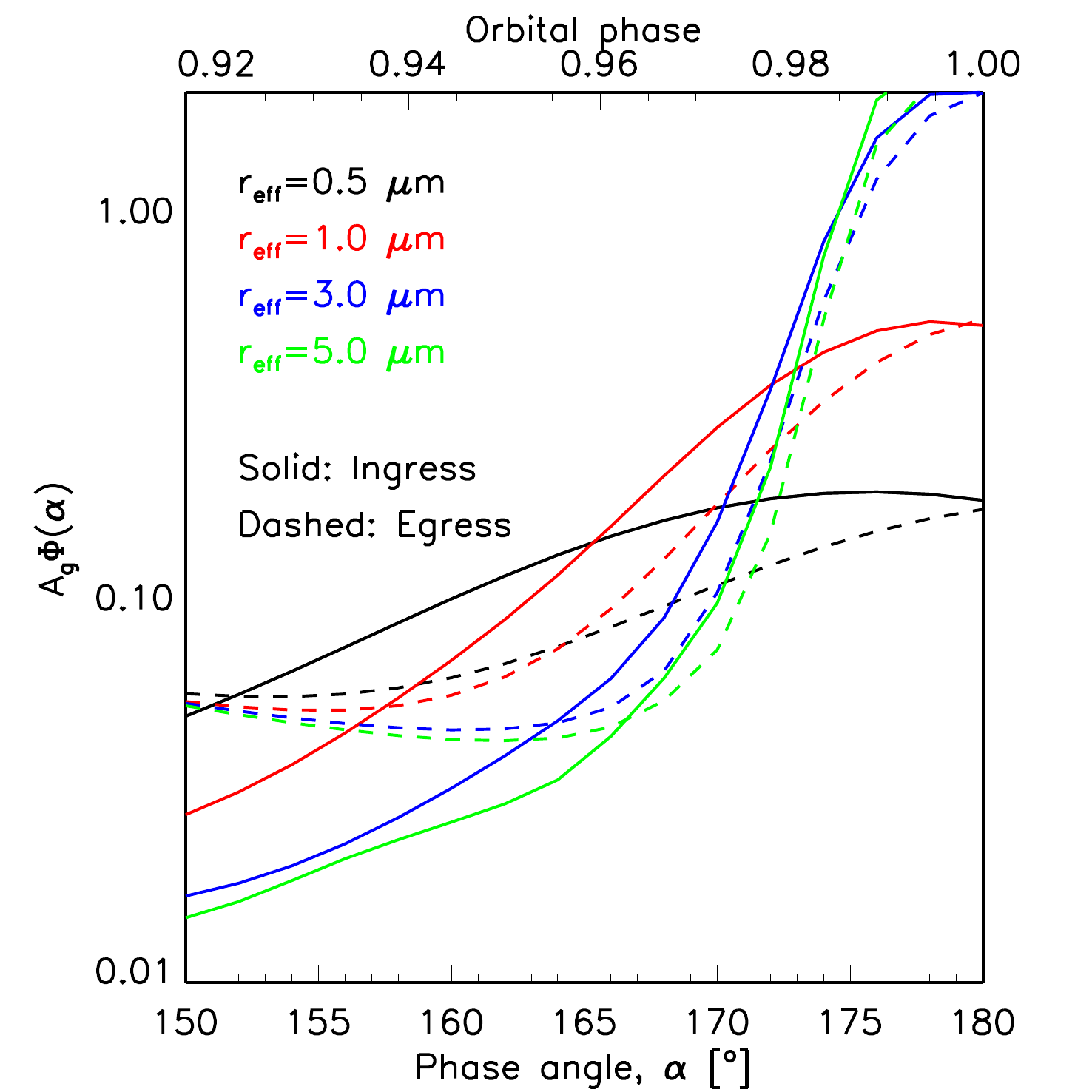}
      \caption{\label{limbscartoon_fig} 
      \textbf{Top}. 
      The cartoon depicts a transiting planet
      at 1st and 4th contacts. 
      The planet's R terminator scatters the incident starlight in the forward direction 
      much more efficiently than the L terminator. This is potentially due to a more 
      extended aerosol layer and/or larger aerosol particles at the R terminator.
      {
      \textbf{Bottom}. Phase curves for a CoRoT-24b-like planet with a hazy terminator (R)
      and an aerosol-free terminator (L). Both ingress and egress curves are shown.
      Contact I and IV are at  $\alpha$$\sim$175.9$^{\circ}$. The simulations take into 
      account the finite angular size of the star.
      }
     }
   \end{figure}

{
We can estimate the importance of this with the simplified model sketched 
in Fig. (\ref{limbscartoon_fig}, Top) as applied to a CoRoT-24b-like planet. In this model,
one of the terminators (L) is aerosol-free and therefore 
forward scattering from it is inefficient, 
whereas the other terminator (R) contains aerosols that scatter efficiently in the forward direction.
The aerosols are vertically distributed with a scale height
$H_{\rm{a,R}}$/$R_{\rm{p}}$=0.035 and have effective $\mbox{<}p_{\rm{a,R}}\mbox{>}$($\Theta$)
as shown in Fig. (\ref{paconv_fig}) for various particle sizes and $\theta_{\star}$=4.1$^{\circ}$.
}

{Figure (\ref{limbscartoon_fig}, Bottom) shows the synthetic phase curves.
The most remarkable characteristic is that the ingress is brighter than the egress. 
The reason for this is that at ingress the hazy terminator 
is seen with a local phase angle that is larger (and the scattering angle smaller)
than the aerosol-free terminator. The difference in phase angles between terminators 
roughly scales as $2R_{\rm{p}}/a$. The magnitude
of this angular resolution element and the details of $\mbox{<}p_{\rm{a,R}}\mbox{>}$($\Theta$) 
 dictate the differences in the ingress/egress curves.
} 

\section{ Blending with other photometric effects}
\label{sec:otherterms_sec}

\subsection{Modulations from stellar tides}

The measured brightness from a close-in planet-star system includes the 
contribution from the planet atmosphere together with modulations due to 
Doppler beaming and tidal ellipsoidal distortion of the star.
The magnitude, period and lag with respect to the orbital motion
of these modulations are sources of information on both the planet and the star  
\citep[e.g.][for a recent review]{shporer2017}.

Assuming that the planet is on a circular orbit and the planet-star system is seen
edge-on, 
the brightness modulation from Doppler beaming is $\propto$sin($\alpha$), 
whereas the modulation due to ellipsoidal distortion is $\propto$cos($2\alpha$)
with additional correcting terms $\propto$cos($\alpha$) and cos(3$\alpha$). 
The theoretical treatment of these phenomena provides expressions 
for the coefficients of each term, 
which depend on e.g. $a$, $R_{\star}$, $M_{\rm{p}}$ and $M_{\star}$ 
\citep{morrisnaftilan1993,loebgaudi2003}. 
The planet atmosphere modulates the planet-star system brightness in two different ways. 
Thermal radiation prevails when the atmosphere is hot and/or the observations are made at
long wavelengths. Reflected starlight dominates at low temperatures and/or short wavelengths. 
With enough photometric precision and multi-wavelength observations, 
it is possible to {disentangle these phenomena, also the two 
atmospheric terms \citep{placeketal2016},} by fitting the observations to models
\citep[e.g.][]{angerhausenetal2015,estevesetal2015}.

The photometric effects described above may blend in the observed brightness signal 
and cause degeneracies in the interpretation of the inferred physical properties
\citep{mislishodgkin2012}. 
Aggravated by moderate signal-to-noise ratios and
a possibly incomplete understanding of each photometric effect, these difficulties
may be at the heart of the 
mass discrepancy reported for some planet-star systems \citep{shporer2017}.
Indeed, it has become apparent that the planet masses inferred from photometric measurements
(through Doppler beaming or tidal ellipsoidal distortions) and from radial velocities 
are at times mutually inconsistent. 

\begin{figure}
   \centering
   \includegraphics[width=9.cm]{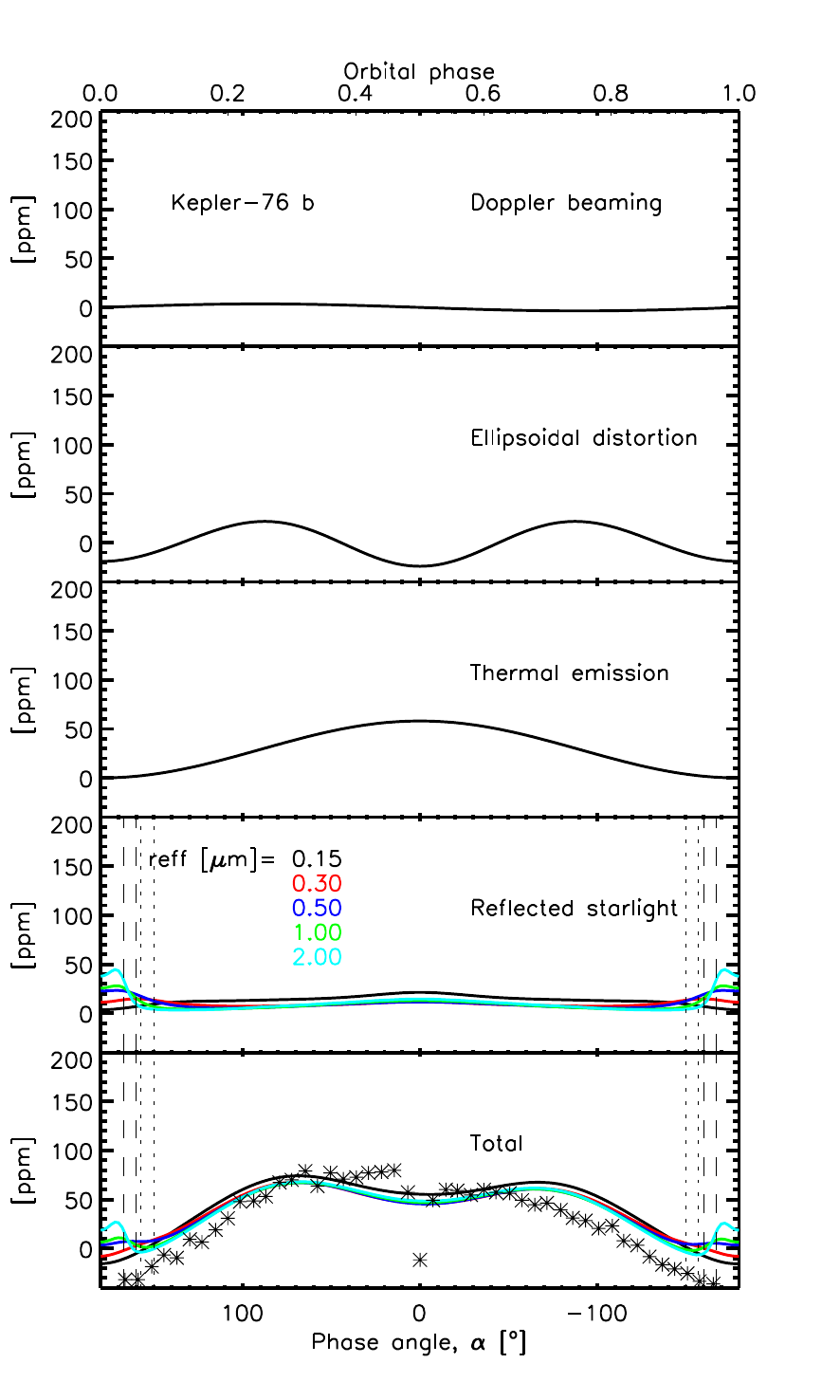}
      \caption{\label{harmonics_fig}
      Photometric modulations for a Kepler-76b-like planet. 
      {
      The Doppler beaming and
      ellipsoidal distortion contributions are}
      based on \citet{estevesetal2015}. 
      {
      We model the thermal emission component as (1+cos($\alpha$))/2,  
      premultiplied by a coefficient of 58 ppm consistent with integrated blackbody radiation 
      between 0.45 and 0.85 $\mu$m for an estimated $T_{\rm{eq}}$=2551 K.
      }
      The reflected starlight calculations
      are based on a FeO aerosol, which explains the low reflectance at 
      small phase angles. {As in the case of CoRoT-24 b 
      (Figs. \ref{corot24_thetas_fig}-\ref{limbscartoon_fig}), the multiple scattering 
      simulations were done considering the finite size of the star, and $\theta_{\star}$$\approx$12.8$^{\circ}$.}
      {The symbols in the bottom plot are the measurements reported by 
      \citet{estevesetal2015} in their Fig. 3, shifted in the vertical by 30 ppm. 
      }
      {The vertical lines in the two bottom graphs represent the positions
      for [$\alpha_{\rm{A}}$, $\alpha_{\rm{B}}$] (dotted) and [$\alpha_{\rm{C}}$, $\alpha_{\rm{D}}$] (dashed)
      as described in the preparation of Table (\ref{derivedparameters_table}).}
      }
   \end{figure}

Figure (\ref{harmonics_fig}) suggests that forward scattering 
{from a horizontally uniform planet}  
will leak into cos($2\alpha$) and higher order {even} cosine harmonics, thereby blending with the
stellar tide modulation. 
The overall effect is a partial cancellation of the ellipsoidal effect, 
{particularly at large phase angles and if the angular size of the star 
is also large, } 
and in turn a reduced planetary mass as estimated from {this photometric effect}. 
We have briefly explored to what extent forward scattering affects the photometric mass 
inferred from the stellar tide modulations in the cases of TrES-2b \citep{barclayetal2012} and 
Kepler-76 b \citep{faigleretal2013}, with quoted ellipsoidal 
semi-amplitudes of 2.8 and 21.5 ppm, respectively. 
We find that 
2$\pi$$R_{\rm{p}}$$H_{\rm{a}}$/$a^2$$\sim$5 (TrES-2b) and $\sim$10 (Kepler-76b) ppm,
{which means that blending of forward-scattered starlight with the 
tidal brightness modulation is a priori possible.}
The fact that the masses retrieved from 
stellar tide modulations 
($M^{\rm{ell}}_{\rm{p}}$/$M_{\rm{J}}$=1.06$^{+0.28}_{-0.23}$ 
for TrES-2b, and 2.1$\pm$0.4 for Kepler-76b) 
and radial velocities for both planets 
($M^{\rm{RV}}_{\rm{p}}$/$M_{\rm{J}}$=1.206$\pm0.045$ 
for TrES-2b, and 2.00$\pm$0.26 for Kepler-76b) are in good agreement
\citep{barclayetal2012,faigleretal2013}
suggests that neither of these planets exhibit significant forward scattering, 
which puts an additional constraint on their atmospheres. 

{It is worth noting the cases of Kepler-12b and -412b, listed on Tables 
(\ref{derivedparameters_table})--(\ref{derivedparameters2_table}), 
whose phase curves have been published by 
\citet{angerhausenetal2015} and \citet{estevesetal2015}. 
Kepler-12b's curve is
distinctly asymmetric with respect to occultation, 
a fact likely attributable to a horizontally non-homogeneous atmosphere. 
Other Kepler planets also show asymmetric atmospheric contributions
\citep{angerhausenetal2015,estevesetal2015}.}
According to our estimates, both planets might exhibit forward 
scattering signals of up to 10-20 ppm. 
A look at the corresponding curves in Esteves et al. (2015) however does not 
reveal clear evidence for forward scattering in the case of Kepler-412b, 
although it hints at a tentative brightness surge at orbital phases close to one 
in the case of Kepler-12b. A thorough analysis incorporating the data from all 
Kepler quarters might provide a more definitive answer.

Clearly, further work is needed
to quantify these contributions and extend the analysis to all planets with 
accurate photometric data available. 
Because the forward scattering signal
scales as $M^{-1}_{\rm{p}}$ and $a^{-5/2}$ (Eq. \ref{FpFstar_eq2}), 
and the amplitude of the ellipsoidal tidal distortion scales as
$M_{\rm{p}}$ and $a^{-3}$, each effect will likely dominate in a different
region of the $M_{\rm{p}}$--$a$ parameter space.

\subsection{Transits}

During a transit, the host star dims by an amount that depends on the 
planet size and its atmospheric structure. {
For an exponential atmosphere described by a single absorber (the conditions explored here),
and omitting the scattering towards the observer of photons having one or more collisions in the atmosphere, 
the planet  appears effectively opaque up to the so-called equivalent height $h_{\rm{eq}}(\lambda)$. 
To a good approximation, the equivalent height 
occurs where the limb opacity of the atmosphere $\tau_{\rm{limb}}$=0.56
\citep{karkoschkalorenz1997, lecavelierdesetangsetal2008}.} 

Part of the starlight that is intercepted by the atmosphere during the transit is 
restored into the forward direction and scattered towards the observer
\citep{brown2001,hubbardetal2001,garciamunozetal2012,garciamunozmills2012,dekokstam2012,robinson2017}.
Under the assumption of an exponential atmosphere, 
Eq. (\ref{F1b_eq}) quantifies how many of these photons reach the observer, 
thereby reducing the transit depth by 2$R_{\rm{p}}$$\Delta$/$R_{\star}^2$
or the equivalent to an annulus area of radius $R_{\rm{p}}$ and width $\Delta$:
\begin{equation}
\frac{\Delta}{H_{\rm{a}}}=\pi \left( \frac{R_{\star}}{a}  \right)^2 \mbox{<}p_{\rm{a}}\mbox{>}(\Theta=0) \varpi_{0,\rm{a}}.
\label{delta_eq}
\end{equation}
{This is also akin to diminishing the transmission-only equivalent height, 
$h_{\rm{eq}}$, by $\Delta$, which means that
the measurable equivalent height during the transit is 
$h'_{\rm{eq}}$=$h_{\rm{eq}}$$-$$\Delta$ rather than $h_{\rm{eq}}$.
}
{
$\Delta$ depends on wavelength through the 
aerosol properties $\mbox{<}p_{\rm{a}}\mbox{>}$$(\Theta$$=$$0)$ and $\varpi_{0,\rm{a}}$.
Typically, the larger the particle radius $r_{\rm{eff}}$ the larger the effective 
 $\mbox{<}p_{\rm{a}}\mbox{>}$($\Theta$=0) (Fig. \ref{paconv_fig}), and 
in turn the impact of forward scattering on the transit depth. 
In an aerosol-rich atmosphere, $\varpi_{0,\rm{a}}$ will depend strongly on wavelength
if there are strong gas absorption bands in the spectral range of interest. Within the 
gas absorption band, $\varpi_{0,\rm{a}}$ can become significantly smaller than in the
continuum, and in turn $h'_{\rm{eq}}$$\approx$$h_{\rm{eq}}$ at the specific wavelengths.
} 

{
Interestingly, the 
angular size of the star enters into Eq. (\ref{delta_eq}) both directly 
($\theta_{\star}$$\approx$$R_{\star}$/$a$) and indirectly through $\mbox{<}p_{\rm{a}}\mbox{>}$($\Theta$=0)
(Eq. \ref{Int4_eq}, Fig. \ref{paconv_fig}). The two effects partially cancel out.
For large orbital distances, the $a^2$ term in the denominator of Eq. (\ref{delta_eq})
dominates and $\Delta$/$H_{\rm{a}}$ becomes small; for small orbital distances, 
the convolution of $p_{\rm{a}}$($\theta$) over an extended solid angle 
results in a reduced  $\mbox{<}p_{\rm{a}}\mbox{>}$($\Theta$=0)
with respect to $p_{\rm{a}}$($\theta$=0).
Figure (\ref{patheta_fig}) incorporates the information presented in Fig. (\ref{paconv_fig})
for $\mbox{<}p_{\rm{a}}\mbox{>}$($\Theta$=0)
and shows that forward scattering will reduce the equivalent height
of the atmosphere by typically less than one scale height, even for the more extreme
configurations ($\theta_{\star}$=20$^{\circ}$, $r_{\rm{eff}}$=10 $\mu$m). 
The connection of $\Delta$ with the particle size is more direct through the analytical
expression of Eq. (\ref{delta_eq}) than in the treatments by \citet{dekokstam2012} and
\citet{robinson2017}, who base their analyses on Henyey-Greenstein parameterizations of 
the aerosols scattering phase function. 
}

\begin{figure}
\centering
\includegraphics[width=8.cm]{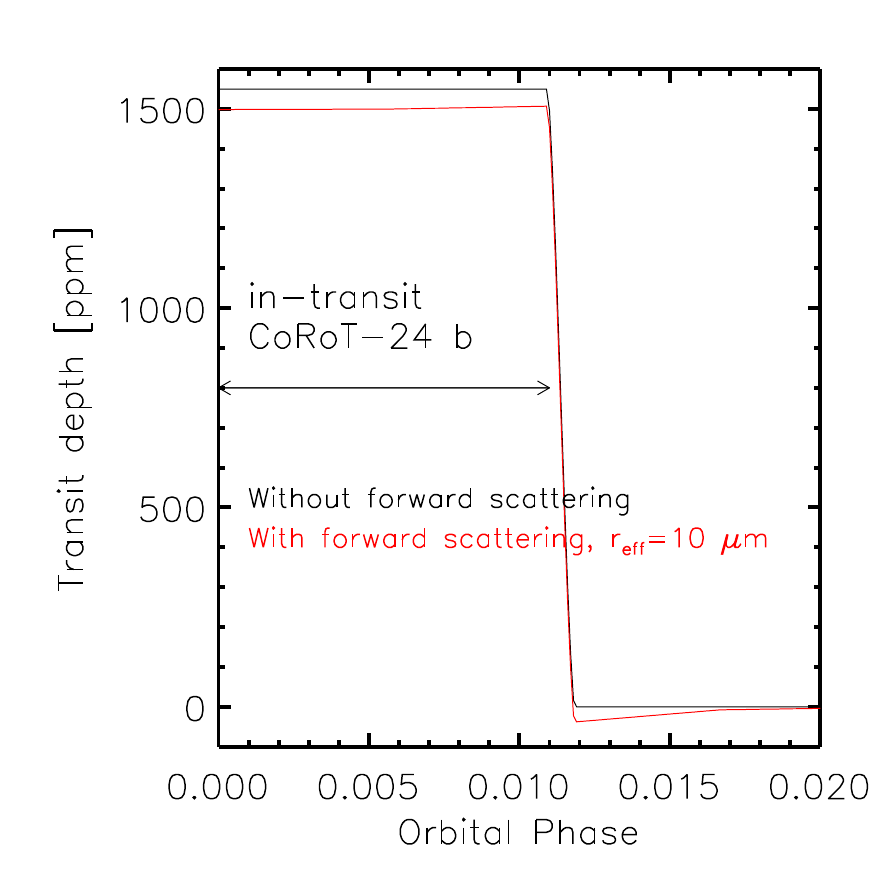}
\includegraphics[width=8.cm]{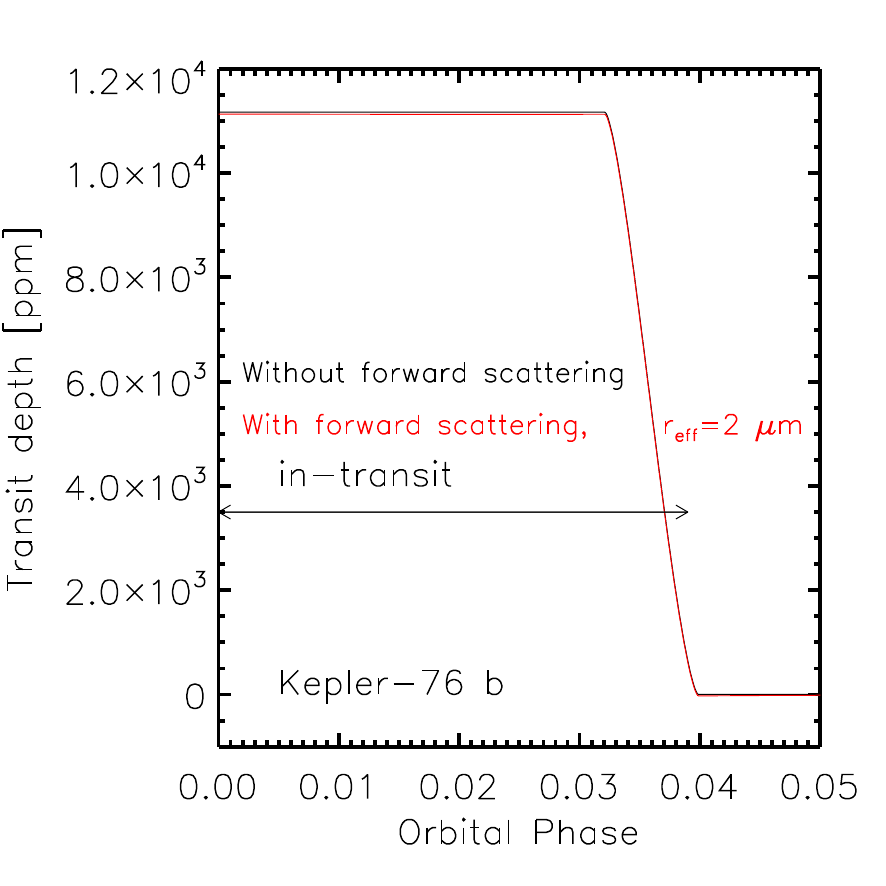}
\caption{\label{transit_fig}
{
Transit depth as a function of orbital phase. For the transit light curve, we assumed the
brightness is uniform over the stellar disk. Forward scattering reduces to some extent
the transit depth. In the two examples shown here, the reduction amounts to 
$\sim$50 ppm (CoRoT-24 b) and $\sim$36 ppm (Kepler-76 b), which is less than the equivalent
to a gas pressure scale height.}
}
   \end{figure}

\begin{figure}
\centering
\includegraphics[width=9.cm]{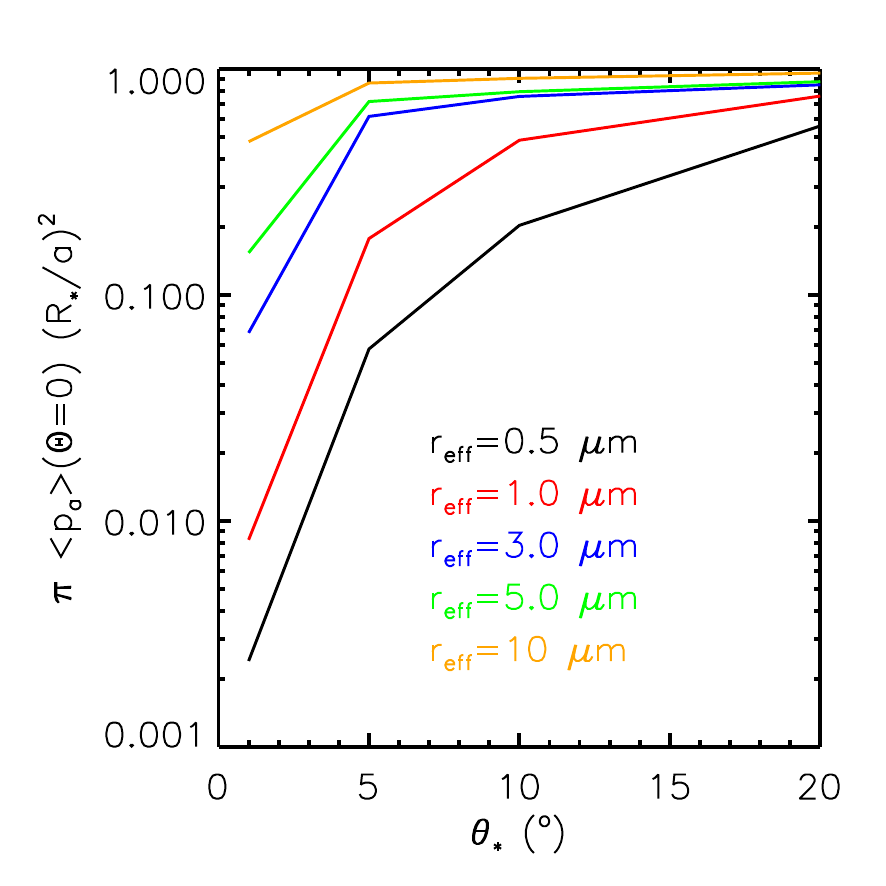}
\caption{\label{patheta_fig}
Forward scattering contribution to the equivalent height of the atmosphere, normalized
by the scale height and adopting $\varpi_{\rm{0,a}}$=1, according to Eq. (\ref{delta_eq}). 
Each curve assumes aerosols of a specific particle radius. 
The values of $\mbox{<}p_{\rm{a}}\mbox{>}$($\Theta$=0) are from Fig. (\ref{paconv_fig}). 
According to these curves, 
forward scattering will reduce the measured equivalent height of the atmosphere by less than
one scale height, even in the most extreme conditions of particle size and star angular 
radius explored here. 
}
   \end{figure}

{For completeness, Fig. (\ref{transit_fig}) shows the transit depth as a
function of orbital phase for both CoRoT-24 b and
Kepler-76 b. For CoRoT-24 b ($\theta_{\star}$$\approx$4.1$^{\circ}$), it is shown 
the case with aerosols of particle size $r_{\rm{eff}}$=10 $\mu$m, which results in a 
correction of the transit depth at mid-transit due to forward scattering 
of about 50 ppm. 
For Kepler-76 b  ($\theta_{\star}$$\approx$12.8$^{\circ}$), the 
corresponding graph shows the case with $r_{\rm{eff}}$=2 $\mu$m, which results in a 
correction of about 36 ppm. Both corrections correspond to a change in 
the equivalent height of less than their estimated gas pressure scale heights.
}

{
We have assumed throughout this work
an effective wavelength $\lambda_{\rm{eff}}$=0.65 $\mu$m. 
Because in Mie theory the diffraction peak of the aerosols is largely dictated 
by the size parameter $x_{\rm{eff}}$=2$\pi$$r_{\rm{eff}}$/$\lambda_{\rm{eff}}$, 
Fig. (\ref{patheta_fig}) can be reworked at other wavelengths by appropriately selecting
the particle radius.
}

{\citet{devoreetal2016} have shown that forward scattering can significantly 
modify the
transit light curve of ultra-short period planets surrounded by dust clouds. 
The difference with our treatment is that these authors assume that the entire (and sizable) 
cloud is uniform
in its dust content, and thus every element of it can scatter the incident
starlight towards the observer. In our treatment, 
the exponential variation of the optical properties reduces
the effective scattering area to a relatively narrow ring around the planet
of width about an atmospheric scale height.
}

\section{Summary}
\label{sec:summary_sec}

A main goal of this work is to raise awareness about the 
diagnostics possibilities of exoplanet brightness measurements at large phase angles. 
As for Saturn's moon Titan, 
a brightness surge when the planet approaches back-illumination 
will provide joint information on atmospheric stratification and aerosol optical properties. 
This is valuable insight difficult to gain by other means. 
{It is unclear how common  this optical phenomenon is, 
but its possibility justifies a dedicated search with existing and
future observations.}

In the framework of exponential atmospheres, 
{we derived an analytical expression for forward-scattered starlight 
in the single-scattering limit. The expression incorporates the effects of the angular
size of the star, one of which is to convolve the aerosols scattering phase function with the
brightness distribution over the stellar disk.
}
Based on this expression, we estimate 
that there are a number of exoplanets with a priori suitable conditions for 
forward scattering. We have refined these predictions with a more elaborate
assessment that considers the shape of the phase curve and the time elapsed during the
brightening phase. Some of these planets potentially exhibit brightness surges of up 
to tens of ppms. 

{At out-of-transit phases, considering the finite angular size of the star tends
to increase the amount of starlight forward-scattered 
towards the observer with respect
to the treatment in the point-like star limit.}
{On the contrary, during the transit the finite angular size of the star
reduces the amount of starlight that reaches the observer
with respect to the point-like star limit. Once the latter effect is
considered, it is seen that forward scattering will modify the 
equivalent height of the atmosphere by less than one scale height in most configurations. 
For future reference, we show how to take into account 
the finite angular size of the star in Backward Monte Carlo radiative transfer models.}
{Our study raises the possibility that, given the appropriate 
atmospheric structure, some low-density planets 
may be easier to detect at large phase angles than in occultation. 
}

{Throughout our treatment, we assumed that aerosols dominate the atmospheric
opacity at the optical radius level. Two additional key assumptions are that the
aerosols are vertically distributed with a scale height equal to the gas
scale height, and that the aerosols are described as having a single 
particle size that we prescribe but do not predict. 
This simplified treatment is similar to the way transit
spectra are often interpreted within retrieval algorithms. 
The reality of exoplanet atmospheres will surely be more complex, but
our simplified approach at least enables a direct comparison with observations
from which to draw physical conclusions. 
} 
{
It may well happen that some of the
planets that are ranked higher as candidates for strong forward scattering
will have no detectable signal because their atmospheres do not form aerosols 
or the aerosol particles at the optical radius level are not large enough. }
Even then, and if observations of high enough precision exist, 
a non-detection will provide 
constraints on the atmospheric structure that can be tested
against microphysical models.  
{This possibility should motivate further studies on the microphysics of aerosols
in the diverse range of conditions found in exoplanet atmospheres 
\citep[e.g.][]{hellingfomins2013,lavvasetal2013,lavvaskoskinen2017,leeetal2016}.
Helpful information that could be obtained from such investigations includes: the
aerosol scale height at the optical radius level, and the particle size of the condensates
dominating the continuum opacity at the corresponding altitudes.
}

Last, 
we emphasize that the starlight reflected by exoplanets varies with 
phase in manners that are not necessarily well described by simple formulations such
as Lambert's law. 
It is challenging to decide when more elaborate descriptions are justified, 
but it is equally important to realize that oversimplified descriptions will likely wash out unique
information on the planet atmosphere.

\section*{Acknowledgements}

AGM gratefully acknowledges conversations with Agust\'in S\'anchez-Lavega 
(Universidad del Pa\'is Vasco/Euskal Herriko Unibertsitatea, Spain)
and Panayotis Lavvas (Universit\'e de Reims Champagne-Ardenne, France).
This research has made use of the Exoplanet Orbit Database
and the Exoplanet Data Explorer at exoplanets.org.
This research has made use of the NASA Exoplanet Archive, 
which is operated by the California Institute of Technology, 
under contract with the National Aeronautics and Space Administration 
under the Exoplanet Exploration Program.
This research has also made use of TEPCat (http://www.astro.keele.ac.uk/jkt/tepcat/),
the Extrasolar Planets Encyclopaedia (exoplanet.eu)
and the Mikulski Archive for Space Telescopes 
(http://archive.stsci.edu/kepler/kepler$\_$fov/search.php).








\appendix

\section{ Starlight forward-scattered by a planet. 
An analytical expression for exponential atmospheres.}
\label{sec:appendixa}

We derive an expression for the starlight forward-scattered by an exponential atmosphere at 
a phase angle of 180$^{\circ}$.  In its general form, the expression considers the
finite angular size of the star. 
A few simplifying assumptions (SAs) enable us to treat
the problem analytically:
\begin{enumerate}
\item \label{SAstratified} The atmosphere is horizontally uniform but vertically stratified.
\item The optical properties of the atmosphere are determined by ubiquitous 
particles that are referred to as aerosols (subscript $_a$)
but that may actually be a mix of gases and condensates.
\item \label{SAexponential} The aerosols extinction and scattering coefficients,  
$\gamma_{\rm{a}}$ and $\beta_{\rm{a}}$, 
drop exponentially in the vertical with the scale height $H_{\rm{a}}$. 
The single scattering albedo $\varpi_{\rm{0,a}}$$=$$\beta_{\rm{a}}$/$\gamma_{\rm{a}}$ is 
altitude-independent.
\item \label{SAptheta} The aerosols scattering phase function
$p_{\rm{a}}(\theta)$ is also altitude-independent. 
$\theta$ is the scattering angle between the incident and exit directions of 
a photon being scattered by an aerosol particle.
\item \label{SAscaleheight} The ratio of the aerosol scale height $H_{\rm{a}}$ and the planet 
radius $R_{\rm{p}}$ satisfies $H_{\rm{a}}/R_{\rm{p}}$$<<$1.
\item \label{SA5} The starlight forward-scattered by the atmosphere is 
dominated by singly-scattered photons. 
The validity of this assumption is tested 
{a posteriori} in Appendix \ref{sec:appendixb}
by comparison with numerical solutions to the multiple-scattering
problem.
\end{enumerate}

\begin{figure}
   \centering
   \includegraphics[width=9.cm]{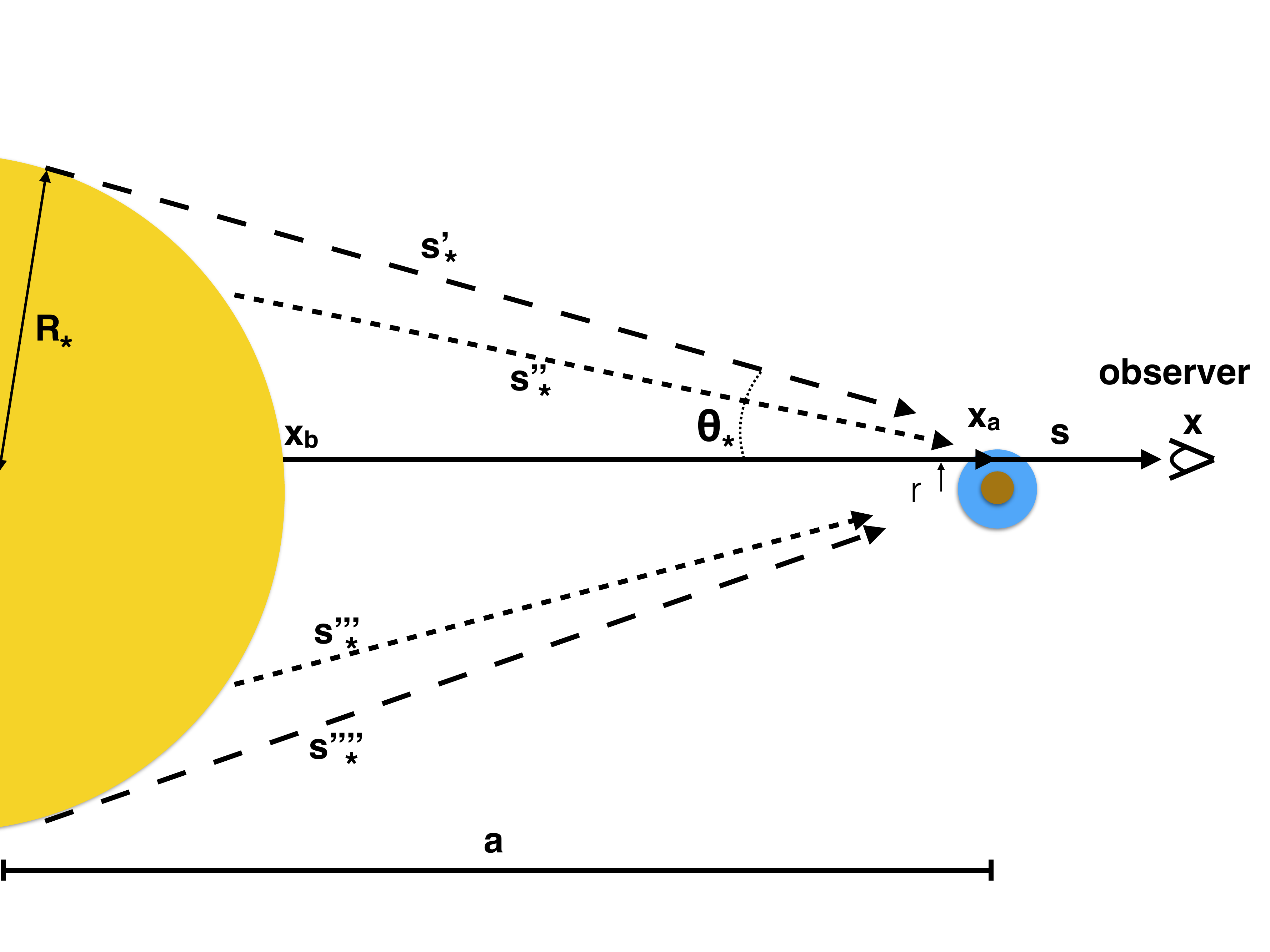}
      \caption{\label{forwardscattering_fig} Mid-transit
      geometry relevant to the derivation of Eq. (\ref{F1b_eq}). The various
      $\mathbf{s_{\star}}$ directions describe radiation rays emerging from the star; 
      their corresponding photons undergo collisions with the atmospheric aerosols 
      at $\mathbf{x_{\rm{a}}}$ and are scattered into direction $\mathbf{s}$ towards the
      observer. $\theta_{\star}$ is the angular radius of the star from the planet.
     }
   \end{figure}

{
Figure (\ref{forwardscattering_fig}) sketches the star-planet-observer
configuration at mid-transit for an impact parameter equal to zero, 
when $\alpha$ is identically equal to 180$^{\circ}$.
}
The radiance at $\mathbf{x}$
in direction $\mathbf{s}$ due to stellar photons that either go through the atmosphere
without scattering (and without altering their trajectories by atmospheric refraction) 
or that are scattered once is:
{
\begin{equation}
I\mathbf{(x,s)}=
t(\mathbf{x,x_{\rm{b}}})
I\mathbf{(x_{\rm{b}},s)}
+\int_{\mathbf{x_{\rm{b}}}}^{\mathbf{x}} d\ell{_{\rm{a}}}
t(\mathbf{x,x_{\rm{a}}}) \beta(\mathbf{x_{\rm{a}}}) \times
\label{Ixs_eq}
\end{equation}}
{
$$
\times
\int_{\Omega_{\star}} d\Omega(\mathbf{s'})
{p_{\rm{a}}(\mathbf{x_{\rm{a}}, \mathbf{s'}})} I\mathbf{(x_{\rm{a}},\mathbf{s_{\star}})}. 
$$}
Here, $t(\mathbf{x,x_{\rm{a}}})$ and $t(\mathbf{x,x_{\rm{b}}})$ are the atmospheric 
transmittances between $\mathbf{x}$ and $\mathbf{x_{\rm{a}}}$, and between 
$\mathbf{x}$ and $\mathbf{x_{\rm{b}}}$, respectively. 
$I\mathbf{(x_{\rm{b}},s)}$ is the outgoing radiance in direction $\mathbf{s}$ 
from the stellar surface element at $\mathbf{x_{\rm{b}}}$.   
For simplicity, we assume that the star emits as a blackbody of 
radiance $B_{\star}$, and omit limb darkening, which means that 
$I\mathbf{(x_{\rm{b}},s)}$$=$$B_{\star}$. 
$d\ell{_{\rm{a}}}$ is the differential arc-length along $\mathbf{s}$ at $\mathbf{x_{\rm{a}}}$, 
and $\beta(\mathbf{x_a})$ is the scattering coefficient of the medium at $\mathbf{x_{\rm{a}}}$.
The second integral represents the starlight that is scattered at $\mathbf{x_{\rm{a}}}$
into direction $\mathbf{s}$ from all directions $\mathbf{s_{\rm{\star}}}$
emerging at the stellar surface. 
The scattering angle $\theta$ is locally defined by the
dot product of the $\mathbf{s}$ and $\mathbf{s}_{\star}$ directions, 
$\cos{\theta}$$=$$\mathbf{s}\cdot\mathbf{s}_{\star}$, and 
is equal to zero in forward scattering.
$d$$\Omega$ is the differential solid angle 
subtended by the star from $\mathbf{x_{\rm{a}}}$, and 
${p_{\rm{a}}(\mathbf{x_{\rm{a}}, \mathbf{s'}})}$ is the corresponding 
aerosol scattering phase function. 
We normalize ${p_{\rm{a}}(\theta)}$ (SA\ref{SAptheta}) so that its 
integral over the 4$\pi$ solid angle is one. 
{This normalization differs from the more conventional approach of 
making the integral of ${p_{\rm{a}}(\theta)}$ over all directions equal to 4$\pi$.} 
In our normalization, 
for isotropic scattering ${p_{\rm{a}}(\theta)}$=1/4$\pi$,
and for Rayleigh scattering ${p_{\rm{a}}(\theta)}$=1/4$\pi$ (1$+$$\cos^2{\theta}$).

For evaluating the second integral of Eq. (\ref{Ixs_eq}), we 
adopt 
$I\mathbf{(x_{\rm{a}},\mathbf{s_{\star}})}$$\approx$$B_{\star}$$t(\mathbf{x_{\rm{a}},x_{\rm{b}}})$,
which assumes that the opacity from $\mathbf{x_{\rm{a}}}$ to $\mathbf{x_{\rm{b}}}$ 
is representative of the opacity from $\mathbf{x_{\rm{a}}}$ in any of the possible
$-\mathbf{s_{\rm{\star}}}$ directions towards the star. 
Also, we take $d$$\Omega$$\approx$2$\pi$$d$($\cos{\theta}$),
which tacitly assumes that all points $\mathbf{x_{\rm{a}}}$ are near the 
planet-star axis. 
With this, the integral becomes:
\begin{equation}
2\pi B_{\star} t(\mathbf{x_{\rm{a}},x_{\rm{b}}}) \int_{\cos{\theta_{\star}}}^{1} d(\cos{\theta}) p_{\rm{a}}(\theta), 
\label{Int2_eq}
\end{equation}
where $\cos\theta_{\star}$=$\sqrt{1-(R_{\star}/a)^2}$. 
As usual, $R_{\star}$ and $a$ are the stellar radius and planet orbital distance, respectively.
The integral of Eq. (\ref{Int2_eq}) has the meaning of an aerosol scattering phase
function averaged over the angular size of the star and, in general, requires numerical evaluation.
We term this general treatment the finite angular size star approach, and is relevant
when the angle $\theta_{\star}$ subtended by the star from the planet is not small. 
In contrast, the so-called point-like star approach is appropriate when the planet-star orbital distance is
large enough that all stellar photons reaching the planet can be assumed to be collimated
and $\theta_{\star}$$\rightarrow$0. 
In this latter approach, Eq. (\ref{Int2_eq}) can be simplified further into:
\begin{equation}
\pi \left(\frac{R_{\star}}{a} \right)^2 B_{\star} t(\mathbf{x_{\rm{a}},x_{\rm{b}}})    
 p_{\rm{a}}(\theta=0).
\label{Int3_eq}
\end{equation}
For convenience, we introduce an aerosol scattering phase function averaged over the stellar angular
size:
\begin{equation}
 \mbox{<}p_{\rm{a}}\mbox{>}(\Theta=0)=2 \left(\frac{a}{R_{\star}}\right)^2 
 \int_{\cos{\theta_{\star}}}^{1} d(\cos{\theta}) p_{\rm{a}}(\theta), 
\label{Int4_eq} 
\end{equation}
and rewrite Eq. (\ref{Int2_eq}) as:
\begin{equation}
\pi \left(\frac{R_{\star}}{a} \right)^2 B_{\star} t(\mathbf{x_{\rm{a}},x_{\rm{b}}})    
 \mbox{<}p_{\rm{a}}\mbox{>}(\Theta=0)
\label{Int4_eq}
\end{equation}
which encompasses both Eqs. (\ref{Int2_eq}) and (\ref{Int3_eq}).
Obviously, in the point-like star limit, 
$\mbox{<}p_{\rm{a}}\mbox{>}(\Theta=0)$$\rightarrow$$p_{\rm{a}}(\theta=0)$, and 
Eq. (\ref{Int4_eq}) reduces to Eq. (\ref{Int3_eq}). {
Both \citet{budajetal2015} and \citet{devoreetal2016} present formulations to
calculate $\mbox{<}p_{\rm{a}}\mbox{>}(\Theta)$ for conditions other than $\Theta$=0,
using either Mie theory or Airy functions in the description of the scattering phenomenon. 
Using $\mbox{<}p_{\rm{a}}\mbox{>}(\Theta)$ rather than $p_{\rm{a}}(\theta)$ 
in the radiative transfer problem when evaluating
the contribution from the star at each photon-atmospheric particle 
scattering collision (through integrals like the second 
one in Eq. (\ref{Ixs_eq})) reduces the radiative transfer problem with a 
finite angular size star to the simpler problem of a point-like star. 
}

\begin{figure}
   \centering
   \includegraphics[width=8.cm]{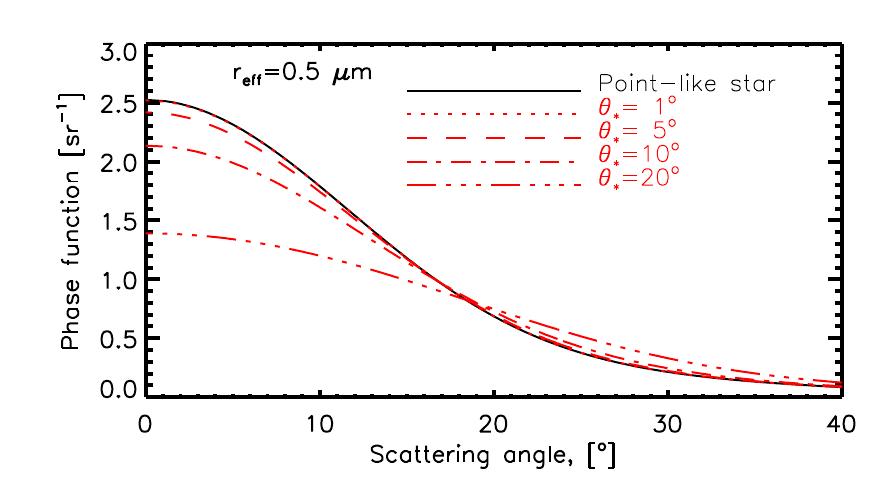}
   \vspace{-0.5cm}
   \includegraphics[width=8.cm]{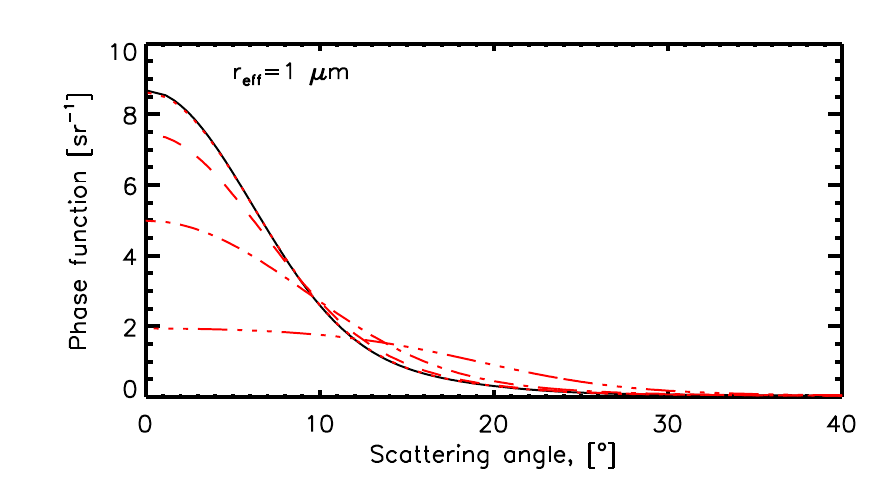}
   \vspace{-0.5cm}
   \includegraphics[width=8.cm]{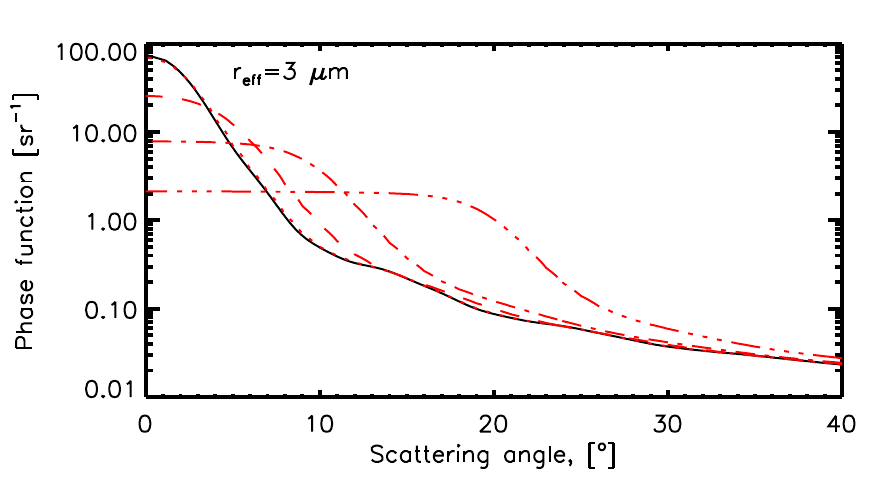}
   \vspace{-0.5cm}   
   \includegraphics[width=8.cm]{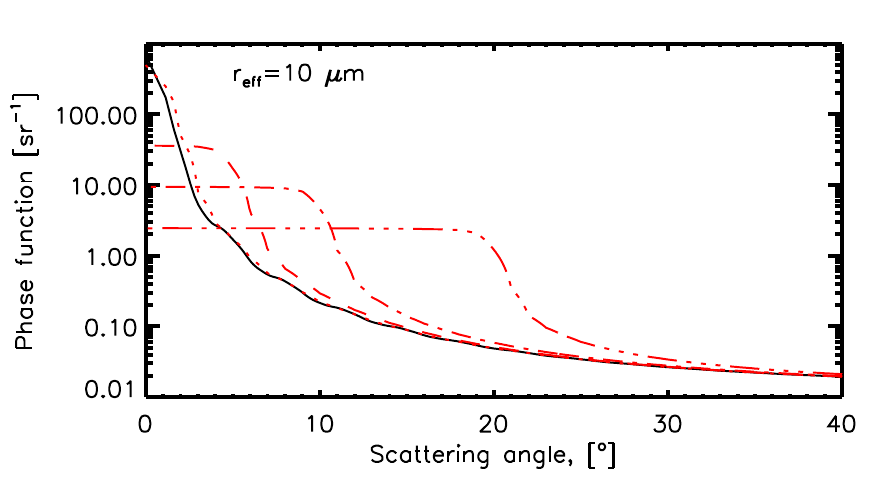}
   \caption{\label{paconv_fig}
   Aerosols scattering phase function in the point-like star limit, $p_{\rm{a}}$($\theta$) 
   (black); and in the finite angular size star approach, $\mbox{<}p_{\rm{a}}\mbox{>}(\Theta)$ (red).
   Each graph considers aerosols of a specific particle size anf FeO composition. 
   Note the different vertical scales for each of the graphs.   
   The various 
   red curves correspond to different star angular radii.
   $p_{\rm{a}}$($\theta$) is calculated from Mie theory. 
   For $\mbox{<}p_{\rm{a}}\mbox{>}(\Theta)$, we convolve $p_{\rm{a}}$($\theta$) with the
   scattering angle for stellar rays from each visible element of the stellar disk.
   Accounting for the 
   finite angular size of the star decreases the
   effective scattering by the aerosols at small scattering angles (typically, angles  $<$$\theta_{\star}$) 
   but increases it at scattering angles somewhat larger. For sufficiently large 
   scattering angles, both $p_{\rm{a}}$($\theta$)
   and $\mbox{<}p_{\rm{a}}\mbox{>}(\Theta)$ merge. 
     }
   \end{figure}

{
Figure (\ref{paconv_fig}) demonstrates $p_{\rm{a}}(\theta)$ and $\mbox{<}p_{\rm{a}}\mbox{>}(\Theta)$
for FeO particles of a few effective radii $r_{\rm{eff}}$ at an effective wavelength
$\lambda_{\rm{eff}}$=0.65 $\mu$m. For $\mbox{<}p_{\rm{a}}\mbox{>}(\Theta)$, we adopted
star angular radii $\theta_{\star}$ of 1, 5, 10, and 20$^{\circ}$.
The comparison of our Fig. (\ref{paconv_fig}) to Fig. 3 of \citet{budajetal2015} (dividing their
results by 4$\pi$) and Fig. 3 of \citet{devoreetal2016} shows that all three 
treatments seem equivalent. 
}

For a spherically symmetric atmosphere, 
the transmittance $t(\mathbf{x,x_{\rm{b}}})$ ($=$$t(\mathbf{x,x_{\rm{a}}}) t(\mathbf{x_{\rm{a}},x_{\rm{b}}})$) 
can be reduced to a function of the 
the minimum distance $r$ from the line of sight to the planet centre
(SA\ref{SAstratified}). 
If $\tau(r)$ is the optical thickness along that chord, 
$t(\mathbf{x,x_{\rm{b}}})$$=$$\exp{(-\tau(r))}$.
And from the definition of optical thickness:
$$
\tau=\int_{\mathbf{x_{\rm{b}}}}^{\mathbf{x}} d\ell{_{\rm{a}}} \gamma(\mathbf{x_a}). 
$$

Following the above, Eq. (\ref{Ixs_eq}) can be rewritten as:
\begin{equation}
I(r)=B_{\star} \exp{(-\tau(r))} 
+ B_{\star} \pi \left(\frac{R_{\star}}{a} \right)^2 
\varpi_{\rm{0,a}}
\tau(r) \exp{(-\tau(r))} \mbox{<}p_{\rm{a}}\mbox{>}(\Theta=0).
\label{Ir_eq}
\end{equation}
Deep into the atmosphere, $\tau(r)$ is large whereas $\exp{(-\tau(r))}$ is small. 
The reverse is true high up in the atmosphere. 

The irradiance measured by the observer from a 
star-to-Earth distance $d$ 
is obtained by integration of  
$I(r)$ over the solid angle subtended by the whole planet-star system:
\begin{equation}
F = \int{I d\Omega} = 2\pi \int I(r) \frac{ r dr}{d^2}.
\end{equation}
According to Eq. (\ref{Ir_eq}), $F$ 
contains the contributions from both unscattered photons and from photons that have 
undergone one scattering collision. 

For the unscattered component: 
\begin{equation}
F_{0}= \frac{2\pi B_{\star}}{d^2} \int_{R_{\rm{0}}}^{R_{\star}} \exp{(-\tau(r))} r dr, 
\end{equation}
where $R_{\rm{0}}$ is a somewhat arbitrary altitude level in the atmosphere such
that $\tau(R_{\rm{0}})$$>>$1. 
Or, if we introduce the planet equivalent cross section to transmission, $\pi r_{\rm{eq}}^2$: 
\begin{equation}
F_{0} = \frac{\pi B_{\star}}{d^2} 
\left( R_{\star}^2 - r_{\rm{eq}}^2 \right)
\end{equation}
For exponential atmospheres, 
\citet{lecavelierdesetangsetal2008} have shown that 
$r_{\rm{eq}}$ (=$R_0$+$h_{\rm{eq}}$) matches the atmospheric level where $\tau(h_{\rm{eq}})$=0.56 that
defines the optical radius of a planet, $R_{\rm{p}}$.

\begin{figure}
   \centering
   \includegraphics[width=9.cm]{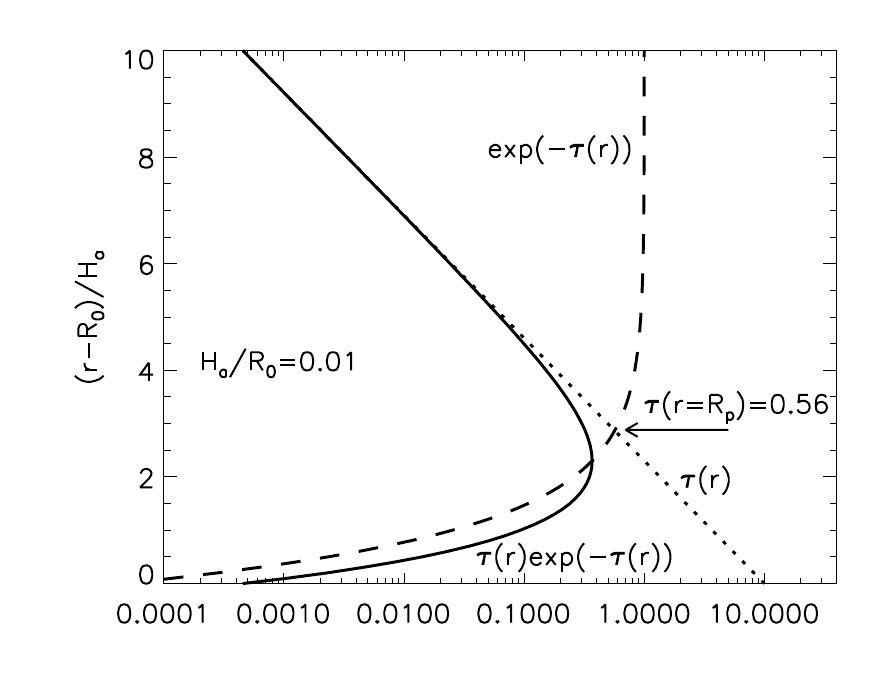}
      \caption{\label{mytau_fig} Most of the contribution to the starlight forward-scattered
      by an exponential atmosphere arises from within a few scale heights above and below 
      the optical radius level, defined by the limb opacity $\tau(r=R_{\rm{p}})$=0.56.
      In the example, $\tau(r)=10\exp(-(r-R_{\rm{0}})/H_{\rm{a}})$. 
      Because, $H_{\rm{a}}/R_{\rm{p}}$$<<$1, the curves for $r \tau(r) \exp{(-\tau(r))}$ and 
      $\tau(r) \exp{(-\tau(r))}$ are nearly undistinguishable in the scale of the plot; only 
      the latter is graphed. 
     }
   \end{figure}

Correspondingly, for the single-scattering component:
\begin{equation}
F_{1}= B_{\star} \pi \left( \frac{R_{\star}}{a} \right)^2 \varpi_{\rm{0,a}} 
\mbox{<}p_{\rm{a}}\mbox{>}(\Theta=0) \frac{2\pi}{d^2}
\int_{R_{\rm{0}}}^{R_{\star}} \tau(r) \exp{(-\tau(r))} r dr.  
\label{F1_eq}
\end{equation}
The occurrence of $\tau(r) \exp{(-\tau(r))}$ suggests that the main contribution to the integral 
arises from $\tau(r)$$\sim$1. 
From SA\ref{SAstratified} and SA\ref{SAexponential}, $\tau(r)=\tau(R_{\rm{0}}) 
\exp{(-(r-R_{\rm{0}})/H_{\rm{a}})}$. 
By virtue of SA\ref{SAscaleheight}, the integral of Eq. (\ref{F1_eq})
converges rapidly in $r$, and it 
is acceptable to take $r$$\approx$$R_{\rm{p}}$ outside of the integration  (Fig. \ref{mytau_fig}). 
The resulting integral is easy to evaluate after realizing that $H_{\rm{a}} d\tau=-\tau dr$, and
results in:
\begin{equation}
 \int_{R_{\rm{0}}}^{R_{\star}} \tau(r) \exp{(-\tau(r))} dr = H_{\rm{a}}
( 1 - \exp(-\tau(R_{\rm{0}}))) \approx H_{\rm{a}}.  
\end{equation}

Introducing these results into $F_1$ and normalizing by the irradiance of the 
unimpeded star at the observer's vantage point:
\begin{equation}
F_{\star} = B_{\star} \pi \left( \frac{R_{\star}}{d} \right)^2,
\end{equation}
leads to:
\begin{equation}
\frac{F_{1}}{F_{\star}} \approx 2\pi \mbox{<}p_{\rm{a}}(\Theta=0)\mbox{>} \varpi_{\rm{0,a}} \frac{H_{\rm{a}}}{R_{\rm{p}}} \left(\frac{R_{\rm{p}}}{a} \right)^2,
\label{F1b_eq}
\end{equation}
which is the analytical expression for starlight forward-scattered towards the observer 
by an exponential atmosphere
under the assumption of single scattering, relative to the net stellar brightness.

{
The orbital distance enters Eq. (\ref{F1b_eq}) in two competing ways.
Since the amount of starlight intercepted by the planet varies as $a^{-2}$, 
close-in planets will in principle appear brighter in reflected starlight. 
However, as the ratio $R_{\star}/a$ becomes larger 
$\mbox{<}p_{\rm{a}}\mbox{>}$($\Theta$=0) samples scattering angles that differ from the
strict forward scattering configuration, and 
$\mbox{<}p_{\rm{a}}\mbox{>}$($\Theta$=0) will typically become much smaller than 
$p_{\rm{a}}$($\theta$=0). The decrease in $\mbox{<}p_{\rm{a}}\mbox{>}$($\Theta$=0) with
respect to $p_{\rm{a}}$($\theta$=0) is more pronounced when $R_{\star}/a$ (or $\theta_{\star}$) 
is large and/or the aerosols exhibit a strong forward scattering peak. }

\section{Multiple scattering for close-in planets. The $\alpha$=180$^{\circ}$ configuration.}
\label{sec:appendixb}

{
\citet{garciamunozetal2017} show that the number of collisions contributing to Titan's
brightness decreases as the star-object-observer phase angles increases (their Fig. 3).
We confirm that for the 
phase curves presented in our Fig. (\ref{phasecurvediversity_fig})  the difference between
the numerical multiple scattering solution at $\alpha$=180$^{\circ}$ 
and the single-scattering estimate based on Eq. (\ref{F1b_eq}) remained below 30\%.
Also, for the phase curves in Fig. (\ref{corot24_thetas_fig}) motivated by the study of CoRoT-24b, 
in both the point-like and finite angular size treatments of the star,
single scattering contributed $\sim$70$\%$ or more to the multiple scattering solution. 
Using the case of CoRoT-24b, we ran a few additional simulations in which we forced the aerosols
to be fully conservative, i.e. $ \varpi_{0,\rm{a}}$=1. In those cases, single scattering
contributed at least $\sim$50$\%$ of the multiple scattering solution at $\alpha$=180$^{\circ}$.
Together, these conclusions suggest that Eq. (\ref{F1b_eq}) generally approximates the
planet-to-star contrast in forward scattering to within a factor of two or better.
}

\section{Backward Monte Carlo calculations with a finite angular size star}
\label{sec:appendixc}

\begin{figure}
   \centering
   \includegraphics[width=9.cm]{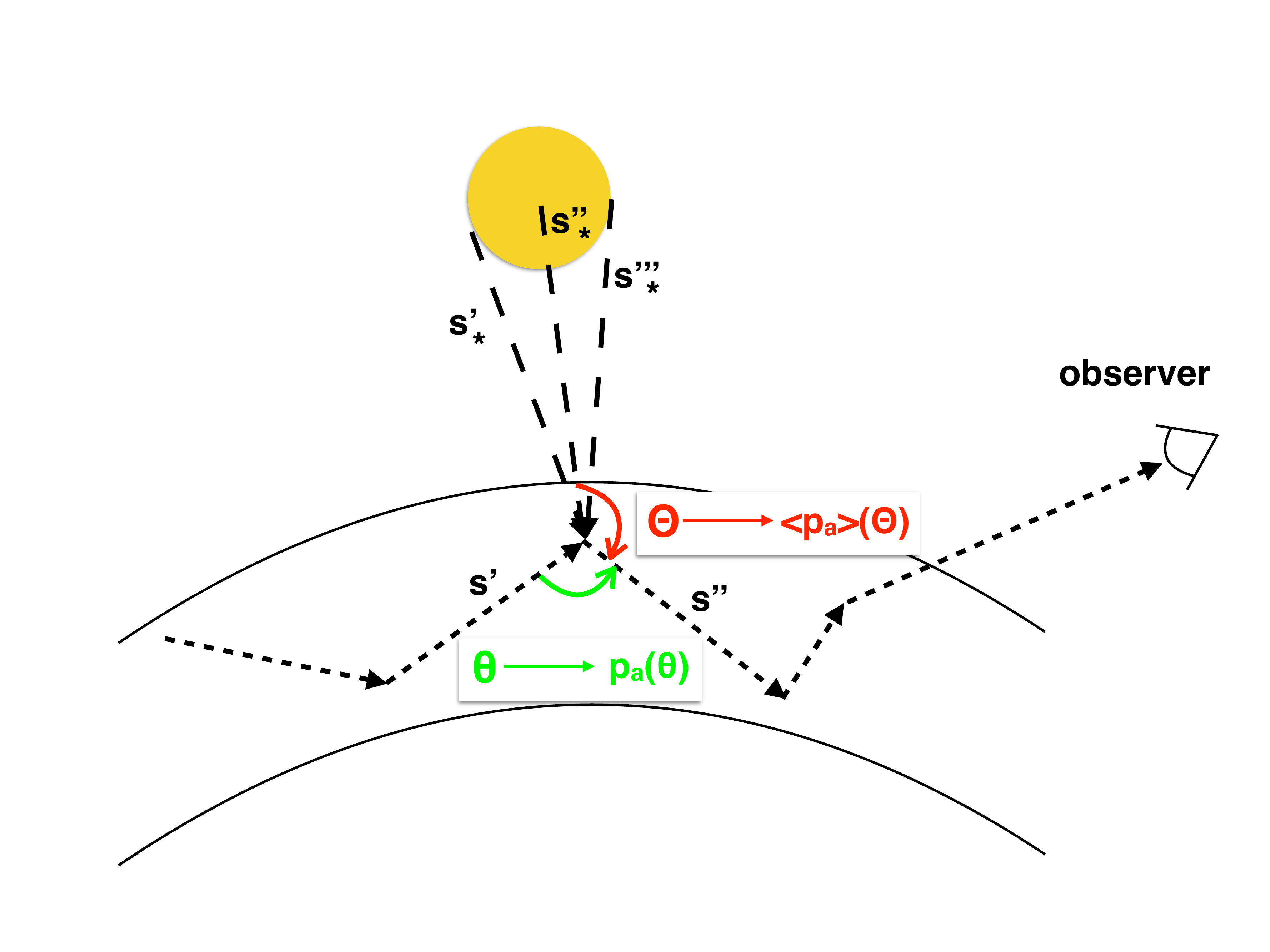}
      \caption{\label{pathsketch_fig} Photon trajectory simulation that shows the 
      definition of $\theta$, $\Theta$ and the corresponding scattering phase functions 
       in the radiative transfer model. 
     }
   \end{figure}

The radiative transfer calculations presented in this work were done with a 
Backward Monte Carlo 
algorithm that has been described elsewhere \citep{garciamunozmills2015}, and used to
investigate the phase curves of exoplanets, Venus and Titan 
\citep{garciamunozisaak2015,garciamunozetal2014,garciamunozetal2017}. 
These previous studies omitted the consideration of the finite angular size of the star,
which can modify the effective scattering geometry at the particle level. 
The effect is likely important for close-in planets at large phase angles because 
atmospheric aerosols can be efficient at scattering in the forward direction. 
Referring to Figs. (\ref{forwardscattering_fig}) and (\ref{pathsketch_fig}), 
a scattering particle sees the incident starlight enter the atmosphere from
a range of directions \textbf{s$_{\star}$} rather than from a single direction. 
This distinction is important since the probability that the incident photon is
re-scattered into another direction 
can be a strong function of the relative angle between the two directions. 
For instance, in the configuration of Fig. (\ref{forwardscattering_fig}), which assumes a perfect
star-planet-observer alignment, 
the angle between \textbf{s$_{\star}$} and \textbf{s} ranges from 0 to $\theta_{\star}$
(the angular size of the star as viewed from the planet). For very close-in
exoplanets, $\theta_{\star}$ can be as large as 20$^{\circ}$. 

Our Backward Monte Carlo algorithm builds the solution to the radiative transfer problem
by tracking simulated photons (or photon packages) from the observer's vantage point through the atmosphere.
At the outset of the simulation, each photon is assigned a `weight' of one. 
As the simulation proceeds and the photon interacts with the medium, its weight
is progressively reduced by amounts that account for the probabilities that the photon
is either absorbed within or escapes from the atmosphere. 
The simulated photon trajectory is terminated when the weight falls below a
user-defined threshold. A tentative photon trajectory is
sketched in Fig. (\ref{pathsketch_fig}). The arrows show the directions for the photon 
displacements, which are the reverse of the directions actually simulated in the Backward
algorithm.
At each photon scattering collision, either within the atmosphere or at the planet surface, 
the algorithm evaluates an integral over solid angle that results into two separate contributions. 
Mathematically, these contributions are expressed as the two terms on the right hand sides of
Eqs. (9)-(10) in \citet{garciamunozmills2015}.

The first contribution accounts for diffuse radiation, i.e. radiation associated with photons that
have had at least one previous collision. 
The relevant scattering phase function to quantify the diffuse radiation 
 re-scattered into the $\mathbf{s''}$ direction is the local 
$p_{\rm{a}}$($\theta$) because the change in photon directions 
from $\mathbf{s'}$ to $\mathbf{s''}$ is independent of the star location.
The corresponding implementation in the algorithm is as described in \citet{garciamunozmills2015}.
The second contribution accounts for stellar photons having 
their first scattering collision. Mathematically 
(Eqs. (9)-(10) in \citet{garciamunozmills2015}, and assuming that all stellar rays are attenuated
by the same amount), this involves an integral over the 
solid angle $\Omega_{\star}$ subtended by the star at the collision location:
$$
\int_{\partial \Omega_{\star}} p_{\rm{a}} (\mathbf{s_{\star}}\cdot\mathbf{s''}) d\Omega(\mathbf{s_{\star}})
$$
which can be pre-calculated and re-written as:
$$
\Omega_{\star} \mbox{<}p_{\rm{a}}\mbox{>}(\Theta),
$$ 
where $\Theta$ is the angle between the $\mathbf{s_{\star}}$ ray passing through the 
star centre ($\mathbf{s''_{\star}}$ in the sketch) and $\mathbf{s''}$. This definition generalizes 
to $\Theta$$>$0 the $\mbox{<}p_{\rm{a}}\mbox{>}(\Theta=0)$ of Eq. (\ref{Int4_eq}). 
Examples of effective scattering phase functions $ \mbox{<}p_{\rm{a}}\mbox{>}(\Theta)$ are shown in 
Fig. (\ref{paconv_fig}).

\bsp	
\label{lastpage}
\end{document}